\documentclass[a4paper,11pt]{article}
\usepackage{pos2}

\usepackage[T1]{fontenc}
\newcommand{\half}{\frac{1}{2}}

\renewcommand{\tilde}{\widetilde} 

\newcommand{\beq}{\begin{equation}}
\newcommand{\eeq}{\end{equation}}
\newcommand{\bea}{\begin{eqnarray}}
\newcommand{\eea}{\end{eqnarray}}
\newcommand{\nn}{\nonumber}

\newcommand{\pL}{\left(} \newcommand{\pR}{\right)} \newcommand{\bL}{\left[} \newcommand{\bR}{\right]}

\usepackage{axodraw2,pix}
\usepackage{graphicx}

\newcommand{\bi}{\begin{itemize}} 
\newcommand{\ei}{\end{itemize}} 
\newcommand{\be}{\begin{equation}} 
\newcommand{\ee}{\end{equation}} 
\newcommand{\pd}[2]{\frac{\partial #1}{\partial #2}}
\newcommand{\pdd}[2]{\frac{\partial^2 #1}{\partial #2^2}}
\newcommand{\dd}[2]{\frac{d #1}{d #2}}
\newcommand{\ddd}[2]{\frac{d^2 #1}{d#2 ^2 }}
\newcommand{\fourth}{\frac{1}{4}}

\newcommand{\Fig}[1]{Fig.~\ref{#1}}

\newcommand{\g}{\lambda_{\phi h}}

\allowdisplaybreaks

\newcommand{\se}[1]{\begin{align}\begin{split} #1 \end{split}\end{align}}

\newcommand{\dmu}{\partial_\mu}
\newcommand{\dnu}{\partial_\nu}
\newcommand{\normal}{\mathbf{\hat{n}}}
\renewcommand\vec{\mathbf}
\newcommand{\vev}[1]{\left\langle #1 \right\rangle}
\newcommand{\tr}{\text{Tr}}

\newcommand{\Tint}[1]{{\hbox{$\sum$}\!\!\!\!\!\!\!\int\,}_{\!\!\!\!\raise-0.9ex\hbox{$\scriptstyle{#1}$}}}
\newcommand{\Tinti}[1]{{{\Sigma}\!\!\!\!\raise0.3ex\hbox{$\int$}_{ \rm ${#1}$}}}
\newcommand{\Tintip}[1]{{{\Sigma'}\!\!\!\!\!\raise0.3ex\hbox{$\int$}_{ \rm ${#1}$}}}

\newcommand{\str}{\ensuremath{\text{str}\,}}

\newcommand{\D}{\mathcal{D}}

\newcommand{\im}{\text{Im}}
\newcommand{\re}{\text{Re}}

\def\Lag{\ensuremath{\mathcal{L}}}

\newcommand\bra[1]{\ensuremath{\langle #1|}}
\newcommand\ket[1]{\ensuremath{|#1 \rangle}}
\newcommand\braket[2]{\ensuremath{\langle #1|#2 \rangle}}

\def\0#1#2{\frac{#1}{#2}}

\title{TASI lectures on Phase Transitions, Baryogenesis, and Gravitational Waves}
\author[a]{Djuna Croon}
\affiliation[a]{Institute for Particle Physics Phenomenology, Department of Physics, Durham University, Durham DH1 3LE, U.K.}
\emailAdd{djuna.l.croon@durham.ac.uk}

\abstract{These lectures, presented at the 2022 TASI summer school, give an introductory overview of first-order phase transitions in the early Universe, baryogenesis, and the resulting gravitational wave phenomenology. 
We introduce thermal field theory via the imaginary time formalism, and comment on the pitfalls of 1-loop calculations and alternative approaches. Then, we discuss how to calculate the false vacuum decay rate in first order phase transitions, of which we give various examples in theories beyond the Standard Model. Baryogenesis is presented via the Sakharov conditions, and how they are met in important classes of examples. Finally, we explore gravitational waves from the early Universe, first reviewing the basics of gravitational wave generation and then focusing on the specific example of first order phase transitions.}

\begin{document}
\tableofcontents
\maketitle

\tableofcontents

\section{Introduction}
\subsection{These lecture notes}
These notes are an (only slightly) extended version of four 75-minute lectures I gave at the TASI 2022 summer school. They cover aspects of early Universe physics, focused in particular on thermal phase transitions, their connection to models of baryogenesis, and their potential gravitational wave phenomenology. Clearly, each of these topics is deserving of its own lecture course, over a term or over several terms. As such these notes should only be considered a fleeting introduction to the topics in question. I have added many suggestions for further reading throughout the text. 

The topics discussed in these notes are areas of high research activity. In this lecture series, I have prioritised developing an intuitive picture of the dynamics in the early Universe over doing full justice to the latest technical developments. 
Any reader interested in pursuing research at the cutting edge is recommended to consult additional educational texts and research papers. 
I would like to highlight two areas in particular in which these notes omit comprehensive detail. 
Firstly, these notes give a very brief introduction to thermal field theory, focusing mainly on the 4-D one-loop imaginary time formalism.
While a helpful starting point, it is known that this formalism of finite temperature field theory is plagued by large theoretical uncertainty \cite{Croon:2020cgk}. Though the formalism is most well-known and used, these flaws cannot be ignored in phenomenological calculations. Dimensional reduction \cite{Ginsparg:1980ef} can be applied to thermal field theory and allows for more accurate calculations in most cases. 
I have included a short description of the spirit of such calculations, but do not attempt to give a complete description of the ongoing research in this area. Pedagogical texts on this topic (including worked examples as well as code packages) exist in the literature, and I have provided examples in the references. 

Moreover, I have not provided a detailed overview on the dynamics of the primordial plasma after bubble nucleation, including the calculation of the bubble wall speed and the full computation of the baryon asymmetry generated across the bubble wall. These notes focus instead on general aspects of baryogenesis and broad classes of models in the literature. The reader is encouraged to consult the many research papers and excellent topical reviews cited in the section. 
A thorough overview of the current status of the technical challenges of these calculations was provided recently in \cite{Asadi:2022njl}.

I hope these lecture notes serve as a valuable first introduction to the captivating subjects of phase transitions and baryogenesis, as well as the ongoing exciting research endeavors dedicated to exploring them. This work connects to fundamental open problems in physics today, and has phenomenological prospects for the near future. The advent of gravitational wave astronomy has delivered a resurgence of interest in these topics. As we will see, there are still plenty of important research questions to answer before potential signals can be reliably understood in terms of early Universe physics. 

\subsection{What do (we think) we know about the early Universe?}
Before we dive into the specialised topics of baryogenesis, cosmological phase transitions, and their (gravitational wave) phenomenology, let us first pause and consider the things we think we already know about the early Universe. Assuming the Universe was radiation dominated for most of the period of interest, 
\se{ H = \frac{1}{2 t} &= \sqrt{\frac{8 \pi}{3} \frac{\rho_{\rm rad}}{M_p^2}}
= \sqrt{\frac{8 \pi}{3} \frac{\pi^2}{30} g_* \frac{T^4}{M_p^2}}
= \sqrt{ \frac{8\pi^3}{90} g_* }\frac{T^2}{M_p}
}
such that time and temperature can be straightforwardly related. Then, we might draw an approximate timeline of the early Universe, assuming the SM degrees of freedom (note that this assumes radiation domination, and therefore stops being accurate at matter-radiation equality):

{\centering
\includegraphics[width=.9\textwidth]{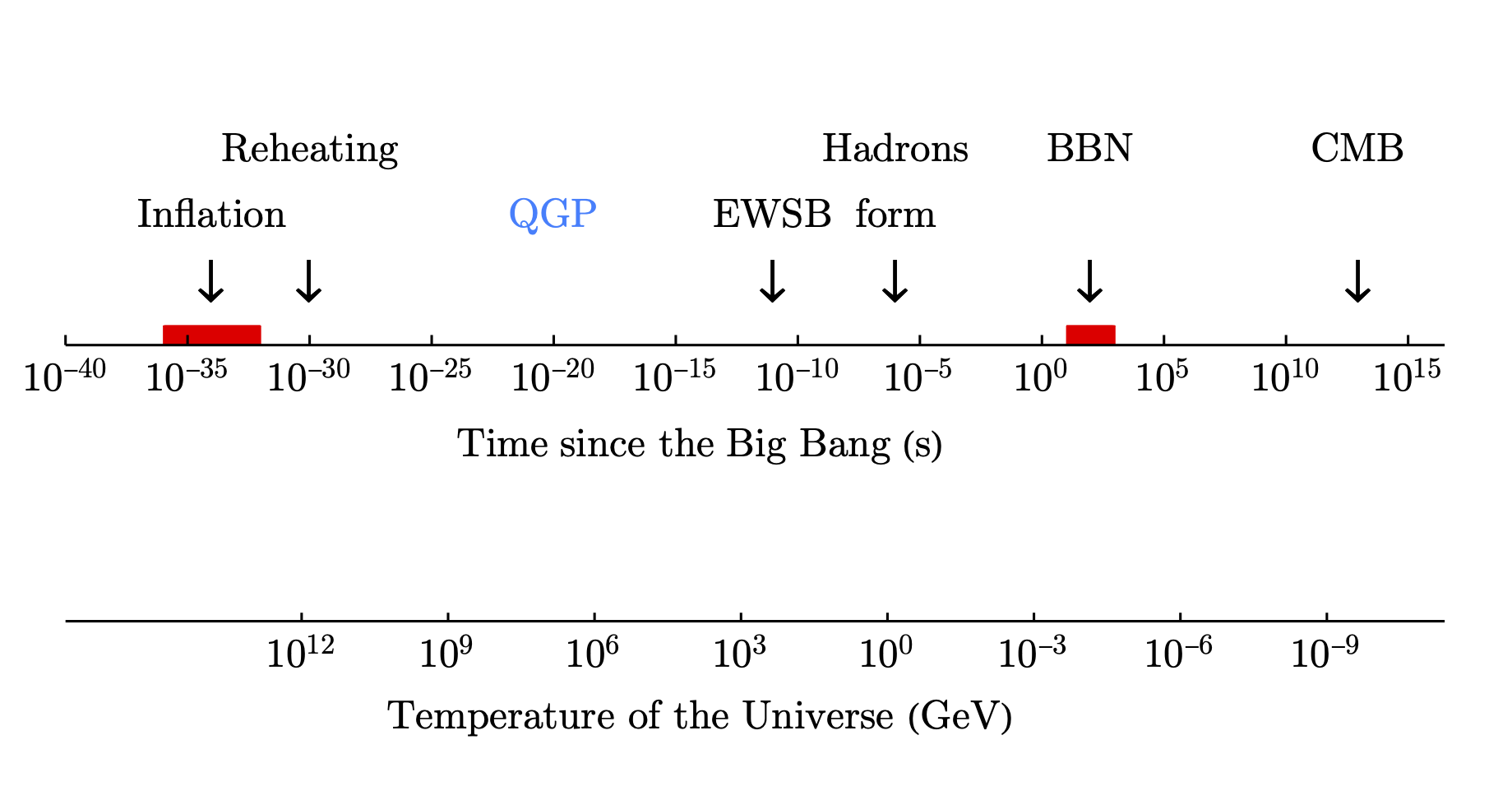}
\par} 

The first event most cosmologists (though not all!) will place on the cosmic timeline is that of cosmic inflation, the exponential expansion of space which provides the initial conditions for structure formation (as well as resolving a number of problems with the Big Bang model it replaced). Unfortunately, neither the theory nor any observations to date pinpoint a scale for cosmic inflation, and in principle it could have taken place at temperatures anywhere between a couple of orders of magnitude below the Planck scale $M_p = 1.2 \times 10^{19}$ GeV and Big Bang Nucleosynthesis (BBN), occuring at $\mathcal{O}(1) $ MeV, the very end of our early Universe timeline.

After inflation, the particle(s) responsible for the process decay to other particles, including those of the Standard Model (SM). If the resulting energetic plasma has a high enough temperature, quantum chromodynamics (QCD) is not confined and quarks and gluons were free.
In the SM, this confinement happens at around $1$ GeV in a phase transition associated with the breaking of the (global) chiral symmetry in the quark sector. This is preceded by the breaking of the electroweak gauge symmetry of the SM, which happens when the Higgs boson obtained its vacuum expectation value (VEV) at around $ 100$ GeV. The latter process gave mass to the fundamental particles, the former accounts for most of the mass of the composite particles. 

During BBN, light elements are formed from the hadrons: hydrogen and helium, deuterium, helium-3, helium-4, and lithium-7. The primordial plasma is still ionized until recombination, many decades later. Then, the electrons bind up into nuclei, and the plasma finally becomes transparent to photons. These first photons constitute the Cosmic Microwave Background (CMB). Note that the conversion timeline above breaks down before this time, as matter-radiation equality happened before recombination. 

Several important events cannot yet be placed on the cosmic timeline. The thermal evolution (including a possible mass mechanism) of the sector containing dark matter is an example. Even the cosmic history of luminous matter cannot be completely determined without answering what biased it over antimatter in the early Universe. 
Part of the problem is that our earliest experimental information comes from the CMB. We simply cannot look back any further with photons, because before recombination the plasma was opaque to photons. Photons released in any big event in the early Universe are therefore reabsorbed, and with them the precious secrets they hold. 

The same is not true for gravitons. Gravitational radiation couples much more weakly then photons, and a ``loud'' source of such radiation in the early Universe can in principle be detected by experiments today. Some of the missing events in our cosmic origins story have been connected to such sources. Therefore, a detection of primordial gravitational waves would revolutionise our understanding of the Universe. 

In these lecture notes, we will delve into the theory of thermal phase transitions, to explore possible answers to these questions. We will also study their phenomenology, with a particular eye to upcoming gravitational wave experiments. Other sources of gravitational waves from the early Universe have been suggested, including inflation and (pre-)heating, scalar field fragmentation, and cosmic defects. All of these topics are growing areas of research, and characterising the phenomenology is a major task ahead of the first generation of space-based interferometers.

As we focus on thermal phase transitions here, we will start with a very brief introduction to the basics of thermal field theory. Then we will study tunnelling and the dynamics of bubble nucleation. We will finish by studying gravitational waves, with some general aspects as well as the spectra of phase transitions in particular. 

\section{Field theory at finite temperature}
To start with, we need to learn more about quantum field theory at nonzero temperature. Some of our intuition about zero-temperature field theory is applicable to finite temperature, but there are some important differences to consider.
Our most important object of interest is the scalar potential, which gets thermal corrections from all particles which couple to it. 
Because we are mostly interested in equilibrium quantities, we will primarily use the imaginary time formalism. At the end, we will address different techniques to deal with quantum corrections. 

The overview in this section is largely based on the textbook \cite{Kapusta:2006pm} and the lecture notes \cite{Quiros:1999jp,Laine:2016hma}.
Since this is a brief overview only, if you are interested in learning more about finite temperature field theory you are encouraged to consider these works (or others such as \cite{das1997finite,Bellac:2011kqa}), as well as the research articles cited in the text. 

\subsection{Imaginary time formalism}
\subsubsection{The partition function}

The object we will be basing our thermal field theory on is the grand canonical partition function (the trace of the density matrix)
\se{
\mathcal{Z}(T) \;\equiv\; \tr [e^{-\beta ( \hat H - \mu \hat N )}] \;, \quad \text{where} \quad \beta \;\equiv\; \frac{1}{T},
\label{eq:partitionfunction}
}
where $\hat{H}$ is the Hamiltonian operator of the theory, $\mu$ is the chemical potential, $\hat N $ is the particle number, the trace is taken over the full Hilbert space, and where we use units $k_B = 1$ (we will generally use particle physics units in these lectures, unless otherwise indicated). Here and in the following we will ignore the role of the chemical potential, usually a good approximation in early Universe physics. 
 One way in which you can see this is that chemical potentials correspond to conserved particle numbers. In the early Universe, even when baryon number is conserved (after electroweak symmetry breaking at $T \sim 100 $ GeV), $\mu_{B}/T \ll 1$. 

The grand canonical partition function can be used to determine all the thermodynamic properties of our early Universe system. It can also be written as a transition amplitude, 
\se{
\mathcal{Z}(T) 
&= \sum_{a} \int d\phi_{a} \bra{\phi_{a}} e^{-\beta \hat H} \ket{\phi_{a}}
}
where all states $a$ are summed over and the eigenbasis of the Schroedinger picture field operator $\ket{\phi_{a}}$ is used. It satisfies:
\se{  \hat\phi(x,0) \ket{\phi} = \phi(x) \ket{\phi} \quad \quad  &\\ 
\braket{\phi_{a}}{\phi_{b}} = \prod_{x} \delta (\phi_{a}(x)- \phi_{b}(x)) \quad \quad &\text{orthogonality} \\ 
\int d \phi \ket{\phi} \bra{\phi} = 1  \quad \quad &\text{completeness}
}
It also has a conjugate momentum $\hat\pi (x,0)$, which satisfies 
\se{  \hat\pi(x,0) \ket{\pi} = \pi(x) \ket{\pi} \quad \quad  &\\ 
\braket{\pi_{a}}{\pi_{b}} = \prod_{x} \delta (\pi_{a}(x)- \pi_{b}(x)) \quad \quad &\text{orthogonality} \\ 
\int  \frac{d\pi}{2\pi} \ket{\pi} \bra{\pi} = 1  \quad \quad &\text{completeness}
}
The overlap is given by,
\se{ 
\braket{\phi}{\pi} = e^{i \int d^{3}x \pi (\vec{x}) \phi (\vec{x})}.
}
Now we can insert the completeness relations for $\pi$ and $\phi$ $N$ times into our expression for $\mathcal{Z}(T)$. Simultaneously, we can split $e^{-\beta \hat{H}}$ into $N$ pieces $e^{-\beta \hat{H}/N} \equiv e^{-\epsilon \hat{H}} $, 
\se{
 \bra{\phi_{a}} e^{-\beta \hat H} \ket{\phi_{a}} =& \int \prod_{i=1}^{N} \frac{d\pi_{i}}{2\pi} d \phi_{i} \braket{\phi_{a}}{\pi_{N}}
 \bra{\pi_{N}} e^{-\epsilon \hat{H}} \ket{\phi_{N}}
 \braket{\phi_{N}}{\pi_{N-1}} 
  \bra{\pi_{N-1}} e^{-\epsilon \hat{H}} \ket{\phi_{N-1}} 
  \\ & \times... \times
   \bra{\pi_{1}} e^{-\epsilon \hat{H}} \ket{\phi_{1}}
   \braket{\phi_{1}}{\phi_{a}}
}
To simplify this, we note that
\begin{enumerate}
\item $\braket{\phi_{1}}{\phi_{a}} = \delta (\phi_{1}-\phi_{a})$
\item $\braket{\phi_{i+1}}{\pi_{i}} = e^{i \int d^{3}x \pi_{i}(x) \phi_{i+1}(x)} $
\item $ \bra{\pi_{i}} e^{-\epsilon \hat{H}} \ket{\phi_{i}} = \bra{\pi_{i}} e^{-\epsilon H(\pi_{i},\phi_{i}) + \mathcal{O}(\epsilon^{2})} \ket{\phi_{i}} 
= \braket{\pi_{i}}{\phi_{i}} e^{-\epsilon H_{i}} 
= e^{-\epsilon H_{i}} e^{-i \int d^{3}x \pi_{i}(x) \phi_{i}(x)}
$
\end{enumerate}
where we have assumed that $N $ is a large number, such that $\epsilon$ is small. Then we have 
\se{ 
\mathcal{Z}(T) = \sum_{a}  \int \prod_{i=1}^{N} \frac{d\pi_{i}}{2\pi} d \phi_{i} \exp\left\{ 
- \epsilon \left( 
\sum_{j=1}^{N} \int d^{3}x \mathcal{H} (\pi_{j},\phi_{j}) - i \pi_{j} \frac{ \phi_{j+1}-\phi_{j}}{\epsilon }
\right)
\right\}
}
where $\mathcal{H}$ is the Hamiltonian density, which integrates to the Hamiltonian. Taking the continuum limit $N\to \infty$, this becomes
\se{ 
\mathcal{Z}(T) = \int \D \pi \int_{\phi(0,x) = \phi(\beta,x)} \D \phi \, \, \exp\left\{ 
\int_{0}^{\beta} d \tau \int d^{3} x \left[ i \pi \pd{\phi}{\tau}
- \mathcal{H} (\pi, \phi)
 \right]
\right\}
}
where all discrete $f_{i}$'s are collected into continuous functions of some variable $\tau$ (i.e. $f(\tau)$), which is restricted to lie in $ \tau \in \{0, \epsilon N \} = \{ 0,\beta \}$.
To evaluate this, we can for example take the Hamiltonian density of a scalar field, 
$$ \mathcal{H} = \pi \pd{\phi}{t} - \mathcal{L} = \half \pi^{2} + \half (\nabla \phi)^{2} + V(\phi) $$
and note that the $\pi$ integral has the form of a Gaussian integral. 

\begin{center}
\fbox{
\begin{minipage}{35em}
Remember, Gaussian integrals are integrals of the form 
$$ \int_{-\infty}^{\infty}  e^{-a x^{2}/2+J x}\,dx= \sqrt{\frac{2\pi}{a}} e^{J^{2}/2a} $$ 
or, for multi-dimensional integrals (of which functional integrals are a special case), 
$$  \int_{-\infty}^{\infty}  e^{- \vec{x}^{\dagger} A \vec{x}/2+\vec{J}^{\dagger}\cdot \vec{x}}\,d\vec{x}= \sqrt{\frac{(2\pi)^{n}}{\det A}} e^{\vec{J}^{\dagger}A^{-1}\vec{J}/2}  $$
see standard textbooks.
\end{minipage}
}
\end{center}
Evaluating the Gaussian $\pi$ integrals, we end up with 
\se{ 
\mathcal{Z}(T) =& \mathcal{N}  \int_{\phi(0,x) = \phi(\beta,x)} \D \phi \, \, \exp\left\{ 
\int_{0}^{\beta} d \tau \int d^{3} x 
\left[ - \half \left(\pd{\phi}{\tau} \right)^{2} - \half (\nabla \phi)^{2} - V(\phi)
\right]
\right\}\\ 
=&  \mathcal{N}  \int_{\phi(0,x) = \phi(\beta,x)} \D \phi \, \, \exp\left\{ 
\int_{0}^{\beta} d \tau \int d^{3} x 
\mathcal \Lag_{E}
\right\}
}
where $\mathcal{N}$ is a field-independent normalisation, which is not relevant: multiplication of $\mathcal{Z}$ by any constant does not change the thermodynamics. 
Thus, we find that we can calculate $\mathcal{Z}(T)$ by wick rotating the usual path integral, identifying $\mathcal{L}_{E} = - \mathcal{L}_{M} |_{\tau = i t} $, restricting the Euclidean time to $\tau \in \{0,\beta \}$, and requiring periodicity of the fields: 
\se{ \mathcal{Z}(T) &= \int_{\phi(0,\vec{x}) = \phi(\beta  ,\vec{x}) } \D \phi \exp \pL - S_E \pR ; \quad
\text{with} \\
 S_E &= \int_0^{\beta  } d\tau \int_V \mathcal{L}_E
 \label{eq:partitionpath}
}
with the subscript $E$ denotes Euclidean signature. Though we used scalar fields to derive it, this recipe holds more generally -- the most important modification for spin-$\half$ fields is that they are anti-periodic instead of periodic as a result of spin-statistics.

A direct consequence of this periodicity is that there is a discrete set of Fourier modes of the theory, called the Matsubara modes \cite{Matsubara:1955ws}. A scalar field can for example be expanded as:
\se{\phi(x) &= T \sum_{\omega_n} \int_{\vec{k}} e^{i k \cdot x} \phi(k) 
= T \sum_{\omega_n} \int_{\vec{k}} e^{i ( \omega_{k} \tau - \vec{k}\cdot\vec{x})} \phi(k) 
\\
\omega_n &= 2n \pi T 
\label{eq:matsubarascalar}
}
where $k = (\omega_n,\vec{k})$. 
It is easy to show that this satisfies the periodic boundary condition in \eqref{eq:partitionpath}.
You may note that there is another choice of Matsubara modes which satisfies the periodic boundary conditions, though in this case they are anti-periodic ($\psi(0,\vec{x}) = -  \psi(\beta,\vec{x})  $). This choice corresponds to fermionic fields, for which the Matsubara frequencies are odd: $\omega_n = (2n+1) \pi T$.  Gauge bosons have even Matsubara frequencies, like scalars. 

Let us first proceed with a real, free scalar field, with Lagrangian 
\se{ \mathcal{L}_E = \half \dmu \phi \dmu \phi +\half m^2 \phi^2 }
The field $\phi$ has expansion \eqref{eq:matsubarascalar}.
In principle, $n$ is an integer which can take both positive and negative values, but because 
\se{\label{eq:realitycond}
\phi(\omega_n, \vec{k})^* = \phi(-\omega_n, -\vec{k})
}
which is necessary such that $\phi(x)$ is real, only the non-negative integers end up being important in this particular case. This also implies of course that the zero mode has to be real. 

Then we can write the partition function in terms of the mode expansion,\footnote{This stems from the fact that the quadratic form is 
\se{ \int_0^{\beta} \! {\rm d}\tau \! \int_\vec{x} \, 
 \phi^{ }_1(\tau,\vec{x}) \, 
 \phi^{ }_2(\tau,\vec{x}) &= 
 T \int_0^{\beta} \! {\rm d}\tau \! \int_\vec{x} \, 
  \sum_{\omega^{ }_n}
 \sum_{\omega^{ }_m}
\int_{\vec{k}}
\int_{\vec{p}}
 \phi^{ }_1(\omega^{ }_m,\vec{p}) \, 
 \phi^{ }_2(\omega^{ }_n,\vec{k}) e^{i (\vec{k}+\vec{p}) \cdot \vec{x}  }
\\ &= T 
  \sum_{\omega^{ }_n}
 \sum_{\omega^{ }_m}
\int_{\vec{k}}
\int_{\vec{p}}
 \phi^{ }_1(\omega^{ }_m,\vec{p}) \, 
 \phi^{ }_2(\omega^{ }_n,\vec{k}) 
\delta_{m,-n} \delta(\vec{k}-\vec{p})
  \\ &=T 
  \sum_{\omega^{ }_n}
\int_{\vec{k}}
 \phi^{ }_1(-\omega^{ }_n,-\vec{k}) \, 
 \phi^{ }_2(\omega^{ }_n,\vec{k})
 \notag}
 in combination with \eqref{eq:realitycond}. This also gives the sign of the $ \vec{k}^{2} + \omega_{n}^{2} $ terms. 
}
\se{\label{eq:Zmodeexp}
\mathcal{Z}(T) &= \int \D \phi 
\exp \pL - T \sum_{\omega_n} \int_{\vec{k}} \half \pL \vec{k}^2 + \omega_n^2 + m^2 \pR |\phi(k)|^2 \pR 
}
This almost looks like our favourite kind of integral: our expression \eqref{eq:Zmodeexp} is a Gaussian integral except for the integration over $\vec{k}$ in the exponent. It turns out that it is easier to compute this expression at discrete volume $V$, and then take the continuum limit at the end.
This momentum is discrete: $  k_i = 2 \pi n_i /L_i $ where $i$ runs over the spatial dimensions, and integrals become products or sums.

We also need to change variables in the integration measure $ \D \phi(x)$ to use our Matsubara mode expansion. 
This change of variables is captured in a Jacobian. We have, 
\se{ \D \phi(x) = \underbrace{\left| \det \frac{\delta \phi(x)}{\delta \phi (k)} \right|}_{\rm Jacobian} \overbrace{d \phi (0,k)}^{\text{zero mode}} \underbrace{\prod_{n\geq 1} d \phi (\omega_n, k)}_{\text{non-zero modes}}
}
such that our expression becomes, writing $\omega^2 \equiv \vec{k}^2 +m^2 $ (suppressing a possible subscript $k$ for convenience, but we should remember that $\omega$ depends on $\vec{k}$) 
\se{\label{eq:ZTfuncint}
\mathcal{Z}(T) &= \int \D \phi 
\exp \pL - \half T \sum_{\omega_n} \frac{1}{V} \sum_{\vec{k}} \pL \omega^2 + \omega_n^2 \pR |\phi(\omega_n, \vec{k})|^2 \pR
\\ &= \int \D \phi \prod_{\vec{k}}
\exp \pL - \frac{T}{2 V} \sum_{n=-\infty}^{\infty}   \pL \omega^2 + \omega_n^2  \pR |\phi(\omega_n, \vec{k})|^2 \pR 
\\ &= \int \D \phi \prod_{\vec{k}}
\exp \pL - \frac{T \omega^2}{2V} |\phi(0, \vec{k})|^2 - \frac{T}{V} \sum_{n=1}^{\infty}   \pL \omega^2 + \omega_n^2 \pR |\phi(\omega_n, \vec{k})|^2 \pR 
\\ &= \prod_{\vec{k}} \left| \det \frac{\delta \phi(x)}{\delta \phi (k)} \right| \int  d \phi (0,\vec{k}) \prod_{n\geq 1} d \phi (\omega_n, \vec{k}) \exp \pL - \frac{T \omega^2}{2V}  |\phi(0, \vec{k})|^2 - \frac{T}{V} \sum_{n=1}^{\infty}   \pL \omega^2 + \omega_n^2  \pR |\phi(\omega_n, \vec{k})|^2 \pR 
\\ &= \prod_{\vec{k}}  \left| \det \frac{\delta \phi(x)}{\delta \phi (k)} \right| 
\underbrace{\sqrt{\frac{2 \pi V}{ T \omega^2}}}_{\text{zero mode}}
\,\times\,
\underbrace{\prod_{n\geq 1} \frac{ \pi V }{T  \pL \omega^2 + \omega_n^2  \pR}}_{\text{nonzero modes}}
}
where we carried out the Gaussian integral in the last line.

Now the determinant term can be found by matching the two computations: the computation using Matsubara modes $\phi(k) $ and the one using the ``full theory'' with $\phi(x)$. I will not do this computation here, but simply quote the result (see e.g. \cite{Laine:2016hma} for more details)
\se{ \left| \det \frac{\delta \phi(x)}{\delta \phi (k)} \right| = \frac{T}{2 \pi} \sqrt{\frac{2 \pi T}{V} } \prod_{n=1}^\infty \frac{T \omega_n^2}{\pi V} }
such that we have 
\se{ \mathcal{Z}(T) &= \prod_{\vec{k}}
\frac{T}{2 \pi} \sqrt{\frac{2 \pi T}{V} }
\sqrt{\frac{2 \pi V}{ T \omega^2}}
\prod_{n\geq 1}\frac{\omega_n^2}{ \omega^2 + \omega_n^2} 
\\ &= \prod_{\vec{k}}
\frac{T}{\omega}  
\prod_{n\geq 1}\frac{\omega_n^2}{ \omega^2 + \omega_n^2} 
\\ &=  
\prod_{\vec{k}} T \prod_{n=-\infty}^{\infty} (\omega_n^2 + \omega^2)^{-\frac{1}{2}} \prod_{n'=-\infty}^{\infty} (\omega_n^2)^{\frac{1}{2}}
\\ &= \exp
\left\{ \sum_{\vec{k}} \bL  \log T - \half \sum_n \log  (\omega_n^2 + \omega^2) + \sum_{n'} \log \omega_n \bR \right\} 
\label{eq:zofT1}
}
where $n'$ implies that the zero mode has been omitted from the sum. 

Alternatively, we can also write the products like expressions we know: 
\se{
 \prod_{n=1}^{\infty} 
 \frac{\omega_n^2}{\omega_n^2 + \omega^2}  &= 
 \frac{1}{\prod_{n=1}^{\infty} 
 \Bigl[ 1 + \frac{(\omega/2\pi T)^2}{n^2} \Bigr]} 
 \\ &= 
 \frac{\omega / 2 T}{ \sinh \left( \omega / 2 T \right) } 
 \\ &=  \frac{ \omega / T}{ e^{\omega/2 T} - e^{-\omega/2 T} } 
 \\ &= \frac{\omega}{T}  \frac{ e^{-\omega/ 2T}}{ 1 - e^{-\omega/ T} }
  \\ &= \frac{ \omega}{T} \exp\biggl\{
 -\frac{1}{T} 
 \biggl[ 
  \frac{\omega}{2} + T \ln\pL 1 - e^{-\omega/T}\pR
 \biggr] 
 \biggr\}
 \label{eq:omegasum}
}
using the identity
$$  \frac{\sinh\pi x}{\pi x} = \prod_{n=1}^{\infty}
 \bL 1 + \frac{x^2}{n^2} \bR = \frac{e^{\pi x} - e^{- \pi x}}{2 \pi x }. $$

Now we are finally ready to take the infinite volume limit:
$$ \frac{1}{V}\sum_{\vec{k}} \to \int \frac{d^3 k}{(2 \pi)^3}$$
to obtain, using \eqref{eq:zofT1},
\se{ 
\mathcal{Z}(T)
&= \exp
\left\{ V \int \frac{d^3 k}{(2 \pi)^3} \bL  \log T - \half \sum_n \log  (\omega_n^2 + \omega^2) + \sum_{n'} \log \omega_n \bR \right\} .
\label{eq:ZTexp1}
}
Alternatively, using \eqref{eq:omegasum}, 
\se{ 
    \mathcal{Z}(T)
    &= \prod_{\vec{k}}
    \frac{T}{\omega}  
    \prod_{n\geq 1}\frac{\omega_n^2}{ \omega^2 + \omega_n^2} 
    \\&= \prod_{\vec{k}} \exp\biggl[ 
   -\frac{1}{T}\pL \frac{\omega^{ }}{2} + T \ln \pL 1 - e^{-{\omega }/T}\pR 
   \pR \biggr] 
   \\ &= \exp\biggl[ -\frac{V}{T} \int \frac{d^3 k}{(2 \pi)^3} 
   \pL \frac{\omega^{ }}{2} + T \ln \pL 1 - e^{-{\omega }/T}\pR 
   \pR \biggr] 
   \label{eq:ZTfinal}
}
where the last expression is again in the infinite volume limit.\footnote{ 
Note that the derivative of the argument of the exponential is proportional to the Bose-Einstein distribution function
\se{ n_{\rm B} (\omega) = \frac{1}{e^{\omega/T}-1}}
since $ d\ln \pL 1 - e^{-x}\pR  /dx= (e^{x}-1)^{-1}.$
As we will see, this implies that the Bose-Einstein distribution function appears in the average energy and the entropy. }

\subsubsection{Scalar field potential}
From the partition function one can derive the following quantities
\begin{itemize}
    \item The free energy \se{F &=  - T \ln\mathcal{Z};}
    \item The average (internal) energy \se{
     E &=  \frac{1}{\mathcal{Z}} \tr [\hat H e^{-\beta \hat H}] 
     \label{eq:avenergy}
    }
    where $\hat{H}$ is again the Hamiltonian operator;
    \item The entropy \se{ S &=-\frac{\partial F}{\partial T}= 
     \ln\mathcal{Z} + \frac{1}{T \mathcal{Z}} \tr [\hat H e^{-\beta \hat H}]
     = - \frac{F}{T} + \frac{E}{T}.
    }
\end{itemize}
We are especially interested in the free energy density, since we want to know how finite temperature affects our scalar field potential. 
The free energy density for our scalar field is given by (using \eqref{eq:ZTfinal})
\se{\lim_{V \to \infty}\frac{F}{V} &=   \int \frac{d^3 k}{(2 \pi)^3} 
   \pL \frac{\omega^{ }}{2} + T \ln \pL 1 - e^{-{\omega }/T}\pR 
   \pR  
\\ &= \int \frac{d^3 k}{(2 \pi)^3} 
   \pL \frac{\sqrt{m^2 + \vec{k}^2}^{ }}{2} + T \ln \pL 1 - e^{-{\sqrt{m^2 + \vec{k}^2} }/T}\pR 
   \pR
\\ &\equiv J_0 (m) + \tilde{J}_B (m,T)
} 
It is immediately obvious that the first term in this integral is UV-divergent. However, it is also independent of temperature, and therefore not our main concern here. The second term is given by\footnote{Usually $J_{i}$ are defined without the factor  $T^4/2\pi^2$, such that $ \tilde{J}_{i} =T^4/2\pi^2 J_{i} $. }
\se{
\tilde{J}_B (m,T) &= T \int \frac{d^3 k}{(2 \pi)^3} \ln \pL 1 - e^{-{\sqrt{m^2 + \vec{k}^2} }/T}\pR
\\ &= \frac{T}{2\pi^2} \int d |\vec{k}| \; \vec{k}^2  \ln \pL 1 - e^{-{\sqrt{m^2 + \vec{k}^2} }/T}\pR
\\ &= \frac{T^4}{2\pi^2} \int d x \; x^2  \ln \pL 1 - e^{-{\sqrt{(m/T)^2 + x^2}}}\pR
\label{eq:JB}
}
This term is not by itself illuminating, but we can get some more insight by doing high and low temperature expansions. For example, in the high-temperature case $m\ll T$,\footnote{In practise, this expansion is numerically reasonably accurate to about $m/T =2$ at quartic order.} 
\se{
\tilde{J}_B (m,T) &= \frac{T^4}{2\pi^2} \pL -\frac{\pi ^4}{45}  + \frac{\pi ^2 m^2}{12 T^2}
 - \frac{\pi |m|^3}{6 T^3}
 - \frac{m^4}{32 T^4}
 \biggl[ 
  \ln\biggl( \frac{ e^{\gamma_E}}{16\pi^2 } \frac{m^2}{T^2} \biggr) - \frac{3}{2}
 \biggr] + \mathcal{O}\pL \frac{m^6}{T^6} \pR
\pR .
\label{eq:JBhighT}
}
where $\gamma_E \approx 0.5772$ is the Euler-Mascheroni constant.
Note that this high-temperature expansion is not a proper Taylor expansion, because the logarithm has a branch cut nearby $ m/T = 0 $, so strictly mathematically speaking the expansion around this point has zero radius of convergence (hence also the absolute value around the cubic term; it should be interpreted as $ |m| = \sqrt{m^2}$). Nevertheless, it helps us build some physical intuition.~\footnote{For a derivation of this expansion, see Ref.~\cite{Laine:2016hma} or \cite{Dolan:1973qd}. The upshot is that the cubic term comes from the Matsubara zero mode. }

We looked at a scalar field, but actually all bosonic fields contribute to the free energy density through a function like \eqref{eq:JB}. For fermions, the derivation is a little different. The difference comes from the fermionic Matsubara frequencies; these are odd in $n$: $\omega_n = (2n+1) \pi T$. 
We will not repeat this calculation, 
but simply quote the result
\se{\pL \lim_{V \to \infty}\frac{F}{V} \pR_{\rm fermions} &=   \int \frac{d^3 k}{(2 \pi)^3} 
   \pL \frac{\omega^{ }}{2} + T \ln \pL 1 + e^{-{\omega }/T}\pR 
   \pR  
\\ &= \int \frac{d^3 k}{(2 \pi)^3} 
   \pL \frac{\sqrt{m^2 + \vec{k}^2}^{ }}{2} + T \ln \pL 1 + e^{-{\sqrt{m^2 + \vec{k}^2} }/T}\pR 
   \pR
\\ &\equiv J_0 (m) + J_F (m,T)
} 
where we have a temperature-independent vacuum energy again, and a temperature-dependent fermionic contribution (note that the derivative of the log gives the Fermi-Dirac distribution $n_{\rm F}(\omega) = (e^{\omega/T} +1)^{-1}  $). 
The high-temperature expansion of this result is given by 
\se{\tilde{J}_F =  \frac{T^4}{2\pi^2} 
\pL -\frac{7 \pi ^4}{360}  + \frac{\pi ^2 m^2}{24 T^2}
 - \frac{m^4}{32 T^4}
 \biggl[ 
  \ln\biggl( \frac{ e^{\gamma_E}}{\pi^2 } \frac{m^2}{T^2} \biggr) - \frac{3}{2}
 \biggr] 
 + \mathcal{O}\pL \frac{m^6}{T^6} \pR
\pR .
\label{eq:JFhighT}
}

\subsubsection{Loop expansion}
At this point you might wonder how we can use our result \eqref{eq:JB} in practice. We are interested in deriving contributions to a scalar field potential, but \eqref{eq:JB} looks field-independent. The crux is that we have been considering a free theory, whereas field-dependent contributions arise in an interacting theory.
In an interacting theory, carrying out the functional integral -- as we did in \eqref{eq:ZTfuncint} -- would not have lead to a closed-form expression. 
However, for small couplings, we can still find an expression for the effective potential in perturbation theory. 

The calculation of the 1-loop effective potential using a path integral can be done in the following way, first introduced by \cite{Jackiw:1974cv},
\se{ V_{\rm eff}^{\rm 1-loop} (\bar\phi) = 
- \half i \int \frac{d^{4}p}{(2 \pi)^{4}} \ln \left( \frac{i \D^{-1} (\bar\phi;p) }{i \D^{-1} (0;p)} \right)
}
where $\D^{-1} $ is the inverse tree-level propagator obtained from a free field Lagrangian. In particular, $\D^{-1} (\bar\phi;p)$ comes from the shift $\phi \to \bar\phi + \varphi $ (where $\bar\phi$ is a constant field), which leads to an effective mass for $\varphi$ in the Lagrangian\footnote{Note that in a gauge theory, this requires fixing a gauge. }
\se{m_{\rm eff}^2 (\bar\phi) \equiv \left. \pdd{V(\phi)}{\phi} \right|_{\phi = \bar\phi}
\label{eq:effmass}
} 
such that
\se{ V_{\rm eff}^{\rm 1-loop} (\bar\phi) = 
- \half i \int \frac{d^{4}p}{(2 \pi)^{4}} \ln \left( \frac{p^{2} - m^{2}(\phi) }{p^{2} - m^{2} } \right).
}
You can check that this expression is equivalent to the Coleman-Weinberg potential derived in an interacting scalar theory. 
This is a result originally derived in zero-temperature field theory, but nothing stops us from applying it to our finite temperature formalism.

In fact, if we had actually calculated the one loop diagrams with the Feynman rules for the imaginary time formalism:
\se{
{\rm Boson\ propagator}&: \quad \frac{i}{p^2-m^2};\
p^{\mu}=\bL 2ni\pi\beta^{-1},\vec{p}\ \bR \\
{\rm Fermion\ propagator}&: \quad \frac{i}{\gamma \cdot p-m};\ p^{\mu}=
\bL (2n+1)i\pi\beta^{-1},\vec{p}\ \bR  \\
{\rm Loop \ integral}&: \quad
\frac{i}{\beta}\sum_{n=-\infty}^{\infty}\int\frac{d^3
p}{(2\pi)^3} \\
{\rm Vertex\ function}&: \quad  -i\beta (2\pi)^3 \delta_{\sum\omega_i}
\delta^{(3)}\pL\sum_i \vec{p}_i\pR 
}
that would have yielded the same result as \eqref{eq:JB} with a shifted mass. 
For example, if we consider a scalar field theory, with $V_{\rm tree} = m^{2} \phi^{2}/2 + \lambda \phi^{4}/4! $, then the sum of all 1-loop diagrams gives the 1-loop potential 
\se{ V_{T \neq 0, \rm 1-loop} (\phi) 
&= - T \sum_{n=1}^{\infty} \sum_{m} \int \frac{d^{3} p}{(2 \pi)^{3}} \frac{(-1)^{n}}{2n} \bL \frac{\lambda \phi^{2}/2}{\vec{p}^{2} + \omega_{m}^{2}+ m^{2}  } \bR^{n}
\\ &= \frac{T}{2} \sum_{m} \int \frac{d^{3} p}{(2 \pi)^{3}}  \ln \bL 1+  \frac{\lambda \phi^{2}/2}{\vec{p}^{2} + \omega_{m}^{2}+ m^{2}  }  \bR
\\ &= \frac{T}{2} \sum_{m} \int \frac{d^{3} p}{(2 \pi)^{3}}  \ln \bL \frac{\vec{p}^{2} + \omega_{m}^{2} + m^{2}_{\rm eff}(\phi) }{\vec{p}^{2} + \omega_{m}^{2} +m^{2}} \bR
\\ &= \frac{T}{2} \sum_{m} \int \frac{d^{3} p}{(2 \pi)^{3}}  \ln \bL {\vec{p}^{2} + \omega_{m}^{2} + m^{2}_{\rm eff}(\phi) } \bR - \ln \bL {\vec{p}^{2} + \omega_{m}^{2} +m^{2}} \bR
\label{eq:1loopSFT}
} 
where $n$ counts the number of vertices, $m$ labels the Matsubara index, and $m_{\rm eff}$ is defined as in \eqref{eq:effmass}. The last term in \eqref{eq:1loopSFT} is field-independent, and the first term can be compared to the field-dependent part of e.g. \eqref{eq:ZTexp1} with $ \omega^{2} = \vec{k}^{2} + m^{2}_{\rm eff} (\phi)$. 

All of this means that while the first term in \eqref{eq:JBhighT} is field-independent, and therefore not of interest to us, the second term becomes field-dependent through the effective mass. For example, if my potential is $V(\phi) = \mu^2 \phi^2 / 2 + \lambda \phi^4/4!$, then $m_{\rm eff}^2 (\phi)  = \mu^2 + \lambda \phi^2/2 $.

Now we understand how to interpret our thermal functions in perturbation theory, I would like to draw your attention to the appearance of the cubic term with the opposite sign in \eqref{eq:JBhighT}. If our potential is indeed $V(\phi) = \mu^2 \phi^2 / 2 + \lambda \phi^4/4!$, you can see that for the right values of the parameters, a temperature-dependent quartic generates an extra minimum at $\phi \neq 0$, separated by a potential barrier: 
\se{ 
V(\phi)_{T \neq 0} 
\sim& \frac{m^{2}_{T\neq 0}}{2} \phi^{2} -  \frac{\lambda_{3}}{3} \phi^{3} + \frac{\lambda_{T\neq 0}}{4} \phi^{4} + \mathcal{O}\pL \frac{\phi^{6} }{T^{2}}\pR 
} 
where $\lambda_{3} \propto T $.
This scalar potential can have two distinct minima. 
You can obviously work out for what coefficients there are two minima by extremising this function.  If we do this, we find that there are three distinct solutions for $\lambda_{3}^{2} > 4 m^{2}_{T\neq 0} \lambda_{T\neq 0} $. 

Notice how there is no effective cubic term in this expansion for the fermionic thermal function. This will be important when we discuss first order phase transitions, because the effective cubic interaction from boson loops can give rise to a potential barrier, whereas no equivalent thing can be said for fermion loops.

\subsection{IR-divergences and non-perturbativity at high temperatures}
Since we rely on perturbation theory to construct our thermal scalar potential, a few remarks about perturbativity at high temperatures should be made. 
To keep track of the perturbative expansion in a theory with a number of couplings, a common approach is to assume a power counting relation between those couplings. For the Standard Model, one might adopt,~\cite{Kajantie:1995dw}
\se{
\label{eq:g_scalings}
    g'^2 \sim
    g_Y^2 \sim
    \lambda \sim
    & g^{2},
}
where $g$ and $g'$ are the weak gauge couplings, $g_{Y}$ is the Yukawa coupling between the Higgs and the top quark, and $\lambda$ is the quartic self-coupling in the Higgs potential. 
 
At high temperatures, loop integrals such as in \eqref{eq:1loopSFT} include a Matsubara frequency sum. The evaluation of this sum implies that in practice, the effective loop expansion parameter involves the bosonic distribution function:
\se{ g^2 \to g^2 n_{\rm B}(E,T) = \frac{g^2}{e^{E/T}-1} \approx \frac{g^2 T}{E} \geq \frac{g^2 T}{m}. }
Then, at high temperatures $ m \ll g^2T$, even a perturbative theory (at zero temperature) can feature non-perturbativity, for example for the bosonic zero mode. This is called Linde's infrared problem \cite{Linde:1980ts}.  Note that this problem exists for bosonic modes only -- the Fermi-Dirac distribution has a different high-temperature expansion.

Let us look at an example. If we evaluate the following diagram (here the dashed lines are a zero mode, whereas the double-dashed lines correspond to non-zero modes), we find \cite{Dolan:1973qd,Kirzhnits:1976ts}\footnote{Note there is a typo in the evaluation of this diagram in \cite{Croon:2020cgk}.}
\se{
\label{eq:daisy}
\TopnSTT(\Lsci,\Asci,\Asaii,N,) &\propto \lambda^{N+1} \Biggl[ \int_p
    \frac{T}{(p^2+m^2)^{N+1}} \Biggr]
    \Biggl[ \Tint{Q}' \frac{1}{Q^2} \Biggr]^N
\\
&\propto g^{2N+2} \Biggl[ \int_p
    \frac{T}{(p^2+m^2)^{N+1}} \Biggr]
    \Biggl[ \Tint{Q}' \frac{1}{Q^2} \Biggr]^N 
\\ &\propto
    g^{2} m T \Bigl(\frac{g T}{m}\Bigr)^{2N}
    \;,
}
where the first integral is over all the internal zero mode propagators and the second integral is over the loops of non-zero modes (indicated by the prime):
\se{
\label{eq:defsint}
\int_p \equiv \Big( \frac{\bar\mu^{2}e^\gamma_E}{4\pi} \Big)^\epsilon \int \frac{{\rm d}^{d}p}{(2\pi)^d}, \quad \quad \text{and} \quad \quad
\Tint{P}' &\equiv T \sum_{\omega_n \neq 0} \int_p.
}
It is obvious that this diagram has an IR-divergence: in the limit $m\to 0$, it goes to infinity for any $N>0$. More generally, when $T \sim m/g $, perturbation theory breaks down. This regime of temperatures is usually relevant for the study of phase transitions, so it would appear that thermal field theory has limited predictivity in the precise context we are interested in. Luckily, there are a few different ways of dealing with this apparent breakdown of perturbativity. We will discuss a few examples below. 

\subsubsection{Daisy resummation}
The IR-divergence of the diagram above can be (partially) cured if the screening mass of the zero mode is taken into account. In a method first worked out by by Parwani and Arnold and Espinosa in Refs. \cite{Parwani:1991gq} and \cite{Arnold:1992rz}, one calculates the thermal corrections of the non-zero modes to find the corrected mass of the zero mode. Then, one can use this corrected mass to compute the contributions of the zero mode,  
\se{\label{eq:screenedzeromode}
m^{2} \to m_T^{2} = m^2 + \# g^2 T^2 
}
where $ \#$ is a numerical coefficient which depends on the multiplicity of the bosonic field. 
With the replacement \eqref{eq:screenedzeromode}, the problematic diagram \eqref{eq:daisy} becomes $ \propto g^3 T^{4}$ independently of $N$, so we can conclude that IR-divergences change the nature of the coupling expansion. Addressing them in this way is called daisy resummation (or sometimes ring resummation), because the diagram looks like a flower. 

Daisy resummation also impacts the effective effective cubic interaction in high-temperature expansion, which is often critical to the order of the transition. We have 
\se{ J_B (m,T) &\ni - \frac{T}{12\pi}  |m|^3 \to - \frac{T}{12\pi}  |m^2 + \# g^2 T^2|^3
} 
such that if the self-energy is large, the corrected expression behaves less as a cubic in $\phi$ and therefore the phase transition may cease to be first order. 

To take into account the daisy resummation, we need to work out the proper form of \eqref{eq:screenedzeromode}, and then self-consistently take it into account in our finite temperature field theory. 
The thermal screening masses (also called Debye masses) are often denoted by $ \Pi$, and are calculated thermal corrections of the non-zero modes to the zero modes of the theory (in a gauge eigenstate basis of the theory, if a gauge theory is used). For more details about the calculation of the screening masses, see e.g. \cite{Croon:2020cgk}.

\subsubsection{Dimensional reduction}
An alternative strategy is that of dimensional reduction. The idea of this method dates back to the 1980s~\cite{Ginsparg:1980ef,Appelquist:1981vg,Nadkarni:1982kb,Landsman:1989be}; it was first applied to the EWPT in the 90s \cite{Kajantie:1995dw} (see also \cite{Farakos:1994kx,Braaten:1995cm,Braaten:1995jr}, and a recent update for the EWPT \cite{Gould:2022ran}).\footnote{For recent reviews of dimensional reduction, see \cite{Croon:2020cgk,Schicho:2021gca} and references therein.} 
Dimensional reduction is a powerful tool when dealing with systems where the physical effects of certain dimensions can be considered negligible on specific scales.
As we have seen, in finite temperature field theory, the behaviour of a system is predominantly determined by its static properties and the dynamics along the `imaginary time' dimension. The latter dynamics effectively decouples on scales $r > \beta $. Then, 
dimensional reduction leaves a lower-dimensional effective field theory which maintains the important physics of the original high-temperature system but is much simpler to deal with.

For the example of a gauge theory with coupling $g$, dimensional reduction takes advantage of a hierarchy of scales 
\begin{equation}
    g^{2}T/\pi \ll
    g^{ }T \ll
    \pi T
    \;,
\end{equation}
to distinguish between a hard scale ($p\sim\pi T$, where $p=|p|$ denotes a momentum scale of particles in the heat bath) at which the theory is perturbative, an ultrasoft scale ($p\sim g^{2}T$) where the theory is non-perturbative, and a soft scale ($p\sim g^{ }T$) in between.
Table \ref{tab:DR} shows the dimensional reduction of a toy model theory containing a scalar, a fermion, and one gauge field.\footnote{The scalar typically sits between the soft and the ultrasoft scale, $m_{\phi}\sim g^{3/2} T / \sqrt{\pi} $, implying it is still just perturbative. We thank Oli Gould for pointing this out. } 

The philosophy of dimensional reduction is to treat perturbative modes perturbatively and nonperturbative modes nonperturbatively. Hence, one integrates out certain modes (fermionic modes and non-zero bosonic Matsubara modes, as well as temporal parts of the gauge modes) perturbatively, whereas the degrees of freedom at the ultrasoft scale can be treated with non-perturbative lattice studies, as was first explored in \cite{Farakos:1994xh,Kajantie:1995kf,Kajantie:1996qd}.
In some cases, however, higher loop perturbative studies of the ultrasoft degrees of freedom also give accurate results \cite{Laine:2018lgj,Niemi:2021qvp,Gould:2021oba,Ekstedt:2022zro} (a mathematica package has been developed to perform matching at the next-to-leading order \cite{Ekstedt:2022bff}).

\begin{table}
\centering
\renewcommand{\arraystretch}{1.}
\begin{tabular}{ccccc}
  \multicolumn{5}{l}{{\sl Start: {\bf $(d+1)$-dimensional theory}}} \\
  \hline
  \textbf{Scale} &
  \textbf{Validity} &
  \textbf{Dimension} &
  \textbf{Lagrangian} &
  \textbf{Fields} \\
  \hline
  {\sl Hard} & $\pi T$ & $d+1$ &
  $\mathcal{L}$ &
  $A_{\mu},\phi,\psi^{ }_{i}$  \\
  &&\multicolumn{3}{l}{$\Big\downarrow$ {\sl Integrate out $n\neq 0$ modes and fermions}} \\
  {\sl Soft} & $g T$ & $d$ &
  $\mathcal{L}_{3d}$ &
  $A_{r},A^{ }_{0},\phi$
  \\
  &&\multicolumn{3}{l}{$\Big\downarrow$ {\sl Integrate out temporal adjoint scalars $A_{0}$}} \\
  {\sl Ultrasoft} & $g^{2}T/\pi$ & $d$ &
  $\bar{\mathcal{L}}_{3d}$ &
  $A_{r},\phi$ 
  \\\hline
  \multicolumn{5}{l}{{\sl End: {\bf $d$-dimensional pure gauge theory}}} \\
\end{tabular}
\caption[Dimensional reduction]{
  Dimensional reduction of a $(d+1)$-dimensional theory with scalar field $\phi$, gauge boson $A_{\mu}$, and fermion $\psi$ based on the scale hierarchy at high temperature. In the first step, one integrates out all hard modes; in the second one also integrates out the temporal adjoint scalar $A_{0}$. At the ultrasoft scale only the spatial gauge field $A_{r}$ and the scalar undergoing the phase transition $\phi$ remains. 
Adjusted from \cite{Croon:2020cgk}.
  }
\label{tab:DR}
\end{table}

\subsubsection{Functional methods}
Lastly, one can use functional methods to deal with renormalisation both at zero and at finite temperature. Such methods do not engage in a loop expansion and are therefore in principle more appropriate in regimes where perturbativity breaks down. 

An example of such a method is the functional renormalisation group (FRG).
The philosophy of this method is to introduce a regulator function $R_{k}$ which ensures fluctuations are taken into account as the theory flows from a UV input scale $k=\Lambda$ to an IR scale $k=0$. 
To do so, the action is supplemented with a mass-like term
\se{ \Delta S_{k}[\phi] = \half \int_{\vec{p}} \phi[-\vec{p}] R_{k}(\vec{p}) \phi[\vec{p}] 
\label{eq:Sk}
}
which ensures fluctuations with $|\vec{p}| \ll k$ are suppressed, while fluctuations with $|\vec{p}| \gg k$ are suppressed by $\partial_{k} R_{k} $ in the flow. The addition of \eqref{eq:Sk} implies that the partition function develops a dependence on $k$.
Then, the average effective action\footnote{
Then, rather than the Helmholtz free energy $F_{k} = - T \ln \mathcal{Z}_{k} $ as above, the average effective action is relevant to FRG.} can be defined in terms of the scale dependent partition function $Z_{k}[J]$, 
\se{ \Gamma_{k} [\phi] = - \ln \mathcal{Z}_{k}[J] + \int_{x} J \phi - \Delta S_{k}[\phi]} 
where $J$ is an external source. 
From this, one can derive a flow equation usually referred to as the Wetterich equation \cite{Wetterich:1992yh} by simply taking the partial derivative with respect to $k$: 
\se{
\partial_{k} \Gamma_{k}[\phi] = \half \tr \bL  \frac{\partial_{k} R_{k}}{\Gamma_{k}^{(2)}[\phi] + R_{k}} \bR
\label{eq:wetterich}
} 
where $\Gamma_{k}^{(2)} $ is the second order functional derivative of $ \Gamma_{k}[\phi]$. 
Note that  this flow equation is formally exact: no expansion was used to derive it. However, in practise one needs to invoke an approximation scheme to solve it. 

The flow equation \eqref{eq:wetterich} can be used in both the zero temperature and the finite temperature regimes. Due to its non-perturbative nature it has been applied to calculate observables in the QCD sector of the SM, with good correspondence to lattice studies.\footnote{See \cite{Dupuis:2020fhh} for a comprehensive recent review of the functional renormalisation group and its applications.}
A method to apply functional renormalisation techniques to the problem of false vacuum decay was first proposed in \cite{Croon:2021vtc} and is still under development. 

\section{First order phase transitions}
Having developed a basic understanding of field theory at finite temperature, and some of the techniques we can apply to calculate corrections to a scalar potential, we now turn to an important application: phase transitions in the early Universe. Different types of phase transitions exist, and can be classified by their order, a measure of how discontinuous the process is. The order of the phase transition is also an important aspect of its phenomenology. 
A first order phase transition is formally defined as a phase transition in which the free energy is non-analytic (its derivatives are discontinuous). Namely, if $ \partial F / \partial T$ is discontinuous, the average energy, defined in the previous section is also discontinuous:
\se{ E &=  \frac{1}{\mathcal{Z}} \tr [\hat H e^{-\beta \hat H}]
\\ &= \frac{T^2}{\mathcal{Z}} \pd{}{T} \mathcal{Z}
\\ &= T^2 \pd{}{T} \log \mathcal{Z}
\\ &= F - T\pd{F}{T}
}
where we have assumed zero chemical potential, as we're doing throughout. The difference in average energy (or the difference in average energy density) is called the latent heat, so the statements that latent heat is released and that the free energy is non-analytic are equivalent. 

\subsection{False vacuum decay}
In field theory, the discontinuity in the derivative of the free energy can be described through the (potential) energy surface of a scalar order parameter, which has a local minimum in which the field is initially located, and a global minimum to which it tunnels. This description can be used at zero temperature as well as at finite temperature -- in the latter case, we use periodicity in Euclidean time in our description of false vacuum decay.

\subsubsection{Reminder: tunneling in quantum mechanics}
To warm up, let's remind ourselves of tunneling through a barrier in quantum mechanics. The semi-classical (i.e. small $\hbar$) approximation to this problem shares many similarities with the most popular way of calculating the false vacuum decay in quantum field theory. 

In classical mechanics, a particle with insufficient energy to surmount a potential barrier would be completely reflected back. In quantum mechanics, there is a non-zero probability that the particle can tunnel through the barrier, appearing on the other side even though its energy is less than the barrier's height. The Wentzel-Kramers-Brillouin (WKB) approximation provides an approximate solution to calculate this tunneling probability.
The key idea behind this approximation is to treat the particle's motion classically outside the barrier, where the potential energy is lower, but quantum mechanically inside the barrier, where the potential energy is higher. The transition between classical and quantum regions occurs in the vicinity of the barrier.

To apply the WKB approximation, we start by dividing the problem into three regions: the incoming region, the barrier region, and the transmitted region. Imagine our barrier region is between $ x_{0}$ and $ x_{1}$. In the incoming and transmitted regions, the particle's motion is described by classical trajectories.
Inside the barrier region, the particle's motion must be treated quantum mechanically. 
Our state $\psi$ satisfies the following equation of motion, the time-independent Schr\"{o}dinger equation,
\begin{equation}
    \ddd{\psi}{x} - \frac{2 m (V(x)-E)}{\hbar^2} \psi=0.
\end{equation}
For slowly varying $V(x)$, the WKB-approximation to the wave-function is
\se{ \psi(x) \propto\frac{1}{\sqrt{k(x)}} \exp{ \pL \pm i \int_{x_{0}}^{x} dx' k(x')  \pR} \quad\text{where}\quad k(x)= \sqrt{2 m (E-V(x))}/\hbar. \label{eq:wkbwave}}
Here $ \int k(x) dx $ is the classical action.
For $E>V(x)$, as is the case in the classical regions, this is an oscillating function, but for $ E< V(x)$, the action is complex and $ \psi $ is exponentially decreasing.\footnote{At the boundary of the classical and tunneling regions, where $V(x_{0}) = E $, the approximation \eqref{eq:wkbwave} clearly breaks down. Expanding around $x_{0}$, in the vicinity of the the boundary the Schr\"{o}dinger equation becomes,
$\ddd{\psi}{x} - U_{1} (x-x_{0})\psi + \mathcal{O} (x-x_{0})^{2}=0,$
the form of which is known as the Airy equation.
By matching the wavefunction \eqref{eq:wkbwave} and its derivative in both regions to the Airy functions at the boundary, one can relate the coefficients. }
The sign inside the exponential depends on the direction of propagation. 

In the WKB approximation, the probability of a particle tunneling through the barrier (versus being reflected back) is essentially determined by the total decrease of the amplitude in the tunneling region. For a particle coming in from $x< x_{0}$, the tunneling amplitude to $x>x_{1} $ is given by
\begin{equation}
    \gamma_{\rm WKB} \propto e^{- \frac{1}{\hbar} \int_{x_0}^{x_1} 
    \sqrt{2 m (V(x)-E)} dx} + \mathcal{O}(\hbar),
\end{equation}
such that the tunneling probability exponentially decreases with both the height and width of the barrier. 

Analogously to tunneling in quantum mechanics, false vacuum decay rate in quantum field theory is associated with an imaginary part of the action. However, path integrals are manifestly real, and therefore recovering the false vacuum decay rate is not a trivial problem. Two main methods have been proposed in the literature: 
\begin{itemize}
    \item The popular semi-classical method accredited to Callan and Coleman~\cite{Coleman:1977py,Callan:1977pt}, which effectively deforms the integration contour;
    \item The direct method, developed in \cite{Andreassen:2016cff,Andreassen:2016cvx}.
\end{itemize}
We will first focus on the former, but also draw a quick comparison to the latter, which may in principle be used in non-perturbative calculations \cite{Croon:2021vtc}. As ever, I encourage the reader to explore the rich topic of false vacuum decay in the wider literature, beyond the cursory review I can offer here. On the topic of phenomenological calculations of first order phase transitions in the early Universe in particular, I highly recommend the recent work \cite{Athron:2023xlk}.

\subsubsection{Semi-classical methods in field theory}
The field theory analogue of the WKB approximation in quantum mechanics was developed by Callan and Coleman in the 1970s~\cite{Coleman:1977py,Callan:1977pt}.\footnote{For some recent reviews, see \cite{Ai:2019dqr,Devoto:2022qen}.}
These two gentlemen proposed that we go to Euclidean space, as we did in the previous section. 
The analytical continuation from Minkowski spacetime to Euclidean spacetime is useful because in Euclidean spacetime, where the action is real and positive-definite, the path integral is convergent and amenable to approximation techniques like the saddle point approximation. A real-time formalism was worked out in Ref. \cite{Ai:2019fri}. 

Let us first discuss false vacuum decay at zero temperature. To use a saddle point approximation, the next step is to expand 
\se{ \phi = \bar\phi +\varphi, 
\label{eq:fieldexp1}}
where $\bar\phi$ is a solution to the classical EOM.
 We use $\hbar = 1$ throughout these lecture notes, but if we were to keep it explicit, the expansion would be $ \phi = \bar\phi + \hbar^{1/2}\varphi$ . 
If $\bar\phi$ minimises the Euclidean action, then it will dominate the path integral. Thus, we can expand the path integral around $\phi = \bar\phi$ as  
\se{\label{eq:bouncepath}
\mathcal{Z} &= \int \D \phi \, e^{- S_E[\phi]}
\\ &= \int_{C} \D \varphi \, e^{- S_E[\bar\phi]- \half S_E'' [\bar\phi] \varphi^2 + ...}
} 
where the linear term vanishes because $\bar\phi$ is a solution to the classical equations of motion (in actuality there is usually more than one solution $\phi_n$, in which case we need to sum over $n$). 
The integration contour $C$ is the steepest descent contour through the saddle point. 
The second order expression is now a Gaussian integral, which we can evaluate to obtain
\se{\mathcal{Z} \sim \mathcal{Z}_{G} = e^{- S_E[\bar\phi]} \pL \det S_E'' [\bar\phi]   \pR^{-1/2} \label{eq:saddlebounce}
} 
(see the note on Gaussian integrals above). As this expression only depends on $ \bar\phi$, in principle our task is simple: just find the solution(s) to the classical equations of motion and substitute them back into \eqref{eq:saddlebounce} to find the free energy and related quantities. Let us do that and then consider how to relate our findings to the false vacuum decay rate. To find the equations of motion we extremise the Euclidean action. For a scalar field, we might have: 
\se{
\nabla^{2}\bar\phi - \frac{dV(\bar\phi) }{d\bar\phi} = 0.
} 
Because solutions that extremise the energy are often expected to be spherically symmetric, and this is indeed also the case for the tunneling solutions \cite{Coleman:1977th}, the only relevant coordinate is the radial one: $ r = (\sum_{i}^{d}x_i^{2})^{1/2}$. Therefore the classical equations of motion (in $d$ dimensions) become,
\begin{align} \label{eq:bounce}
    \ddd{\bar\phi}{r} + \frac{d-1}{r}\dd{\bar\phi}{r} &= \frac{dV(\bar\phi) }{d\bar\phi}.
\end{align}
At zero temperature, the relevant number of dimensions is obviously $d=4$: the field configuration describing tunnelling has O(4) symmetry. At finite temperature, for length scales $r > \beta $, the relevant number of dimensions is $d=3$.

The decision of \emph{which} saddle points to consider is an important aspect of the semi-classical approximation for false vacuum decay. In particular, if all saddle points are taken into account, the path integral is manifestly real. As we expect false vacuum decay to be associated with an imaginary contribution, we have to be selective. So, let us discuss the solutions of \eqref{eq:bounce} for a typical situation we would like to consider: a double well potential. First of all, there are two trivial solutions, which are the static field values corresponding to the bottoms of the wells: for these field configurations all terms vanish: $\partial_{r}^{2} \phi = 0$, $\partial_{r} \phi = 0$ and $V' = 0 $. We can call these solutions $ \phi_{\rm FV}$ and $\phi_{\rm TV}$ for the less deep (false vacuum) and deeper vacuum (true vacuum) respectively. 
Then there are a number of other solutions, which depend on the boundary conditions we impose. We are interested in field configurations that interpolate between $\phi_{\rm FV}$ and some field value on the other side of the barrier (note that this need not be $\phi_{\rm TV} $), so a natural choice is to impose $\phi_{\rm FV}$ at $ r=\pm \infty$. As we are only interested in reaching the other side of the barrier, and not at which field value exactly this happens, we may not want to impose a similar Dirichlet boundary condition at the origin. We can instead impose finite energy at the origin to give us the second boundary condition, $\partial_{r} \phi|_{r=0} = 0$. 
Thus, it seems reasonable that the boundary conditions of a metastable state of the false vacuum would be,
\se{\label{eq:bounceBC}
    \bar\phi(r \rightarrow \pm \infty)&= \phi_{\rm FV}, \\
    \left.\dd{\bar\phi}{r}\right|_{r=0}&=0.
}
Indeed, you can find \eqref{eq:bounce} and \eqref{eq:bounceBC} in many papers describing cosmic phase transitions. They describe the \emph{bounce} $\bar\phi_{b} $, an O($d$) field configuration which starts (and ends) in the false vacuum, and reaches the other side of the potential barrier (say $\bar\phi = \phi_e$) with zero velocity ($ \dmu \bar\phi = 0 $). But they also still describe the static solution $\bar\phi = \phi_{\rm FV}$, so we are stuck with that saddle point too. 

Now, how do we find the false vacuum decay rate from $\mathcal{Z}$? A simple way to get intuition for that (courtesy of \cite{Laine:2016hma}),\footnote{For a more rigorous argument, see the original paper by Callan and Coleman \cite{Callan:1977pt}.} is to imagine a metastable (zero-temperature) state and time-evolving it forward. We have,
\se{ \ket{\phi(t, \vec{x})} &= e^{- i E t} \ket{\phi(0, \vec{x})} = e^{- i (\re(E) + i \im(E)) t} \ket{\phi(0, \vec{x})} 
\\ &\to \bra{\phi(t,\vec{x})} \ket{\phi(t,\vec{x})} = e^{2\im(E) t  } \bra{\phi(0,\vec{x})} \ket{\phi(0,\vec{x})}
}
such that the decay rate is given by $\Gamma = -2 \im(E) $ -- as expected, metastability is associated with imaginary energy (eigenvalues). 
Analogously, for a thermal state, we might expect something like $\Gamma = -2 \im(F) $. 
This would mean that 
\se{ 
\Gamma &= - 2 \im(F)
\\ &= 2 T \im\bL  \ln \pL \mathcal{Z}_0 + \int \D \varphi \, e^{- S_E[\bar\phi_{b}]- \half S_E'' [\bar\phi_{b}] \varphi^2 + ...} \pR \bR 
\\ &\sim 2 T \im \bL \ln \mathcal{Z}_0 \bR + 2 \frac{T}{\mathcal{Z}_0}
\im\bL  \int \D \varphi \, e^{- S_E[\bar\phi_{b}]- \half S_E'' [\bar\phi_{b}] \varphi^2 + ...} \bR
 \\ &\sim \frac{T}{\mathcal{Z}_0}
\im\bL
 \pL \det S_E'' [\bar\phi]   \pR^{-1/2} e^{- S_E[\bar\phi_{b}]}  \bR
\\ &\equiv A(T) e^{- S_E[\bar\phi_{b}]} .
\label{eq:gammaThermal}
}
where we have restricted ourselves to saddle points with the boundary conditions \eqref{eq:bounceBC},
$ \mathcal{Z}_0 \sim \mathcal{Z}_{G}[\phi_{\rm FV}]$ is the contribution from the solution $ \bar\phi = \phi_{\rm FV}$ which stays in the false vacuum, which we expect to dominate over $  \mathcal{Z}[\bar\phi_{b}]$  (such that we could expand the log). For the other saddle point, we have evaluated the Gaussian integral, including a factor $1/2$ which we will comment on below.\footnote{This seemingly innocent factor of 2 was derived in \cite{LANGER1967108}, elaborated on in \cite{Callan:1977pt} and more recently in \cite{Andreassen:2016cvx}. }
We have also grouped together the prefactor $A(T)$: 
\se{
A(T) = \frac{T}{\mathcal{Z}_0}
\im\bL
 \pL \det S_E'' [\bar\phi_{b}]   \pR^{-1/2}  \bR
}

At finite temperature, a Euclidean action we may be interested in is the scalar field action\footnote{One should be careful to avoid double-counting, as per the discussion in \cite{Croon:2020cgk,Croon:2021vtc} }
\begin{equation}\label{eq:SE}
    S_E = \int_{0}^{\beta} d \tau \int d^3x_E \left[(\partial_{x_E} \phi)^2 + V(\phi,T) \right]
\end{equation}
in which the subscript $E$ denotes Euclidean and the integral is over Euclidean space. The coordinates in Euclidean space are $\vec{x}$ as before, and $ \tau = i t$.
As mentioned above, on length scales large compared to $\beta $, the system behaves like it is approximately 3-dimensional. In this case, 
\se{ \label{eq:SEoverT}
 S_E [\phi] 
&= \int_\tau \int_\vec{x} L_E 
\\ &=  \pL \frac{1}{T} \pR \int_\vec{x} L_E 
\\ &= \frac{S_{E,3}}{T}
}
and thermal fluctuations assist in the phase transition (as mentioned, in this case one should also take $d=3$ in \eqref{eq:bounce}). In the following I will often just use $ S_E$ to mean $ S_{E,3}$, but it will be clear from context that we are in a thermal situation. That's also the situation we are most interested in in the early Universe. 
It is usually the case that the nucleation rate \eqref{eq:gammaThermal} increases exponentially as the temperature decreases, because the ratio $S_E/T$ decreases with decreasing temperature. If the barrier disappears, $S_E/T$ tends to zero. 

We have yet to determine the prefactor $A(T)$ in \eqref{eq:gammaThermal}. If we were interested in the decay rate per unit volume $\Gamma/\mathcal{V}$, which we indeed often are, dimensional analysis tells us that it should be $A(T)/\mathcal{V} \sim T^4 $, which is indeed an approximation that is sometimes made -- the argument is that because it features in the argument of the exponential,  $S_E/T $ is far more important and therefore dimensional analysis is sufficient. However, an understanding the origin of the prefactor lends some insight to false vacuum decay rate calculations in general. 
Namely, in our exploration of the prefactor we will discover what makes the free-energy we are calculating imaginary, leading to a nonzero false vacuum decay rate. 

Let's go back to our field expansion. We may expand our field \eqref{eq:fieldexp1} further into an eigenbasis of an operator: 
\se{ \phi = \bar\phi + 
\sum_i \zeta_i \varphi_i.
\label{eq:fieldexp2}
}
Given the expansion \eqref{eq:bouncepath}, good choice is the operator $ S_{E}''[\bar\phi] = \pL -\Box + V''[\bar\phi]\pR $, which characterises the size of fluctuations.
Note that in this case the result of our Gaussian integral in \eqref{eq:saddlebounce} is equivalent to 
\se{\mathcal{Z} &= e^{- S_E[\bar\phi]} 
\prod_i \pL \lambda_i   \pR^{-1/2}
\label{eq:gaussianEVs}
} 
where $\lambda_i$ are the eigenvalues of the operator $S_{E}''[\bar\phi]$:
\se{
S_{E}''[\bar\phi] \varphi_i = \pL -\Box + V''[\bar\phi]\pR \varphi_i = \lambda_i \varphi_i.
}

The operator $ S_{E}''[\phi]$ has an interesting set of eigenvalues. In particular, it (in principle) has zero modes ($ \lambda = 0$). The zero modes are given by $\partial_\mu \phi $, which we can easily check
\se{
S_{E}''[\bar\phi]\dmu \phi = \dmu (S_{E}'[\bar\phi])=0
}
where $S_{E}'[\bar\phi] = 0 $ by the definition of $\bar\phi$. A seemingly trivial point is that if the field $ \phi$ is space-time independent -- such as the false vacuum saddle point ($ \bar\phi = \phi_{\rm FV}$) -- no zero modes exist. 

Since the bounce solution has four zero modes, it is clear that this is not the ground state, as that must be unique. So there must be at least one lower energy state: a negative mode $\lambda_{-}$. In an analogy with simple harmonic oscillators (SHO), we can motivate that there is just a single negative mode.
The spectrum of a SHO contains a ground state without nodes, and a first excited state with one node. As our zero mode has one node (where the bounce happens, at $ \bar\phi (r=0) \equiv \bar\phi_e$, as there $ \dmu\bar\phi = 0$), it must be the first excited state. 
The existence of the negative mode also implies that $F$ is imaginary. This can be seen directly from \eqref{eq:gaussianEVs}: the single negative mode under the square root renders the entire product imaginary.

We should make a few comments about this negative mode, and its implications for the path integral. First, its existence implies that the Gaussian integral around the bounce solution \eqref{eq:bouncepath} is divergent when the integration contour $C$ is real. This is the first hint that to get something sensible, we must perform an analytic continuation of the integration contour.
The negative mode arises from the fact that the bounce solution is not a minimum of the Euclidean action, but a saddle point: among the family of solutions which penetrate the barrier -- which includes besides the bounce also configurations which return to the false vacuum before or after reaching the other side of the barrier, and the limiting case $\phi_{\rm FV}$ -- the bounce is a maximum. This means that if our integration contour includes all saddle points in this family, the bounce gives a subdominant contribution.

So, it turns out that to compute the false vacuum decay rate, the integration contour must be chosen carefully. The appropriate contour is complex, and along the steepest descent contour through the false vacuum saddle point: $C_{F}$. This corresponds to the physical situation we have in mind when calculating the false vacuum decay rate: choosing this contour implies that the resonance associated with the false vacuum dominates, even if $\phi_{\rm FV} $ is not the dominant saddle point (it is easy to see that $\phi_{\rm TV} $ will give a greater contribution). Integrating along this contour results in a complex path integral, with the imaginary part associated with the decay rate: $\Gamma \propto \text{Im} \ln \mathcal{Z} \sim \text{Im}\mathcal{Z}/\text{Re}\mathcal{Z}$. However, the imaginary part comes from a part of the contour far away from the FV saddle point, such that the \emph{Gaussian} path integral evaluated on the false vacuum saddle point is real -- $\im \bL \ln\mathcal{Z}_{G}[\phi_{\rm FV}] \bR = 0$.

As shown in \cite{Andreassen:2016cvx}, the imaginary part of the path integral over $C_{F}$ is equivalent to \emph{half} of the steepest descent contour passing through the bounce saddle point (modulo a sign).\footnote{Note that this makes \eqref{eq:gammaThermal} rather subtle, as it would be incorrect to state $ 2 T \im \bL \ln \mathcal{Z}_0 \bR + 2 \frac{T}{\mathcal{Z}_0}
\im\bL  \int \D \varphi \, e^{- S_E[\bar\phi_{b}]- \half S_E'' [\bar\phi_{b}] \varphi^2 + ...} \bR
\sim 2 T \im \bL \ln\mathcal{Z}_{G}[\phi_{\rm FV}] \bR + 2 \frac{T}{\mathcal{Z}_0}
\im\bL  \mathcal{Z}_{G}[\phi_{b}] \bR $ -- this would imply ignoring an imaginary contribution which is of the same order as the bounce contribution.}
This complex contour $C_{b}$ leads to a finite path integral despite the negative eigenvalue, with an imaginary part:
\se{ \text{Im} \mathcal{Z}_{G} &= \text{Im}\int_{C_{b }} \D \varphi \, e^{- S_E[\bar\phi]+ \half |\lambda_{-} | \varphi^2 }
= \half e^{- S_E[\bar\phi]} \pL \det |\lambda_{-} |    \pR^{-1/2} 
 .
 }
 This choice of contour is the justification for restricting our analysis to the bounce and the FV saddle points (note that we still need the latter to compute $\text{Re}\mathcal{Z} = \mathcal{Z}_{0}$), which we imposed through the boundary conditions \eqref{eq:bounceBC}. A more detailed analysis of the appropriate integration contour in can be found in \cite{Andreassen:2016cvx}.

With the right choice of integration contour (or equivalently the boundary conditions \eqref{eq:bounceBC} for the saddle points), \eqref{eq:gammaThermal} is our false vacuum decay rate. We can clarify this expression some more by analysing the zero modes. 
Let us first separate them out of \eqref{eq:fieldexp2}:
\se{\phi = \bar\phi +  N \zeta^\mu \dmu \phi +
 \sum_i' \zeta_i \varphi_i
}
where the prime indicates that the zero modes are to be left out (and as before we have set $\hbar=1$). 
Here $N$ indicates that the modes are not properly normalised yet, because 
\se{\braket{\dmu \phi}{\dnu \phi } &= \fourth \delta_{\mu \nu} \int d^4x (\partial_\rho \phi)^2 
\\ &= \delta_{\mu \nu} S_E [\phi]
} 
where the last line is using that for solutions to the EOM, multiplying by $\partial_{\mu} \phi $ a sort of virial theorem holds: $\half (\dmu \bar\phi)^2 = V(\bar\phi) $.
So, the normalisation of the zero modes should be $ N = \sqrt{2 \pi/S_E [\bar\phi_{b}]} $ if we choose modes to be normalised such that $\braket{\phi_{i}}{\phi_{j}} =  2 \pi \delta_{ij}$.
Moreover, the zero modes, being space-time translation of the bounce, are proportional to the fourth root of the space-time volume $\mathcal{V}^{1/4}$. 

All in all, for four space-time dimensions (and thus four zero modes), we find 
\se{ \Gamma
&\sim  \frac{T}{\mathcal{Z}_0}
\im\bL \pL \det S_E'' [\bar\phi]   \pR^{-1/2} e^{- S_E[\bar\phi]}  \bR \\ 
&\sim  T \left| \frac{  \det S_E'' [\bar\phi_{b}]   }{\det S_E'' [\bar\phi_{\rm FV}]   } \right|^{-1/2} e^{- S_E[\bar\phi]}\\ 
& \sim  T \mathcal{V} \pL \frac{S_E [\bar\phi_{b}]}{ 2 \pi } \pR^{2} 
 \left| \frac{  \det' S_E'' [\bar\phi_{b}]   }{\det S_E'' [\bar\phi_{\rm FV}] }  \right|^{-1/2}  e^{- S_E[\bar\phi]}  
 }
where once again the prime denotes the zero modes are taken out of the determinant, and where we have written $\mathcal{Z}_0 = \pL \det S_E'' [\bar\phi_{\rm FV}]   \pR^{-1/2}$. 
At high temperatures, there is effectively one fewer zero mode (i.e. not $\partial_{\tau} \phi $). Therefore, 
\se{
- 2 \im F \sim  T \mathcal{V} \pL \frac{S_E [\bar\phi_{b}]}{ 2 \pi T} \pR^{3/2}  \left| \frac{  \det' S_E'' [\bar\phi_{b}]   }{\det S_E'' [\bar\phi_{\rm FV}] } \right|^{-1/2} 
   e^{- S_E[\bar\phi]/T}
}
It also turns out the relation for $\Gamma$ does not correspond exactly to our naive guess at high temperatures, but needs to be modified to \cite{Affleck:1980ac} 
\se{ \Gamma &= - \frac{\sqrt{|\lambda_{-}|}}{\pi T} \im F 
\\ &= \frac{\sqrt{|\lambda_{-}|}}{2\pi} 
 \mathcal{V} \pL \frac{S_E [\bar\phi_{b}]}{ 2 \pi T} \pR^{3/2}  \left| \frac{  \det' S_E'' [\bar\phi_{b}]   }{\det S_E'' [\bar\phi_{\rm FV}] } \right|^{-1/2} 
   e^{- S_E[\bar\phi]/T}
\label{eq:classgrowth}
}
where $\lambda_{-}$ is the eigenvalue of the negative mode. This expression captures the classical growth of the unstable mode at temperature $T$.\footnote{The expression \eqref{eq:classgrowth} assumes the calculation can be done in equilibrium. Non-equilibrium physics is expected to enter the unstable (negative mode), leading to modifications in the nucleation prefactor \cite{LANGER1967108}. See also \cite{Gould:2021ccf,Ekstedt:2022tqk}. }

From this, we can derive the thermal parameters of the phase transition:
\begin{itemize}
\item  The nucleation temperature (or onset temperature) of the phase transition needs to be defined by convention. For example, it may be defined as the temperature at which one can expect one bubble per Hubble volume:
$
\Gamma \times H^{-1} = H^{3} .
$
Solving for $S_{E}/T$, this implies $ S_{E}/T \sim -3 + \log A(T)/T^{4} - 4 \log T/M_{p}$.
One may also be interested in the temperature of percolation, after the growth of the nucleated bubbles and closer to the completion of the phase transition. This rate will also depend on the dynamics after nucleation. 
\item The parameter $\beta$ parametrises the rate of change of the nucleation rate, and is usually defined as 
$\beta = (1/\Gamma) \times {\rm d} \Gamma /{\rm d}t .$
This has the dimensions of a rate, and is usually normalised to the Hubble rate, $\beta/H$.
\item The parameter $\alpha$ gives the latent heat at the nucleation temperature, usually normalised to the radiation energy density in the plasma at the nucleation time. Different definitions exist, but it is usually defined in terms of the trace anomaly: $\alpha = (\Delta V - 1/4 \, {\rm d}\Delta V/{\rm d}\ln T )/\rho_{\rm rad} $ where $\Delta V $ is the potential difference between the vacua, evaluated at the nucleation temperature.
\end{itemize}
We will encounter these parameters further on in these notes, when characterising the phenomenology of the phase transition.

\subsubsection{The direct method}
The method we just studied is the most commonly applied method of calculating false vacuum decay, and it relies on a saddle point approximation and a particular analytic continuation. There is however an alternative method, developed by Andreassen, Farhi, Frost and Schwartz in 2016 \cite{Andreassen:2016cff,Andreassen:2016cvx}. Let's have a brief look at how this works. 

First, we need to go back to the problem we wanted to address. We are interested in the decay of a field initially localised in the false vacuum, $\phi_{i} = \phi_{\rm FV}$, which decays to the other side of the barrier, a (generally multi-dimensional) region which we can call $R$:

 {\centering
\includegraphics[width=.35\textwidth,angle=270]{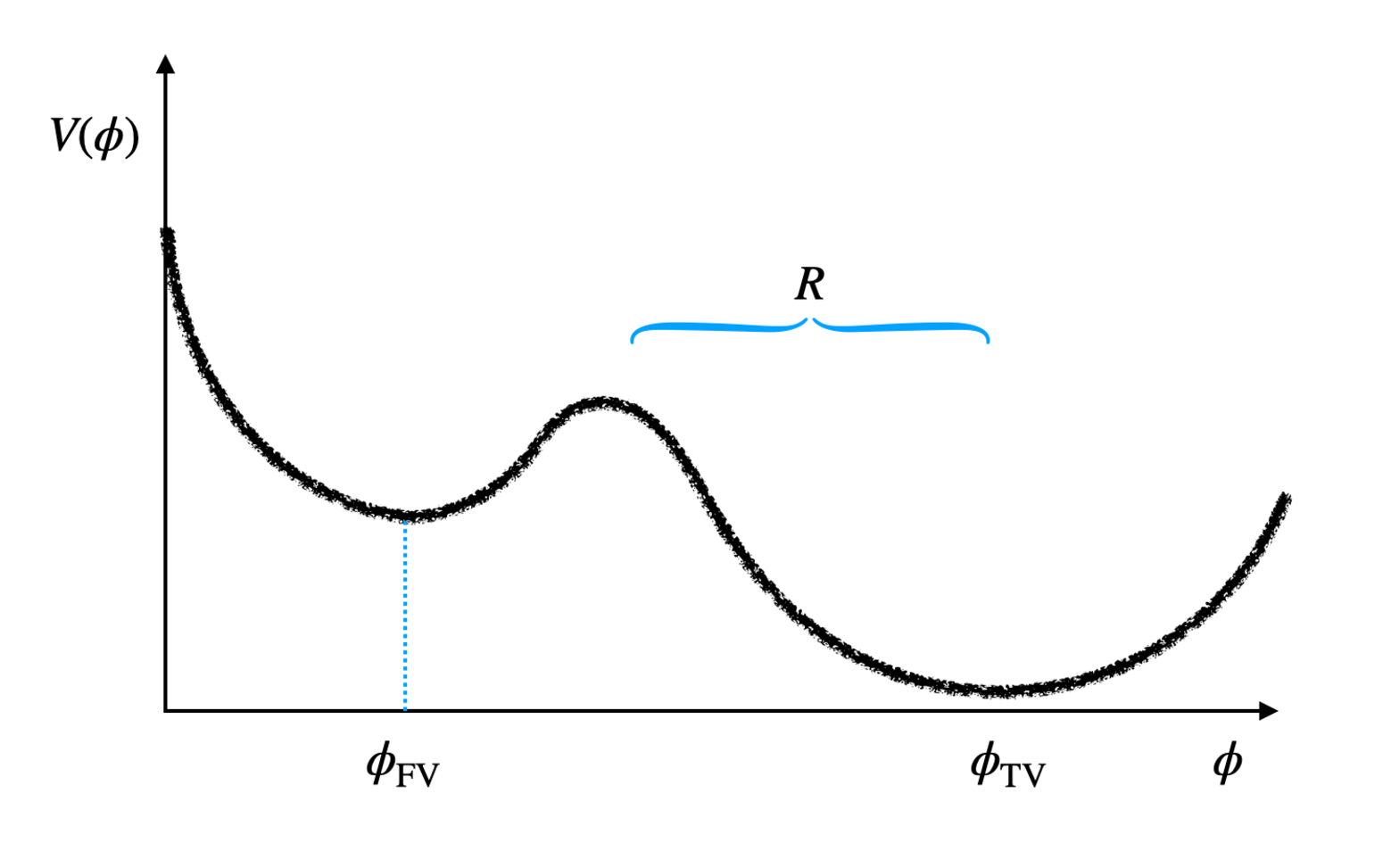}
\par} 

This decay rate is given by
\se{\Gamma &= \lim_{\substack{t/t_{\rm slosh} \to \infty \\ t/t_{\rm NL} \to 0}} \frac{1}{P_{\rm FV} (t) } \dd{P_{R}(t)}{t}
\\ P_{R }(t) &= \int_{R} d\phi_{f} | \braket{\phi_{f}; t }{\phi_{\rm FV}; 0} |^{2}
\label{eq:direct1}
}
with the latter the probability that the state originally at $\phi_{\rm FV} $ is found somewhere in $ R $ at time $t$. Here $t_{\rm slosh}$ characterises the time-scale for movement within the region of the false vacuum, and $t_{\rm NL}$ is the timescale for the transmitted wave-function to start propagating back to the false vacuum. Thus, \eqref{eq:direct1} is applicable for timescales long compared to movements within the neighbourhood of the false vacuum, but short compared to timescales where the wave-function might be transmitted back. This may seem trivial, but is actually a very important insight underpinning false vacuum decay rate calculations. 
Namely, the more naive limit ${T \to \infty}$ would pick out a saddle point called the shot -- which starts and ends at $\phi_{\rm FV}$ but spends most of its time on the other side of the barrier, and will therefore be associated with the energy of the true vacuum -- unless the path integral contour is deformed, as we did in the previous subsection. If the shot dominates, one finds that the false vacuum decay rate is zero.

After tunneling, when our state emerges on the other side of the barrier, it will do so on a surface in $\phi$ which we can call $\Sigma$.
The bounce conserves energy, and therefore it will appear on the surface $\Sigma$ with $U[\phi_{\Sigma}]= U[\phi_{\rm FV}] $. Here $U$ includes not only the potential energy but also the gradient energy:
\se{ U[\phi ] = \int d^{3} x  \bL \pL\nabla \phi \pR^{2} + V(\phi) \bR }
Then, the two-point function in \eqref{eq:direct1} can be written as a Feynman propagator, which we can split in two parts, before and after it crosses the surface $\Sigma$: 
\se{ | \braket{\phi_{f};t}{\phi_{\rm FV}; 0} |^{2} &= | D_{F}(\phi_{\rm FV}, 0; \phi_{f},t) |^{2} 
\\ &= \int_{\Sigma} d b \int_{0}^{t} dt' D_{F} (\phi_{\rm FV}, 0; b, t')D_{F} (b, t'; \phi_{f},t) \delta (t_{b}[\phi]- t')
\\ &= \int_{\Sigma} d b \int_{0}^{t} dt' D_{F} (\phi_{\rm FV}, 0; b, t')D_{F} (b, 0; \phi_{f},t-t') \delta (t_{b}[\phi]- t')
\label{eq:directtwopoint}
}
where 
\se{ D_{F} (b, 0; \phi_f,t) &=  \mathcal{N} \int^{\phi(t) = \phi_f}_{\phi(0) = b} \D \phi e^{i S[\phi]}
} 
where $\mathcal{N}$ is just a prefactor. 
Here $ t_{b}[\phi]$ gives the time at which $\phi$ first crosses the point $b$ on the surface $\Sigma$, on the other side of the barrier; the delta function ensures that this happens \emph{at least once}. 

The two-point function \eqref{eq:directtwopoint} can be simplified by a number of manipulations detailed in \cite{Andreassen:2016cff,Andreassen:2016cvx}. The result is not convergent for real $t$, a problem that can be dealt with through the use of an imaginary integration path, or equivalently through analytic continuation to Euclidean space. 
The analytic continuation of the delta function requires a careful analysis which we will not repeat it here. We simply quote the result:
\se{\Gamma = \frac{\mathcal{N}\mathcal{N}^\star}{P_{\rm FV}(t)}
\im \left( \int_{\Sigma} d b \int_{\phi(-\tau)=\phi_{F}}^{\phi(\tau)=\phi_{F}} \D \phi  e^{-S_{E}[\phi] } \delta( \tau_b[\phi]) \right) , 
}
where the integral in the brackets needs to be evaluated for real $\tau$, and then analytically continued to $ \tau \to i t$ to yield something purely imaginary. 

The final expression for the decay rate $\gamma_{\text{FV}}$ may be written as
\begin{align}
    \Gamma &=
    \lim_{T\rightarrow \infty} \left\lvert 2 \, \text{Im} \left(\frac{\int \D \phi \, e^{- S_E[\phi]} \delta(\tau_b [\phi])}{\int \D \phi \, e^{- S_E[\phi]}}\right)_{\tau = i t}\right\rvert
    \\
    &= \lim_{T\rightarrow \infty} \left\lvert 2 \, \text{Im} \left(\frac{1}{\tau} \frac{\int_{\text{hits } b} \D \phi \, e^{- S_E[\phi]}}{\int \D \phi \, e^{- S_E[\phi]}}\right)_{\tau = i t}\right\rvert
    \label{eq:Gammadirect}
\end{align}
where the path integrals have boundary conditions $\phi(\pm \tau) = \phi_F$. The meaning of $\text{Im}()_{\tau = i t}$ is to evaluate the content inside the parentheses for real $\tau$, analytically continue to imaginary $\tau = i t$, and then take the imaginary part.  

The direct method for calculating false vacuum decay offers an appealing feature -- it provides a non-perturbative description of the decay process. Unlike the semi-classical approach, which relies on a saddle point approximation and careful selection of the integration contour, the direct method should in principle allow for the determination of the false vacuum decay rate without such approximations.
 However, one still needs to apply boundary conditions to the path integral, as well as the condition that the numerator hits the point $b$. A saddle point approximation, as we relied on extensively in the previous section, is the easiest way of doing this. In this case, as discussed in Ref.~\cite{Andreassen:2016cvx}, there are three saddle points which satisfy the boundary conditions $\phi(\pm \tau) = \phi_F$: the (constant) false vacuum solution, the bounce, and the shot. The latter is a field profile which rapidly transitions to the true vacuum, stays there for a long time, and then rapidly transitions back. 

Only the bounce and shot cross $b$ and contribute to the numerator of \eqref{eq:Gammadirect}. To see which one dominates, we can examine the $\tau$-dependence of the two saddle points. The shot rapidly transitions from the false vacuum to the true vacuum and back, resulting in a $\tau$-independent term (from the rapid transition) plus a term linear in $\tau$ proportional to the true vacuum energy. On the other hand, the bounce also has a quick transition to the turning point and back, but spends the most amount of time in the false vacuum, therefore contributing -- besides a $\tau$-independent term -- a term linear in $\tau$ proportional to the false vacuum energy:
\begin{align}
    S_\text{shot} &=\tau E_\text{TV} + S_S^0
    \\
    S_\text{bounce} &=\tau E_\text{FV} + S_B^0.
\end{align}
Importantly, these need to be evaluated for imaginary $\tau$ -- it is real time we are interested in, after all. In this case, the terms linear in $\tau$ are both purely imaginary and play no role. In contrast, the two pieces $S_S^0$ and $S_B^0$ remain purely real when rotated back to imaginary $\tau$. Since the shot moves faster than the bounce, $S_S^0> S_B^0$. Therefore, the shot is exponentially suppressed with respect to the bounce, and the latter dominates the path integral. 

Thus it can be seen that the direct method provides an alternative route towards the calculation of the false vacuum decay rate. It has been shown that applying the saddle point approximation to \eqref{eq:Gammadirect} yields an equivalent result to the semi-classical method. 

\subsection{Bubble expansion}
We will briefly comment on what happens to the critical bubbles after they have nucleated. At this point, there are several forces on the bubble wall: the latent heat of the phase transition results in an outward pressure (which we may call the vacuum pressure), bubble wall tension, and friction on the bubble wall, caused by interactions with the particles in the plasma. 

If these forces are balanced at some point before the bubble wall has traversed the inter-bubble distance, an asymptotic bubble wall speed is reached. What that bubble wall speed is has phenomenological consequences both for how the latent heat is dissipated and for processes which are concurrent with the phase transition, such as baryogenesis. 

\begin{figure}
     \centering
     \includegraphics[width=.9\textwidth]{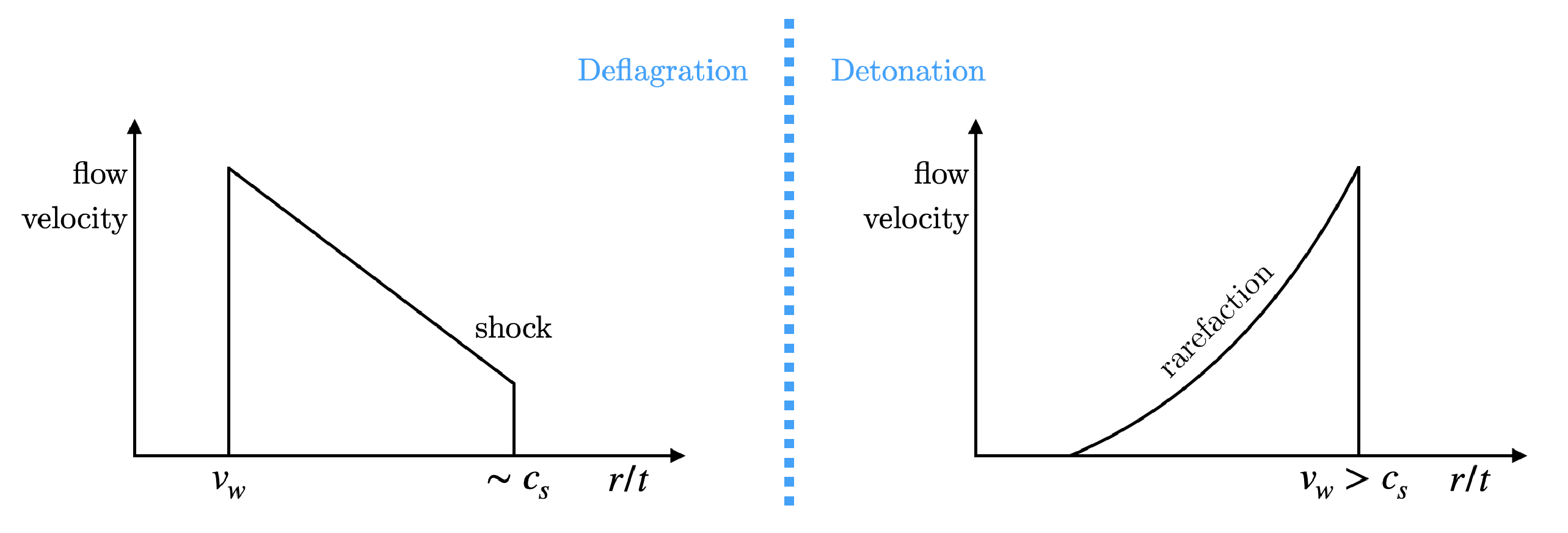}
     \caption{Fluid profiles for deflagrations and detonations \cite{Kamionkowski:1993fg}.}
     \label{fig:detdef}
 \end{figure}

Qualitatively, one can distinguish two different scenarios: the detonation scenario, in which the wall speed of the bubbles $v_w$ (sometimes called the wall velocity, despite the assumption that bubbles expand spherically and only the magnitude is therefore relevant) is larger than the speed of sound $ c_s = 1/\sqrt{3}$ in the relativistic plasma, and deflagrations, where it is smaller. The situations are different in fluid flow profiles, as illustrated in Fig.~\ref{fig:detdef}. Note that in the rest frame of the medium, the fluid velocity needs to be zero both at infinity and in the center of the bubble. In the deflagration case, a shock front of cosmological fluid travels in front of the bubble wall. In the detonation case, the fluid flow profile does not form a shockfront, but rather a rarefaction behind the bubble wall. The classic reference \cite{Kamionkowski:1993fg} contains a more in-depth discussion and plots of the fluid profile.

For gravitational wave generation, it is important to know whether the bubble wall velocity $v_w \to 1$, and moreover, the Lorentz factor $\gamma_w = (1-v_{w}^2)^{-1} \to \infty $. Such phase transitions are often referred to as runaway transitions. Implicitly, this means that most of the energy released in the phase transition has gone into accelerating the bubble walls, and has not been leaked to the plasma via friction. In this scenario the contribution to the gravitational wave spectrum coming from the bubble walls dominate over the processes in the plasma. 
Up until relatively recently, it was believed that runaway transitions could be realised in generic models of phase transitions. 
However, this insight was based on an incomplete estimation of the transition radiation of gauge fields in the bubble wall. Namely, the calculation of the 1-to-1 pressure for relativistic bubble walls, done in 2009 by Bodeker and Moore, found that
\se{ P_{\rm thermal } < P_{\rm vacuum }, \quad \quad \text{1-to-1 pressure \cite{Bodeker:2009qy} }, \label{eq:BM1}}
(where $P_{\rm thermal } $ is the pressure generated by interactions with the plasma)
was possible and therefore runaway transitions could occur in generic theories of particle physics. 
However, it turns out that when a gauge boson gains a mass in the phase transition (as is the case when the phase transition results from the spontaneous breaking of a gauge theory), a bunch of soft gauge bosons may be emitted in the process. A repeated calculation by the same authors in 2017 took into account an extra soft boson or multiple soft bosons in the final state, and found that 
\se{ P_{\rm thermal} \propto \gamma_w, \quad \quad \text{1-to-2 or 1-to-n pressure \cite{Bodeker:2017cim} }, \label{eq:BM2} }
and therefore a terminal velocity is always reached. Even more recently, other authors have used a different approach towards the 1-to-n resummation of all diagrams. In these works, it was found that the scaling of $P_{\rm thermal}$ scales with an even higher power of $\gamma_w $: 
\se{ P_{\rm thermal} \propto \gamma_w^{2}, \quad \quad \text{1-to-2 or 1-to-n pressure \cite{Hoche:2020ysm} }. \label{eq:Turneretal}}
There is no current consensus on what the appropriate scaling of $P_{\rm thermal}$ with $ \gamma_{w}$ should be (see also e.g. \cite{Azatov:2020ufh,Gouttenoire:2021kjv,DeCurtis:2022hlx,Lewicki:2022nba,Laurent:2022jrs}). However, in both cases \eqref{eq:BM2} and \eqref{eq:Turneretal} runaway phase transitions are not realised for models of spontaneous symmetry breaking of a gauge theory, which includes many important scenarios such as EWSB.

\subsection{Examples of first order phase transitions}\label{sec:models}
In the SM, there are two phase transitions in the early Universe: the phase transition related to the breaking of the electroweak symmetry, and that related to the confinement of the QCD sector. 
Alas, with the measured values of masses in couplings, we do not expect a first order phase transition took place in either instance. 
While indeed the weak gauge bosons and the Higgs self-coupling introduce an effective cubic coupling in the Higgs potential at high temperatures, this term comes with a coefficient that is too small to separate the vacua at finite temperature and that at zero temperature.\footnote{In fact, technically the transition is not even second-order: there is no discontinuity in the thermodynamic observables. This type of smooth change in the ground-state is called a crossover.}
For the case of QCD, confinement leads to a spontaneous breaking of the global chiral symmetry. The order of this process is somewhat more subtle to study (as we will see below), but through non-perturbative lattice simulations the consensus is that this transition was also not first-order. 
Nevertheless, there are many theories beyond the SM (BSM) in which a first order phase transition features, some of which are extra interesting because they also address a big open question in particle cosmology. In this section we will study some examples.

\subsubsection{The Electroweak Phase Transition} \label{sec:EWSB}
Electroweak symmetry breaking (EWSB) is an example of a broad class of phase transitions, related to the spontaneous breaking of a gauge symmetry. As EWSB happened in the context of the SM, and because it can be related to baryogenesis (as we will study in the next section), it is natural that the electroweak phase transition (EWPT) is a well-studied topic. 

We should begin our study of the EWPT by computing the thermal corrections to the potential of the Higgs boson $\phi $:\footnote{For a detailed overview of this computation, see \cite{Croon:2020cgk}.}
Using the 1-loop formalism and daisy resummation we introduced in the section above, we have: 
\se{
V_{{ \rm eff}}(\phi,T,\bar\mu) &=
    V_{{ \rm tree}} +
    V_{\rm 1-loop}
    \;.
}
where $\bar\mu $ is the renormalisation scale.
Because the Higgs is an SU(2) doublet, the contributions to the 1-loop potential in the direction that develops the VEV include both the physical Higgs field and the (would-be) goldstone modes. Of course, there are also corrections from gauge bosons and fermions which couple to the Higgs. 
The one-loop correction can be further divided into a zero-temperature Coleman-Weinberg piece and a thermal piece \se{V_{{ \rm 1-loop}} = V_{{ \rm CW}} + V_{\rm T} + V_{{ \rm daisy}}.}
We are mostly interested in the thermal piece here, which we can evaluate using the thermal functions we derived above: 
\se{V_{\rm T}  =& 
\frac{T^{4}}{2 \pi^{2}} 
\sum_{i \ni \phi, \chi, W, Z} c_{i} J_{B}(m_{i})  +  
\frac{T^{4}}{2 \pi^{2}} 
\sum_{j \ni t} c_{j} J_{F}(m_{j})
\\ =& \frac{T^{4}}{2 \pi^{2}}  \pL 
    J_{{ \rm 1-loop}}(m_\phi) - 4 N_{c} J_{{ \rm 1-loop}}(m_t) \pR  \\ &
    + \frac{T^{4}}{2 \pi^{2}}    \underbrace{ \pL 2 J_{{ \rm 1-loop}}(m^+_2)
    + 2 J_{{ \rm 1-loop}}(m^-_2) 
    + J_{{ \rm 1-loop}}(m^+_1)
    + J_{{ \rm 1-loop}}(m^-_1) \pR }_{\text{goldstone modes}} 
    \nn \\ &
    + \frac{T^{4}}{2 \pi^{2}}  \underbrace{\pL (D-1)\Big(
    2 J_{{ \rm 1-loop}}(m_W)
    + J_{{ \rm 1-loop}}(m_Z)
    + J_{{ \rm 1-loop}}(m_\gamma) \Big) \pR }_{\text{gauge bosons}}
    \;.
} 
with the field-dependent mass eigenvalues, given by 
\se{
\text{Higgs boson} & \quad 
m_\phi^2 =
    m_h^2
    + 3\lambda \phi^2
    \;,\\
\text{goldstone modes}&
\begin{cases}
(m_1^\pm )^2 &=
    \frac{1}{2}\left( m_\chi^2
    \pm \sqrt{ m_\chi^2 (m_\chi^2 - \xi_{1}{g'}^2 \phi^2 - \xi_{2} \g^{2} \phi^2)} \right)    \;,\\
(m_2^\pm )^2 &= (m_3^\pm )^2 =
    \frac{1}{2} \left( m_\chi^2 
    \pm \sqrt{m_\chi^2  (m_\chi^2 - \xi_{2}g^{2} \phi^2)} \right)  \;,
\end{cases} 
\\
\text{gauge bosons}&
\begin{cases}
m_W^2 &=
    \frac{1}{4}g^{2} \phi^2
    \;,\\
m_Z^2 &=
    \frac{1}{4} \Big( {g'}^2+g^{2} \Big) \phi^2
    \;,\\
m^2_{\gamma} &= 0
\;, 
\end{cases}
\\
\text{top quark} & \quad  m_t^2 =
    \frac{1}{2} g_Y^{2} \phi^2
    \;,
}
where
$m_\chi^2= m_h^2 + \lambda  \phi^2 $ is the Goldstone mode mass eigenvalue in the Landau gauge. Here $g$ is the SU(2) gauge coupling, $g'$ the U(1) gauge coupling, and $g_{Y}$ the top Yukawa coupling. As the latter dominates over the other Yukawa couplings, it is usually sufficient to consider the top quark only. 

At high temperature, the bosonic contributions to the thermal potential can introduce a barrier between the symmetric and the broken phase, as can be seen by expanding the thermal functions at high temperatures as we a saw above: 
\se{
J_{B} \pL \frac{m}{T} \pR \sim \frac{1}{24} \frac{m^{2}}{T^{2}} - \frac{1}{12\pi^{2}} \frac{|m|^{3}}{T^{3}}.
} 
It turns out that in the SM, the resulting cubic is too small to lead to a first order phase transition (which would require something like positive Lagrangian parameter $m^{2} $ and $\lambda_{3}^{2} > 4 m^{2} \lambda$). If the Higgs mass was smaller, it might have been. A lattice study based on the dimensionally reduced method introduced earlier on in these notes demonstrated that this would be the case for $ m_{h} \lesssim 72$ GeV \cite{Kajantie:1995kf,Kajantie:1996mn,Laine:1998jb,Laine:2000xu}.

While the EWPT is not a first order transition in the SM, it can be in models with an extended scalar sector. This includes theories with extra scalar singlets, doublets, or triplets coupling to the Higgs boson. We can demonstrate that the order of the phase transition can be changed through a simple example in which the theory has a heavier scalar singlet, coupling to the Higgs boson like 
\se{ \Lag \ni  \tilde{m} \phi^{2} s + \lambda_{\phi s}\phi^{2} s^{2}.
} 
At scales (much) below the mass of the singlet, this leads (besides corrections to the mass and the quartic) to an effective dimension-6 operator in the Higgs potential, which is now of the form,\footnote{A complete basis of operators includes besides this potential operator also derivative couplings, see e.g. \cite{Buchmuller:1985jz,Grzadkowski:2010es,deBlas:2014mba,Marzocca:2020jze}. }
\se{ V(\phi ) = \half m^{2} \phi^{2} + \fourth \lambda \phi^{4} + \frac{\phi^{6}}{\Lambda^{2}}.
} 

With this potential, it is even possible to have a tree-level barrier, as we can accomodate a positive mass squared parameter $m^{2} $ and a negative self coupling parameter $\lambda $. 
Note that this does not mean the effective self coupling is negative in the Higgs vacuum, just like the negative Higgs mass parameter does not mean the physical Higgs mass is negative in the SM.
Fixing the physical Higgs vev and Higgs mass in the zero temperature vacuum, the phase transition now depends on the parameter $\Lambda$. In this effective theory, previously studied in \cite{Grojean:2004xa,Bodeker:2004ws,Delaunay:2007wb,Cai:2017tmh,Chala:2018ari,Croon:2020cgk}, this parameter controls the strength of the phase transition, including its potential to lead to observable phenomenology. 

\subsubsection{Confinement and chiral symmetry breaking}\label{sec:confinement}
The EWPT is an example of a phase transition related to the breaking of a gauge symmetry. But this is not the only type of phase transition that we know to have occurred in the early Universe: the confinement of QCD is another kind. 
Above the confinement scale, the SM has a global $SU_{L}(N_{f}) \times SU_{R}(N_{f})$ symmetry, where $N_{f}$ is the number of dynamical flavours in the primordial plasma.
The confinement of free quarks into hadrons lead to the breaking of this preceding chiral symmetry to its diagonal subgroup:
$$ SU_{L}(N_{f}) \times SU_{R}(N_{f}) \to SU_{V} (N_{f}).$$ 
For SM QCD, at the temperature of confinement $ T_{\rm QCD} \sim 100 \text{MeV}-\text{GeV} $ (the exact temperature is a target of lattice simulations; the most recent result is $157$ MeV \cite{Borsanyi:2020fev}), the number of dynamical quarks in the plasma is $N_{f}=2$ or $N_{f} = 3$, depending on how you count the strange quark with mass $m_{s} \sim  95 \rm \, MeV $ (sometimes this is denoted as $N_{f} = 2+1 $). 

The order of the chiral symmetry breaking phase transition has been the topic of much debate. Low- and high energy effective theories become unreliable in the precise regime of the phase transition. 
An old and much cited argument by Pisarski and Wilczek (PW) \cite{Pisarski:1983ms} is based on the linear $\Sigma$-model -- a low energy theory of the quark bilinears in QCD which form a condensate $ \Sigma_{ij} \sim \langle {\bar \psi_{Rj}} {\psi_{Li}} \rangle  $. The field $\Sigma_{ij} $ can be decomposed into the pions and mesons (the light composite states) of QCD:
\begin{equation}
\Sigma_{ij}
    = \frac{ \varphi + i \eta' }{ \sqrt{ 2 N_F} } \delta_{ij} 
        + X^a T^a_{ij}
        + i \pi^a T^a_{ij} \ .
\label{eq:sigma-decomposition}
\end{equation}
In this model, one studies a $\Sigma $ field Lagrangian invariant under $SU_{L}(N_{f} \times SU_{R}(N_{f}) $: 
\se{
V(\Sigma) =  m_{\Sigma}^{2} \tr\bL \Sigma \Sigma^{\dagger} \bR - \mu_{\Sigma} \pL \det \Sigma + h.c. \pR + \frac{\lambda}{2} \pL \tr\bL  \Sigma \Sigma^{\dagger} \bR \pR^{2} + \frac{\kappa}{2} \tr\bL  \Sigma \Sigma^{\dagger}  \Sigma \Sigma^{\dagger} \bR 
}
where the $\mu_{\Sigma}$ term is generated by instanton effects, and gives the $\eta'$ a mass.\footnote{See \cite{Croon:2019iuh} for some further discussion.}
Then, chiral symmetry breaking corresponds to the real diagonal direction obtaining a vacuum expectation value, $\vev{\varphi} \sim f_{\pi} $.

PW derived the $\beta$ functions for the couplings $\lambda$ and $\kappa$ in this theory, in $4-\epsilon $ dimensions. Analysing these $\beta$ functions, they found that no infrared stable fixed point exists for $N_{f} \geq 3$. In RGE analyses, IR fixed points are associated with continuous (second order) phase transitions \cite{Bak:1976zza}.
Close to the fixed point, all physical quantities can be expressed as a power law in $(T-T_{c})/T_{c} $. For a first order (non-continuous) phase transition, no such scaling behaviour is observed; the thermodynamic variables instead exhibit discontinuities. 
The conclusion was therefore that for $N_{f} \geq 3$ the phase transition must be first order, realised by fluctuations. 

Needless to say, and also recognised by PW at the time, the argument is not water-tight. The linear $\Sigma$ model is a low-energy effective theory whose validity should break down right around the confinement scale, due to the appearance of free quarks in the spectrum. 
More importantly, the $4-\epsilon$ expansion might not be appropriate for an effectively 3-dimensional theory at finite temperature. And then the anomaly term $\mu_{\Sigma}$ needs to be taken into account away from the limit $N_{c} \to \infty$. 
Lattice studies have sought to shed light on this issue, with some studies in the 90s agreeing with PW \cite{Iwasaki:1995ij}, but some more recent studies not finding evidence for a first order phase transition up until $N_{f} = 6$ \cite{Cuteri:2021ikv}. 
For SM QCD, the consensus is that the transition is a crossover (at zero chemical potential, see e.g. \cite{HotQCD:2018pds,Aoki:2009sc,Aoki:2006we}). However, is still unclear whether a modified version in which the chiral symmetry is larger at the time of breaking does lead to a first order transition, a possibility that has been studied by model builders in the literature (e.g. \cite{Ipek:2018lhm,Croon:2019ugf,Croon:2019iuh,Berger:2020maa}). 

Lastly, let us comment on the possibility that $N_{f} = 0$, such that the phase transition is pure-Yang Mills. This phase transition obviously has little to do with chiral symmetry breaking, and is simpler to study on the lattice. In this case, the evidence points to a first order phase transition \cite{Svetitsky:1982gs,Kaczmarek:1999mm,Alexandrou:1998wv,Aoki:2006we,Saito:2011fs}.

\subsubsection{Hidden sectors}
A possibility which has become a popular topic of study in the previous few years, is that a phase transition in a hidden sector took place in the early Universe. For example, a dark Higgs mechanism is a plausible way of generating mass in a hidden sector. 
Unlike the Higgs sector of the Standard Model, which has been extensively studied and constrained by collider experiments, we have very limited knowledge about the masses and couplings in such a hidden sector. As a result, there is a higher degree of freedom for speculation and exploration in understanding its properties and implications.

Hidden sector phase transitions have garnered significant attention due to the possibility of producing detectable gravitational wave signals. 
It is an attractive idea that even with a secluded hidden sector -- with no (appreciable) non-gravitational interactions with the SM -- there is still a discovery potential via gravitational waves, although the information that can be gleaned from such signals is typically relatively limited, as we will see in the last section.
With our results from the previous sections, we can intuit some of the general rules of thumb for the strength of the phase transition, which relates to its observability: 
\begin{itemize} 
\item A significant latent heat needs to be released for the phase transition to lead to an observable gravitational wave signal, which implies that $\Delta V $ should be of order $T^{4}$ at the nucleation temperature; 
\item For a thermally induced barrier, the more bosons in the theory (which gain appreciable mass in the phase transition), the greater the effective cubic term in the potential and the stronger the phase transition; 
\item The shallower the order parameter potential (i.e. the larger the ratio $ v/\Lambda$, where $ V \propto \Lambda^{4}$), the stronger the transition \cite{Croon:2018erz}. 
\end{itemize}
Indeed, a gravitational wave signal from a first order phase transition could only be observable if a large fraction of the energy in the primordial plasma participated in the phase transition. Though this does not directly point to dark matter, many proposals nevertheless connect the hidden sector phase transition to an appreciable relic abundance.

\section{Baryogenesis}
It is almost self-evident that our Universe must contain more matter than anti-matter, for matter and anti-matter can mutually annihilate. The matter-antimatter asymmetry of the Universe (or the BAU, the baryon asymmetry of the Universe) can be measured in two different ways. The first is using the light element abundances formed during BBN, as these processes are sensitive to the baryon-to-photon ratio -- in particular, more $^{4}$He is formed for higher ratios (see, e.g. \cite{Fields:2019pfx}). The second is using the positions of the peaks in the CMB (most recently by the Planck collaboration \cite{Planck:2018vyg}), as the baryon-to-photon ratio reduces the sound speed, moving the peaks to smaller angular scales. The two measurements agree.\footnote{Note, however, the lithium problem: a tension between the observed abundance of $\rm ^{7}Li$ in the universe and CMB data.
}

As the baryon number density redshifts, it is convention to normalise the BAU to something else that redshifts at the same rate, to obtain a constant number. There are two different ways of doing this. The first is normalised to the photon number density:
\se{
\eta = \frac{n_{B} - n_{\bar{B}}}{n_{\gamma}} \sim 6.1 \times 10^{-10}
} 
(sometimes $n_{B} $ is used to denote the net baryon number) and the second is normalised to the entropy density: 
\se{
\eta = \frac{n_{B} - n_{\bar{B}}}{s} \sim 8.6 \times 10^{-11}.
}

The BAU could have formed across a vast range of scales in the early Universe, only bracketed by BBN -- as obvious from the above, to be consistent with the observed light element abundances today -- and the end of inflation, such that the BAU is not diluted.\footnote{Additionally, there are constraints on isocurvature perturbations in the CMB from rolling fields during inflation.} However, as we will see, in many models the BAU must have formed before EWSB (at $T\sim 100$ GeV), because the weak sphaleron processes form an essential component. 

There is no general theory to explain production of the BAU, or \emph{baryogenesis}; a vast number of proposals exist. In this section we will study the general aspects that such theories must necessarily address, and then we move on to study a few notable examples. Some recommended reviews of baryogenesis, including lectures given at previous instalments of TASI, are \cite{Riotto:1999yt,Dine:2003ax,Cline:2018fuq}.

\subsection{The Sakharov conditions}
Andrei Sakharov, in 1967, formulated three conditions which theories that explain the BAU must (generally) satisfy. The Sakharov conditions continue to be widely employed nowadays, and they offer a practical way of framing our discussion. Models of baryogenesis must:
\begin{enumerate}
    \item Violate baryon number conservation. This one is probably very obvious: if a theory conserves baryon number, it cannot create more baryons than anti-baryons.
    \item Violate C-- and CP symmetry. The former is such that the reactions producing baryons are not balanced by the reactions that produce anti-baryons. The latter is such that the theory does not produce the same number of left-handed baryons and right-handed anti-baryons (and vice versa).
    \item Out-of-equilibrium reactions. If the theory establishes equilibrium, thermal processes are balanced by their reverse processes, including the processes producing baryons. There are exceptions to this condition however, an important one being spontaneous baryogenesis which we will discuss below. 
\end{enumerate}
It is the last condition which gives the special role a first order phase transition can play in a theory of baryogenesis. In a first order phase transition, the baryon asymmetry can be created around the expanding bubble walls. There are other ways, however, the simplest being out-of-equilibrium decays. 

As certain Sakharov conditions are fulfilled within the framework of the SM, it is not immediately obvious that the BAU is a mystery. Therefore, it is instructive to investigate the extent to which these conditions are met within the SM:
\begin{enumerate}
\item The Standard Model features baryon number violation via the Bell-Jackiw chiral anomalies. Classically, the electroweak Lagrangian conserves baryon number, i.e. $ \partial^{\mu} j_{\mu}^{B} = 0 $, as quarks always appear in bilinears: 
\se{
j_{B}^{\mu} = \frac{1}{3} \sum_{i} \bar{Q}_{L}^{i} \gamma^{\mu} Q_{L}^{i} + \bar{u}_{R}^{i} \gamma^{\mu} u_{R}^{i} + \bar{d}_{R}^{i}\gamma^{\mu}d_{R}^{i}
\label{eq:baryoncurrent}
} 
However, quantum mechanically, this changes to 
\se{ \partial^{\mu} j_{\mu}^{B} = \frac{g^{2} c}{16 \pi^{2}} F^{\mu\nu a} \tilde{F}^{a}_{\mu\nu}
} 
where $c \propto \hbar$ is a constant ($c\to 0$ in the classical limit) and where $F_{\mu\nu}^{a}$ (with $\tilde{F}^{a}_{\mu\nu} = \epsilon_{\mu \nu \rho \sigma} F^{\rho \sigma a} $) is the electroweak gauge field strength. 

It turns out that \emph{sphalerons}, solutions to the electroweak field equations, mediate non-perturbative processes which can change anti-leptons into baryons and vice-versa. An example of a sphaleron process would be: 
$$ \bar\nu_{e}\bar\nu_{\mu}\bar\nu_{\tau} \to u_{L}d_{L}d_{L} c_{L}b_{L}d_{L} t_{L}b_{L}b_{L}.  $$
This process has lepton number $\Delta L = 3$, baryon number $\Delta B = 3$, $\Delta (B-L) = 0$ -- sphalerons violate $B+L$ and conserve $B-L$.
Sphalerons only couple to left-handed quarks as they are part of the weak sector of the SM. 

In the broken phase, sphaleron transitions have a rate 
\se{\Gamma_{\rm sph} \propto e^{-E_{\rm sph}(T)/T}}
where $E_{\rm sph}(T) $ is the sphaleron energy, which depends on the masses of the weak gauge bosons (and therefore on the vev of the Higgs field $\phi $). The transitions are in equilibrium for $\Gamma_{\rm sph} > H $, which happens before EWSB. In equilibrium, sphalerons realise 
\se{ 3 \mu_{B} + \sum_{i}^{3} \mu_{i L} = 0 } 
where the sum is over all lepton generations. 
Thus, they communicate a lepton asymmetry to a baryon asymmetry and vice versa. After EWSB, the sphaleron energy becomes large, and the transitions are exponentially shut off. 

\item The weak sector of the SM violates C-symmetry (in particular in the absence of right-handed neutrinos). The SM also features CP violation through the phases in the CKM and PMNS matrix. Because in general these phases are not invariant under redefinitions of the quark fields, a popular invariant to use is the Jarlskog invariant \cite{Jarlskog:1985ht}: 
\se{\notag J = c_{12} c_{13}^{2} c_{23} s_{12} s_{13} s_{23} \sin \delta \sim \mathcal{O}(10^{-5}) \quad \quad\quad \quad \text{CKM matrix}
} 
where the $c$'s and $s$'s are cosines and sines of angles in the CKM matrix, and $\delta$ is the complex phase. In the PMNS matrix there is currently only an upper limit on $J$, though the phase might be measured in future experiments such as DUNE. 

We can estimate the amount of CP violation realised by the complex phase in the CKM matrix as $ J \times (m_{c}- m_{u})(m_{t}-m_{c})(m_{t}-m_{u})(m_{s}-m_{d})(m_{b}-m_{s})(m_{b}-m_{d})/T^{6} \sim 10^{-16} \ll \eta $ for $T = 150 \, \rm GeV $.
Thus, the amount of CP violation in the CKM matrix is not large enough to realise baryogenesis within the SM.

\item In the previous section, we have already established that we do not expect a first order phase transition in the context of the SM, but we might investigate whether out-of-equilibrium decays are expected to happen. 
In the early Universe, particles of mass $m$ and (dominant) coupling $g$ feature out-of-equilibrium decays if 
\se{ \Gamma \sim \frac{g^{2} m}{4 \pi \gamma}  < H } 
where $\gamma = \vev{E} /m \sim 1$ for $m>T$. Assuming radiation domination, this implies an upper limit for the involved coupling of 
\se{ g < \sqrt{\frac{4 \pi m}{\sqrt{g_{*}} M_{p} } } }
if we conservatively assume the decay happens when $m\sim T$. This means that for particles with masses around the weak scale, the couplings would have to be extremely small to lead to out-of-equilibrium decays.
\end{enumerate}

Thus, it is not possible to explain the BAU within the SM, and its existence can be taken a piece of evidence that new physics is needed. 
BSM theories can realise baryogenesis in a variety of different ways, often taking advantage of one (or more) of the conditions already present in the SM. In the following we will discuss a few important examples. 

\subsection{Models of baryogenesis}
After examining the general conditions of baryogenesis, we will now delve into a few specific examples to illustrate the concept further. It is important to note that these examples are not intended to provide an exhaustive analysis. Nonetheless, they will help develop a better understanding of how the Sakharov conditions can be met in the early Universe and how new physics can be employed to explain the Baryon Asymmetry of the Universe (BAU).

\subsubsection{Electroweak baryogenesis}

In theories of electroweak baryogenesis (EWBG),\footnote{For reviews, see e.g. \cite{Morrissey:2012db}.} the Sakharov conditions for the generation of the matter-antimatter asymmetry are satisfied at the electroweak (EW) scale. 
The EWPT is a first order process, in which bubbles with $\vev{h} = v_h$ are nucleated in a background with $\vev{h}=0$. As the EWPT is not first order in the SM, models of EWBG typically include an extended scalar sector (including an extra singlet, for example, or a second Higgs doublet). 

Besides the first order phase transition, the CP violation of the SM also needs to be enhanced. 
Usually, the CP-violating source term is given by the complex spatially varying mass term of a fermionic field:
\se{
\mathcal{L} \ni - m(z) \bar\psi_L \psi_R - m^*(z) \bar\psi_R \psi_L  \quad \text{with} \quad m(z) = |m(z)|e^{i\theta(z)} = m_R(z) + i m_I(z)
}
where $z$ is the coordinate perpendicular to the bubble wall.
In this case, when the fermion is decomposed into a left-handed and a right-handed chiral component, the Dirac equation will split into two coupled EOM with opposite signs in front of the $\partial_{z} $ propagation term, meaning that the chiral components propagate differently in the $z$ direction. 
Note that electric dipole moment (EDM) constraints, in particular of the electron, severely constrain what CP violating sources are still allowed.  

The SM does already contain processes which can violate baryon number: the EW sphaleron transitions. EWBG takes advantage of these transitions, which only occur for $ T \gtrsim v_h$, so only outside of the expanding bubbles.  

 \begin{figure}
     \centering
     \includegraphics[width = 0.6\textwidth]{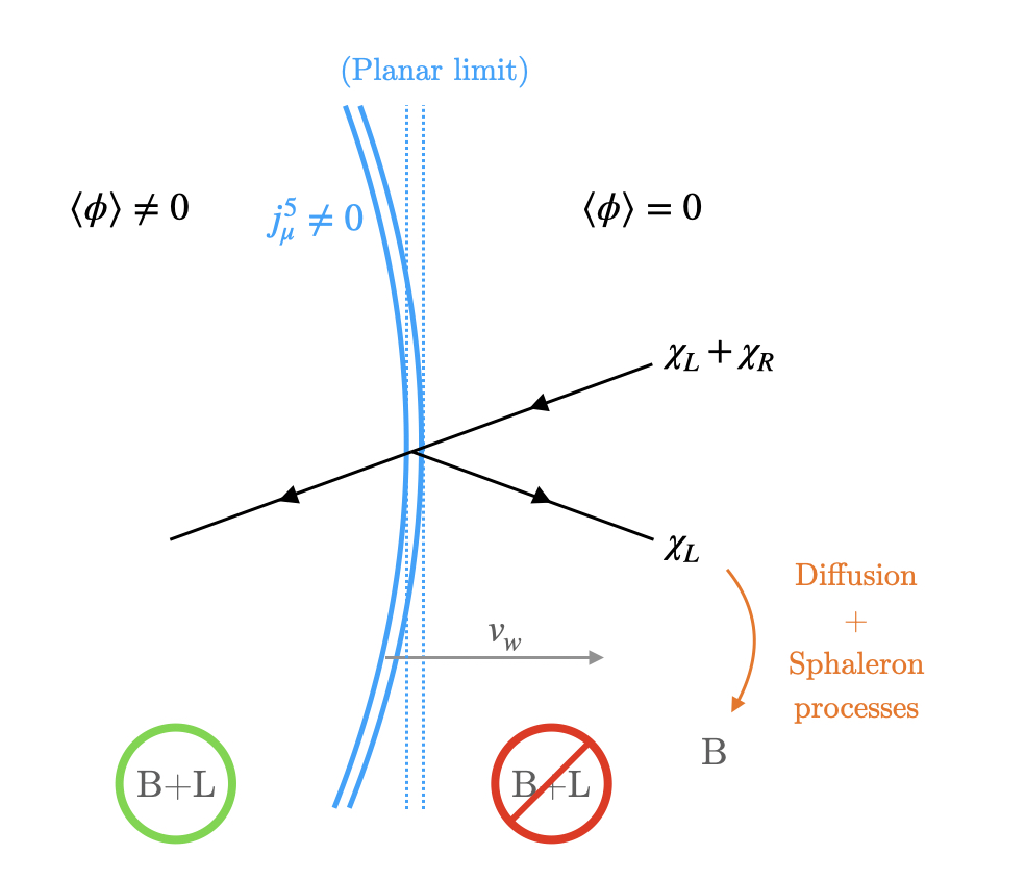}
     \caption{ }
     \label{fig:EWBGillustration}
 \end{figure}

The moving bubble walls separate particles in the plasma, see \Fig{fig:EWBGillustration}. After scattering with the bubble wall, the particles diffuse in the plasma. 
The typical diffusion time can be estimated by equating the diffusion length $d_{\rm diff}= \sqrt{ \bar{D} t}$ (where $\bar{D}$ is an effective diffusion constant, from Fick's law), with the distance the wall has moved in a particular time interval $ d_{\rm wall} = v_w t$:
\se{ \sqrt{ \bar{D} t} = v_w t \quad \to \quad t_{\rm diff} = \frac{\bar{D}}{v_w^2}.
\label{eq:tdiff}
}
The effective diffusion time $ t_{\rm diff} $ determines which processes need to be taken into account when considering the CP violating number densities in front of the bubble wall. For example, besides the CP-violating source, there may be interactions which dilute the number density.

In front of the bubble walls, then, the EW sphalerons convert the CP asymmetry (in particular, the assymmetry in left-handed particles over their anti-particles $n_{L}- \bar{n}_{L} $, as sphalerons are configurations which interpolate between distinct SU(2)$_{L}$ vacua) into a baryon asymmetry: 
\se{ \dmu j_{B}^{\mu} = - \frac{N_{f}}{2} \bL  c_{1} n_{B} + c_{2} n_{L}\bR
}
where $c_{1,2}$ are constants which depend on the weak sphaleron rate. 
The weak sphaleron rate is much slower than both the timescales of interactions with the bubble wall producing a net $n_{L}$ and the diffusion ahead of the bubble wall, so the steps are effectively decoupled (one first computes $n_{L}$ and then evaluates the sphaleron production of $n_{B}$).

The EW sphalerons are active before EWSB: the transitions are proportional to $\Gamma_{\rm sph} \propto {\rm exp}(- E_{\rm sph}(T)/T) $ where $E_{\rm sph} (T)$ is the sphaleron energy (which grows as $T$ falls), which is proportional to the masses of the EW gauge bosons. The sphalerons are in equilibrium if $ \Gamma_{\rm sph} > H$. Therefore, the sphaleron rate is suppressed if $ v_h /T \gtrsim 1$, in the interiors of the bubbles. This needs to be the case, because otherwise the net baryon number created in front of the expanding bubble walls is erased again inside the bubbles. 
The sharp turnoff of sphaleron processes across the wall solidifies the BAU.

\begin{figure}
     \centering
     \includegraphics[width = 0.95\textwidth]{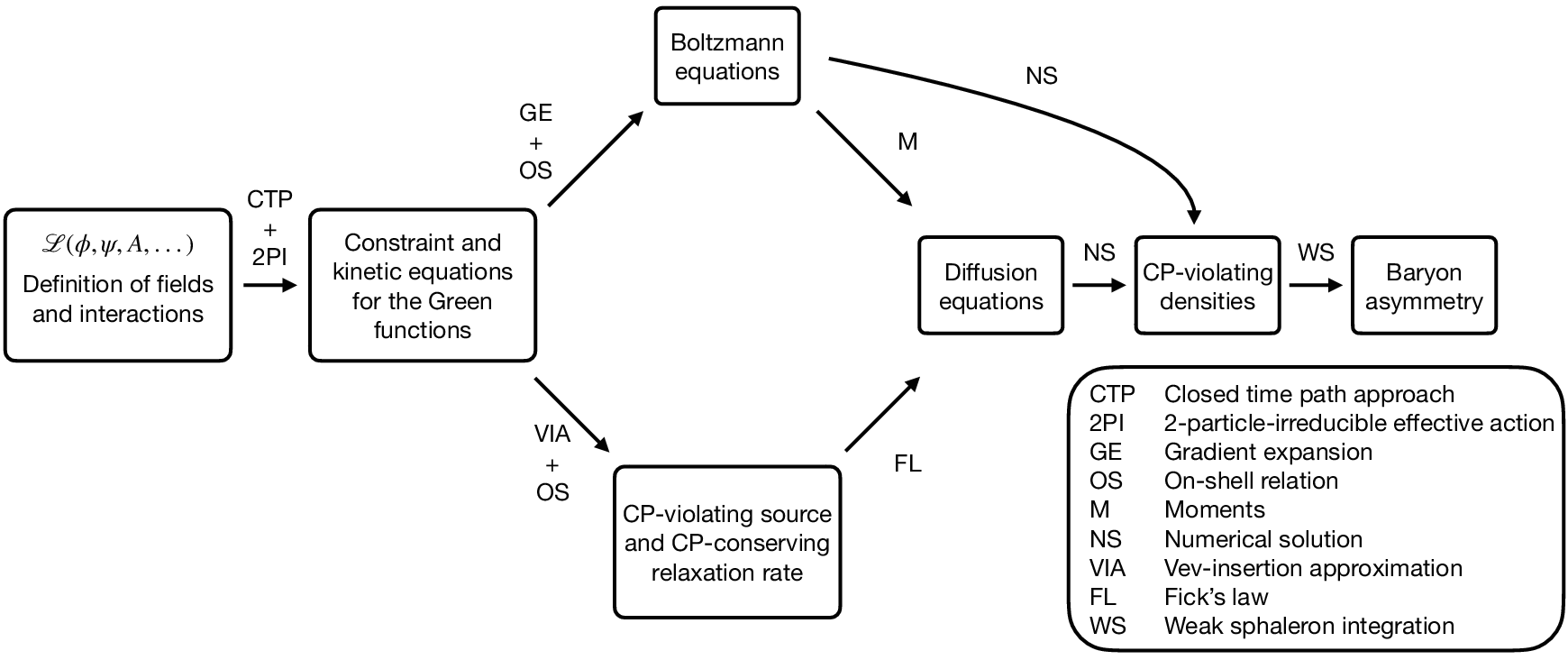}
     \caption{Schematic overview of the computation of the baryon asymmetry in EWBG calculations. From \cite{Asadi:2022njl}. }
     \label{fig:EWBScheme}
\end{figure}

Figure \ref{fig:EWBScheme} gives an overview of the calculation of the BAU in EWBG. As is seen in this figure, there are two different methods in which the CP-violation in front of the bubble wall is established. There is the well-established semi-classical method \cite{Kainulainen:2001cn,Prokopec:2003pj, Prokopec:2004ic} -- the top route in the diagram.
This method relies on the typical de Broglie wavelengths of the particles being small compared to the bubble wall width ($L_{dB} / L_{w} \ll 1 $), and an on-shell (OS) relation.
The alternative approach to derive the diffusion equations uses the so-called VEV-insertion approximation (VIA) \cite{Huet:1994jb, Huet:1995sh, Riotto:1995hh, Lee:2004we}, in which the VEV-dependent part of the mass is treated as a perturbation on top of the thermal masses. The VIA relates the self-energy and the Green's function. Then, the CPV source and relaxation rate enter into the diffusion equations obtained from Fick's law.  
For $\mathcal O(1)$ couplings between the CP-violating particle and the scalar field the VEV expansion should break down, though the VIA is still a topic of active debate outside this regime \cite{Postma:2019scv, Cline:2020jre, Postma:2021zux,Cline:2021dkf, Kainulainen:2021oqs,Postma:2022dbr}.

\subsubsection{Spontaneous baryogenesis}
The mechanism called spontaneous baryogenesis \cite{Cohen:1987vi} is an interesting counter-example to the universal validity of the Sakharov conditions, as it can happen in thermal equilibrium. There are a lot of different implementations of the mechanism, which share a few basic features. Namely, there is a term in the Lagrangian of the form
\se{ \mathcal{L} \ni \frac{\dmu \phi}{f}  j_{B}^{\mu}
\label{eq:LagSB}
} 
where $\phi$ is a scalar particle, usually a pseudo-Nambu Goldstone boson of the breaking of a global $U(1) $ symmetry (could be gauged baryon number), $f$ is some scale (if $\phi$ is a pNGB it is the corresponding decay constant), and where $j_{B}^{\mu}$ is the baryon current \eqref{eq:baryoncurrent}.  

Then, one assumes $\phi$ is spatially homogeneous, for example as it is smoothed out by inflation, which means
\se{ \dmu \phi \sim \partial_{t} \phi  \quad\quad \to \quad \quad \dmu \phi j_{B}^{\mu} \sim \partial_{t} \phi \, j_{B}^{0} = \partial_{t} \phi \, n_{B} }
recognising that the $0$-component of a current is a number density, and
\se{n_{B} \equiv  \sum_{i} n_{i} - \bar{n}_{i} .} 
Then, if $\phi$ starts to roll, it acts like a baryon chemical potential
\se{ \frac{\dot{\phi}}{f} = \mu_{B} \quad\quad \to \quad \quad \mathcal{L} \ni \mu_{B} n_{B}.
}
and breaks CPT dynamically. 
We can further write
\se{
n_{i} - \bar{n}_{i} &= g_{i} \int \frac{d^{3}p}{(2 \pi)^{3}} \bL  \frac{1}{e^{(p - \mu_{i})/T} +1 }-  \frac{1}{e^{(p + \mu_{i})/T} +1 }    \bR
\\ &= \frac{g_{i}}{6} \mu_{i} T^{2} \pL 1 + \mathcal{O} \pL \frac{\mu_{i}^{2}}{T^{2}} \pR \pR.
} 
In the presence of B-violating interactions which freeze out at some $T_{f}$, another necessary ingredient for spontaneous baryogenesis, the final baryon number is thus given by,
\se{
n_{B} \propto \left. \frac{\dot\phi}{f} T^{2} \right|_{T=T_{f}}.
}
Note that eventually the system falls to its ground state which has $\dot\phi = 0$, such that if the baryon number violating interactions do not freeze out before this, eventually the Universe would relax to $B = 0$.

A variation to spontaneous baryogenesis is QCD baryogenesis \cite{Ipek:2018lhm,Croon:2019ugf}, which relies on an early cosmological period of confinement. Such a period could be realised by a modification to the gluon kinetic term, i.e. 
\se{ -\fourth \pL \frac{1}{g_{s,0}^{2}} + \frac{S}{M_{*}} \pR G_{\mu\nu}^{a} G^{a\mu\nu}
}
where $S$ is a SM singlet, $M_{*}$ some mass scale, and $g_{s,0}$ the strong coupling in the absence of $S$. Now, one can derive the running of the effective strong coupling, which leads to a confinement scale 
\se{\Lambda_{\rm QCD} (v_{S}) = \Lambda_{0} e^{\frac{24 \pi^{2}}{2 N_{f}-33} \frac{v_{S}}{M_{*}}}
} 
where $\Lambda_{0}$ is the QCD scale for $v_{S}=0$ (note that this does not need to coincide with the SM value; we may live in a vacuum with $v_{S} \neq 0 $). Now for $v_{S} < 0 $ QCD confines earlier, and may even confine before EWSB. If this is the case, QCD confinement could be a first order phase transition, as all the SM quarks are light before it happens. 
Moreover, confinement \emph{causes} EWSB, as the quark yukawa couplings give a linear term in the Higgs potential, 
\se{ V (h) \ni y \vev{\bar{q} q} h \sim y \Lambda^{3}_{\rm QCD} h .}
This means that the sphaleron processes are active before confinement, but freeze out at the high confinement scale. 
 
Now, the QCD axion may play the role of the field $\phi$ in \eqref{eq:LagSB}. In QCD axion models, the strong sector of the SM may feature CP violation before confinement. At the QCD scale, the axion potential switches on, and dynamically relaxes the CP-violating $\bar\theta$-angle to zero. The action potential can be related to the $\vev{G \tilde{G}} $ condensate as
\se{ \frac{\alpha_{s}}{8 \pi} \vev{G \tilde{G}} = m_{a}^{2} f_{a}^{2} \sin \bar\theta }
Then, the CP-violation in the strong sector could be communicated to the weak sector through the $\eta'$ meson, which talks to both $ G \tilde{G}$ and $W \tilde{W}$. 

Finally, the resulting chemical potential can then be estimated as 
\se{\mu_{B} \propto \frac{1}{f_{\pi}^{2} m_{\eta'}^{2}}  \dd{}{t} \bL \sin\bar\theta \,  m_{a}^{2} f_{a}^{2} \bR 
} 
in particular as $m_{a}(T)$ evolves, this will be non-zero.

Of course, we live in a vacuum with $\Lambda_{\rm QCD} \sim 1 $ GeV, which we know from e.g. meson masses. Therefore, the period of early confinement needs to be followed by either a deconfinement phase transition, or a relaxation processes. Details of how that may occur in a minimal model can be found in \cite{Croon:2019ugf,Sagunski:2023ynd}.

\subsubsection{Affleck-Dine baryogenesis}
The Affleck-Dine mechanism provides a way of generating the BAU during and after inflation \cite{affleck1985new}. The mechanism requires the existence of flat directions with non-zero baryon number: scalar fields charged under $B $ (or $L$, or $B-L$) with a shallow potential, which are typically motivated in the context of supersymmetry. 
In its simplest form, the Affleck-Dine field is a complex scalar field, with a potential that looks like
\se{ V(\phi) = \half \mu^{2}\phi^{2} + \fourth \lambda \phi^{4} + \frac{1}{4}\lambda'  \left( \phi^{4} - \phi^{*4} \right).
}
Clearly, the last term breaks the U(1) symmetry otherwise associated with the field. If $V(\phi)^{1/4} \ll H$ during inflation, the field is stabilised by Hubble friction: the field is not in thermal in equilibrium. If this happens at nonzero $ \langle \phi \rangle$, it is clear CP is also broken.

As inflation ends and the universe reheats, the complex $\phi$ field begins to oscillate around its potential minima. Due to the $\lambda' $ term, the Affleck-Dine field experiences a complex phase evolution as it oscillates. This rotation gives rise to the creation of a nonzero baryon number density (the zeroth component of the baryon current carried by $\phi$), 
\se{ n_{B} = i c_{B} \left(\dot\phi^{*} \phi - \phi^{*}\dot\phi  \right)}
where $ c_{B}$ is the $B$ charge of the Affleck-Dine field. 
As the amplitude of oscillations redshifts away, the first few oscillations give the dominant contribution to the generated baryon asymmetry. 

For high scale inflation, the electroweak sphalerons are still active after the creation of baryon number by the Affleck-Dine mechanism. In this case, the sphalerons will redistribute some of the created baryon number into lepton number \cite{Harvey:1990qw}. However, as the electroweak sphalerons only violate $B+L$, a net asymmetry remains. 

\subsubsection{Leptogenesis}
To finish this section, let us briefly consider a class of models of baryogenesis where the departure from thermal equilibrium comes from decays. 
In theories of leptogenesis, decays of heavy sterile neutrinos create a lepton number asymmetry, which can then be converted to a baryon number by the electroweak sphalerons \cite{Fukugita:1986hr}. In these scenarios, the heavy sterile neutrinos can play a double role, as they can also function to explain the light neutrino masses via a type-1 seesaw model. The SM Lagrangian is supplemented a Majorana mass for the sterile neutrinos $N$ as well as by an interaction\footnote{I am using simplified notation here, for example suppressing generation indices. For a more complete treatment, please refer to reviews such as \cite{Luty:1992un,Buchmuller:2004nz,Davidson:2008bu}. }
\se{
\mathcal{L} \ni - \half M_{N} N^{2}- \lambda L H N  + h.c. 
} 
where $ \lambda$ is a coupling, $L$ a left-handed SM lepton, $H$ the Higgs doublet, and $N$ the sterile neutrino. This interaction can give rise to the following decays, 
$$ N \to LH \quad \quad \text{or} \quad\quad N \to \bar{L} \bar{H}. $$
the first of which violates lepton number by $\Delta L = 1 $, and the second violates it by $\Delta L = -1$. It is clear that if these processes occur at the same rate, the net lepton number generated is zero. However, CP-violation comes from the interference between the tree-level decay diagram and the 1-loop diagram, such that  
$\Gamma (N \to HL )< \Gamma (N \to \bar{H}\bar{L}) $. In particular, in the scenario where there are three right handed neutrinos with $M_{1} \ll M_{2}, M_{3} $, the interference between the different diagrams implies, 
\se{
\epsilon &= \frac{\Gamma (N_{1} \to HL )- \Gamma (N_{1} \to \bar{H}\bar{L} )}{\Gamma (N_{1} \to HL )+ \Gamma (N_{1} \to \bar{H}\bar{L} )} \\
&\propto \frac{\sum_{a=2,3} \im\bL (\lambda^{\dagger} \lambda)_{a1}^{2}\bR \frac{M_{1}}{M_{a}}}{(\lambda^{\dagger} \lambda)_{11}}.
} 

The produced lepton asymmetry is not the end of the story, as the decay products can thermalise and give rise to so-called \emph{washout} processes, include lepton-number-violating scatterings, decays, and inverse decays, which can reduce the produced asymmetry. It is possible to set constraints on the masses in couplings in the neutrino sector from the avoidance of washout and the necessity of out-of-equilibrium decays . 

The resulting baryon number can be expressed in terms of the final baryon number, noting that the EW sphalerons will drive $B + L \to 0 $. This implies $ B_{i} = (B_{i}+L_{i})/2 + (B_{i}-L_{i})/2 \to (B_{f}-L_{f})/2 = B_{f}$ such that if leptogenesis produces an initial $L_{i}$, the resulting baryon number after sphaleron processes is $B_{f} \sim -L_{i}/2 $ (a more detailed calculation shows a slight departure from this relation, see \cite{Harvey:1990qw}).

\section{Gravitational waves from the Early Universe}
Much of the present interest in primordial first order phase transitions can be explained by the prospect of detecting a stochastic gravitational wave background in the near future. As we will see, first order phase transitions represent one of the early Universe phenomena that could give rise to such a spectrum. If this signal were to be detected, it would provide us with the earliest known probe of the Universe. In this section, we will explore the anticipated phenomenology by first revisiting the fundamental aspects of gravitational wave sources and detection, followed by an examination of gravitational wave backgrounds.
 
\subsection{Gravitational wave basics}
The simplest way to get some intuition for gravitational waves is through linearized gravity, 
\se{g_{\mu \nu} = \eta_{\mu \nu} + h_{\mu \nu},
\label{eq:lineargrav}} 
where $\eta_{\mu \nu}$ is the Minkowski metric of flat space and $h_{\mu \nu}$ is a small perturbation. Here the Greek indices run over space-time dimensions: $\mu,\nu=0, 1,2,3$.
From Einstein's field equations (to linear order in $ h_{\mu \nu}$), one can obtain a wave equation for the (massless) metric perturbation $h_{\mu \nu}$, where the source term is given by the \emph{energy-momentum tensor} of the GW-source:\footnote{
This is actually a wave equation for the trace-reversed metric perturbation $ \tilde{h}_{\mu \nu} = h_{\mu \nu} - \eta_{\mu \nu} h/2$ where $h = h^\mu_\mu$ (note that this means that $\tilde{h} = -h$) and I have dropped the tilde, but for our purposes the difference is unimportant.
See e.g. Refs.~\cite{Flanagan:2005yc,Maggiore:2007ulw} for more detailed introductions to this equation and the solutions discussed in this section.}
\se{
\label{eq:metricwaveeqn}
\Box h_{\mu \nu} = - 16 \pi G_{N} T_{\mu \nu},
}
where the $\Box$ operator is the d'Alembertian:
$\Box=\partial_\mu\partial^\mu $.\footnote{Contrary to some of the literature on gravitational waves, I will continue to use particle physics units $c=1$ but $G_{N} \neq 1 $.} To derive this equation, one has to assume the Lorentz gauge, $ \partial^\nu h_{\mu \nu} = 0 $.

One thing that we can immediately examine is the what degrees of freedom live in $h_{\mu \nu}$. Because $h_{\mu \nu}$ is a symmetric rank 2 tensor, it has in principle 10 independent components. However, not all of these correspond to physical degrees of freedom: four of these are gauge degrees of freedom (corresponding to coordinate transformations). In fact, in deriving the wave equation, we have already had to assume the Lorentz gauge, fixing four components. 
If we are interested in the propagation of the wave, we solve the wave equation \emph{outside the source}, $\Box h_{\mu \nu} = 0 $. This can be used to determine four more components, so we should end up with $ 10- 2 \times 4 = 2$ propagating degrees of freedom. Typically, one fixes the gauge further into the transverse-traceless gauge (TT, meaning $ h_{\mu}^\mu = 0$, $h_{0 \mu}=0$, and $ \partial^i h_{ij} = 0$) and the two independent components are the $+$ and $\times$ polarisations of the wave. 
We can find an example: if we consider $h_{\mu\nu}$ to be a monochromatic wave traveling in the $z$-direction, it will take the following form in the TT gauge:
  \begin{equation}
    h_{\mu\nu}^{\rm TT} = \left[ h_{+} \left(
      \begin{bmatrix}
        0 & 0 & 0 & 0 \\
        0 & 1 & 0 & 0 \\
        0 & 0 & -1 & 0 \\
        0 & 0 & 0 & 0
      \end{bmatrix}
      \right)
      +
      h_{\times} \left(
            \begin{bmatrix}
        0 & 0 & 0 & 0 \\
        0 & 0 & 1 & 0 \\
        0 & 1 & 0 & 0 \\
        0 & 0 & 0 & 0
            \end{bmatrix}
            \right)
            \right] e^{ikz}.
  \end{equation}
The TT components are actually the only radiative components of $h_{\mu \nu}$; the fact that further components obey a wave equation in the Lorentz gauge is a gauge artifact. 

Just as in electrodynamics, the linear wave equation \eqref{eq:metricwaveeqn} can be solved by a retarded Green's function: 
\se{\label{eq:hgreens}
h_{ij}^{\rm TT}(t,\mathbf{x}) = 4 G_{N} \Lambda_{ij,kl} \int d^3 x' \frac{1}{| \mathbf{x} - \mathbf{x}'|} \, T_{kl} \pL t - |\mathbf{x}-\mathbf{x}'|, \mathbf{x}' \pR 
}
where $\Lambda_{ij,kl} = P_{ik}P_{jl} - P_{ij}P_{kl}/2$ with $P_{ij} (\hat{\vec{n}})= \delta_{ij} - n_{i}n_{j} $ (where $\hat{\vec{n}}$ is a unit vector in the direction of propagation, e.g. $\hat{\vec{n}} = (0,0,1) $ in our case above) is the transverse-traceless projector. As this solution is in the TT gauge, it only features spatial
dimensions ($i,j = 1,2,3$); the temporal contribution can be related to it using energy-momentum conservation.

The solution to the wave equation \eqref{eq:hgreens} is not in general not very insightful. However, in many cases, we can learn a lot from its limiting behaviour. For example, if the distance $r$  to the source is large compared to the typical size of the source, we can replace $|\mathbf{x} - \mathbf{x}'| \approx \mathbf{x} = r $:
\se{\label{eq:hgreens2}
h_{ij}^{\rm TT}(t,\mathbf{x}) = \frac{4 G_{N}}{r}  \, \Lambda_{ij,kl} \int d^3 x' \, T_{kl} \pL t - r,\mathbf{x}' \pR .
}
The replacement in the time argument of $T_{kl}$ corresponds to the assumption that the motion inside the source is non-relativistic (i.e. $v\ll 1$).
With a bit of massaging (taking advantage of the conservation of the energy-momentum tensor to write $\int d^{3} x T_{ij} = \int d^{3} x \partial_{t}^{2} (T_{00} x_{i}x_{j})/2 + \text{total derivatives}$), \eqref{eq:hgreens2} can be recognized as the first term in the expansion in spherical harmonics: the quadrupole (i.e. $h_{ij} = h_{ij}^{\text{[quad]}} + \text{...}$). It can be rewritten as
\se{
  \label{eq:GWquad}
    h_{ij}^{\text{[quad]}} =& \frac{2 G}{r} \, \Lambda_{ij,kl} \, \ddot{Q}_{kl} \quad \quad \text{where} \quad \quad Q_{ij} = \int d^3 x \, \rho(t,\mathbf{x})\left( x_i x_j - \frac{1}{3} r^2 \delta_{ij} \right)
}
which, as we anticipated, depends only on spatial dimensions.
$Q_{ij}$ is the \emph{quadrupole moment} tensor of the source (the
dots denote time derivatives), and $\rho(t,\mathbf{x})$ is a time-varying mass density distribution.

We made some assumptions in this derivation, most importantly the validity of the linearised version of the Einstein equations. This linearised version is not always valid; in particular, as gravitational waves cary energy, $h_{\mu \nu}$ itself should affect the metric. 
However, it turns out that it is straightforward to generalise the derivation, and all that it will cost is a replacement of the stress-energy tensor by $ T_{\mu \nu} + t_{\mu \nu}$, where $t_{\mu \nu}$ is a pseudo-tensor (constructed from $h_{\mu \nu}$) which takes into account the backreaction of $h_{\mu \nu}$ on the metric.
In situations where self-gravity dominates, $t_{\mu \nu}$ describes the gravitational binding energy. Thus, the quadrupole formula can be applied to situations like binary mergers.

Let us consider the result \eqref{eq:GWquad} for a moment. Unlike electromagnetism, which features dipole radiation, gravitational waves are generated by accelerated sources with nonzero quadrupole moments (or, in principle, higher order multipoles).
One way to explain the difference is using the fact that in electromagnetism you have opposite charges (positive and negative electric charge), whereas in gravity, the charge (mass) is positive definite. 
Dipole moments describe the distribution of charge away from some centre in some direction, so if the charges have equal sign and the center of mass is picked as the center, they vanish. In contrast, quadrupole moments do not depend directly on the sign of the charges.\footnote{In general relativity, the monopole describes the total amount of mass.}

There are a few practical lessons we can learn from the quadrupole formula for a single source \eqref{eq:GWquad}. An obvious one, which nonetheless gives some good first intuition, is that spherically symmetric systems do not generate gravitational waves. Moreover, static or uniformly moving systems do not generate gravitational waves either: acceleration is needed. 

To get some intuition for the types of sources that generate observable gravitational waves, we can make a very naive estimate of the size of $h_{\text{quad}}$. Using dimensional analysis, $\ddot{Q} $ should have mass dimension one. Let us take it to be $\ddot{Q} = 60 \, M_{\odot}$, the mass of a very large black hole. For $r$, let us take $r = 140 \times 10^6 \,\text{ly}$, the distance to the first binary NS merger measured by LIGO -- one of the more nearby ones. These two numbers imply together that $h_{\text{quad}} \sim 10^{-19}$: a very small, dimensionless number, despite the large ``source'' we chose. The reason is of course that Newton's constant makes for a very small coupling constant. 
This is why only gravitational waves from enormous and relatively nearby sources can be observed. 

On that topic of detection, let us make one last remark about \eqref{eq:GWquad}. Gravitational waves get weaker with $1/r$, not with $1/r^{2}$, like the gravitational force itself. The total \emph{energy} of the quadrupole radiation would fall off with $1/r^{2}$, but that is not what gets detected. Unlike photon detectors, which absorb EM radiation, gravitational wave detectors measure the deplacement of test masses by the action of a gravitational waves. Therefore, the signal is proportional to the amplitude of the wave, and only falls off as $1/r$. This allows future gravitational wave detectors to look back at very high redshifts compared to EM telescopes.

\subsection{Stochastic gravitational wave backgrounds}
The results in the previous section apply to transcient signals: binary mergers of compact objects which are ``loud'' enough to be individually observable, and perhaps other sources such as supernova explosions. While it is still true that for an observable signal, an energetic changing quadrupole moment is needed, gravitational waves from the early Universe fall into a different category: that of stochastic gravitational wave backgrounds (SGWB). This category includes gravitational waves from first order phase transitions, as we studied earlier in these notes. Much like the CMB, such gravitational wave sources are observable via their power spectrum.\footnote{Note that binary mergers also contribute a stochastic gravitational wave background, which is the sum of all binary merger signals which are cannot individually be resolved.}

To compute the SGWB power spectrum, we will use two-point correlators. For example, to find the 
 energy in a stochastic gravitational wave background (SGWB), one might expect to need to compute the two-point correlator of the energy-momentum tensor (and in particular its TT part): $\langle T_{ij}(x) T^{ij}(y)\rangle$.
It is useful to use a plane wave decomposition, 
\se{
h_{ij}(t,x) &= \int_{-\infty}^{\infty} d f \int_{S^2} d^2 \Omega_k \;  \; h_{ij}(f,k) \; e^{2 \pi i f (t - k\cdot x /c) }
\\ &= \sum_{A = +, \times} \int_{-\infty}^{\infty} d f \int_{S^2} d^2 \Omega_k \; e^{A}_{ij} \; h_A(f,k) \; e^{2 \pi i f (t - k\cdot x /c)}
}
For a stochastic background, the Fourier coefficients $ h_A(f,k)$ are random variables; their ensemble averages are more interesting. 
The $e^{A}_{ij}$ are again polarisation tensors. 
Since our stochastic background will consist of waves with all possible propagation directions (labeled by $k$), we can no longer use the transverse-traceless gauge to assume $i,j = 1,2$. Instead, the indices need to go from $1$ to $3$.

We are interested in cosmological sources, such that our stochastic background is stationary (in time), approximately Gaussian (i.e. all n-point functions can be expressed in terms of two-point functions $\langle h_A h_{A'} \rangle $), and isotropic. 
Most cosmological stochastic backgrounds are also unpolarised, although there are some examples which have definite polarisation (gauge boson production after axion inflation, for example).
The SGWB due to first order phase transitions is not expected to be polarised, such that the two point function of Fourier coefficients $ h_A (f,k)$ can be characterised by the so-called spectral density $S_h$:
\se{
\langle h^*_A(f,\normal) h_{A'}(f',\normal') \rangle = \overbrace{\delta (f - f')}^{\text{stationarity}} 
\overbrace{\frac{\delta^{(2)}(\normal,\normal')}{4 \pi}}^{\text{isotropy}} \overbrace{\delta_{A A'}}^{\text{unpolarised}} \half S_h (f)
\label{eq:twopointhA}
}
where the $4 \pi$ is just a normalisation choice, such that integrating \eqref{eq:twopointhA} over $\normal$ and $\normal'$ leads to $ \delta (f, f') 
\delta_{A A'} S_h (f)/2$ (as $ d^2 \normal = d \cos \theta d \phi$). The same goes for the factor $1/2$: a normalisation choice. 

The correlator \eqref{eq:twopointhA} implies that 
\se{ \langle h_{ij}(t) h^{ij} (t) \rangle = 4 \int_0^\infty d f S_h(f)
} 
where the $4$ follows from the normalisation of the polarisation tensor. 
In experiments, the spectral density $ S_h$ can be compared with the strain noise $S_n$. 
An equivalent definition that is sometimes used is the signal characteristic strain amplitude $h_c$, which for a stochastic gravitational wave background is defined by
\se{ \langle h_{ij} h^{ij} \rangle = 2 \int_{f = 0}^{f = \infty} d \log f h_c^2 (f)
}
such that $ S_f$ and $h_c$ are related by 
\se{
S_f = \half \frac{h_c^2}{f }.
}

For cosmological sources we are often interested in the energy density of the gravitational wave background. This is given by\footnote{Fun fact: for some time, it was controversial whether gravitational waves carried energy or not. If you're curious, look into the ``sticky bead'' gedanken experiment. }
\se{
\rho_{\rm GW} &= \frac{1}{32 \pi G_N} \left\langle \dot{h}_{ij}^2(t, \mathbf{x}) \right\rangle. }
For cosmological sources, the energy density in the SGWB is often written in terms of the critical energy density $\rho_{c}$ using the parameter $\Omega_{\rm GW}(f)$,
\se{
\rho_{\rm GW} &= \int_{f = 0}^{f = \infty} d \log f \,  \dd{\rho_{\rm GW}}{\log f}
\\ &\equiv  \rho_c \int_{f = 0}^{f = \infty} d \log f \,  \Omega_{\rm GW}(f)
}
where $ \rho_c = 3 H_0^2 / 8 \pi G_N$ is the critical energy density. This definition of $\Omega_{\rm GW}(f)$ can be a little confusing; as can be seen, it is the quantity which, \emph{when integrated over} $\log f$, yields the fractional energy density in gravitational waves $\Omega_{\rm GW} = \rho_{\rm GW} / \rho_{c} $.

\subsection{Gravitational waves from phase transitions}
As we have seen, first order phase transitions are described by the nucleation of spherical field configurations, ``bubbles'', of true vacuum in a background of false vacuum. Critically sized bubbles subsequently expand. However, we have also seen that spherically symmetric configurations cannot generate gravitational waves. It is only when expanding bubbles collide, and when the accompanying plasma shells collide, that gravitational waves are sourced \cite{Witten:1984rs,Hogan:1986qda}. 

Thus, there are three main sources of gravitational wave generation during (and after) a first order phase transition: the collisions of the walls of the bubble, the plasma shells which move at the speed of sound, and the turbulence in the plasma which results from the decay of this motion. 
Which contribution to the spectrum is most important depends on how the latent heat of the phase transition is released. The collisions of the scalar field shells themselves carry most energy in the so-called runaway transitions discussed in an earlier section, where $ \gamma_{w} \to \infty$. Away from this limit, the latent heat goes into acoustic motion of the plasma, which has a finite lifetime before it decays into turbulence. 

The calculation of the GW spectra has been approached both analytically (e.g. \cite{Jinno:2016vai,Hindmarsh:2016lnk,Hindmarsh:2019phv,Cai:2023guc}) and through simulations (e.g. \cite{Hindmarsh:2013xza,Hindmarsh:2015qta,Hindmarsh:2017gnf}). For the scalar field contribution, an oft-employed approximation is called the envelop approximation, and considers the bubble walls to be infinitesimally thin and to disappear immediately upon collision. In hydrodynamical simulations, correlators of the $T_{\mu \nu }$ can be extracted from the relativistic plasma during and after the transition. 

We have previously looked at two different scenarios for bubble expansion: deflagration and detonation, distinguished by the bubble wall speed compared to the speed of sound in the plasma. 
The most relevant situation for gravitational wave production is the detonation case, both because this scenario produces the largest spectra and because this scenario is most likely realised for interesting scenarios.\footnote{There is friction on the bubble wall from interactions with the plasma, and transition radiation. However, this is not typically large enough of a break to realise $v_w < c_s$.} 
The envelope approximation works particularly well for detonations, because the kinetic-energy density is concentrated in a thin shell near the bubble wall.\footnote{A fully analytic calculation the reader is referred to \cite{Jinno:2016vai}.} Moreover, because of the supersonic wall speed, the collisions do not affect the propagation of the parts of the bubble wall which have not collided yet. 

Even without doing a detailed calculation of the gravitational wave spectrum, it is possible to make a few estimates. This is because the typical frequency of a gravitational wave signal that originated in the early Universe today is largely determined by cosmological redshift. In other words,
\se{f_0 &= \frac{a(t_{*})}{a(t_0)} f_*
\\ &\sim 7.8 \times 10^{-16} \pL\frac{106}{g_* (T_*)} \pR^{1/3} \pL \frac{100 \, \rm GeV}{T_*}\pR f_*
}
where the subscript $0$ indicates a value today and the $*$ indicates the value at emission. Here $g_*$ is the number of degrees of freedom. The second line follows from the conservation of entropy during radiation domination (s.t. $ g_*(T_*) T_*^3 a_*^3= g_{*,0} T_0^3 a_0^3 \to a_* = (T_0/T_*) (g_0/g_*)^{1/3} a_0$).

The typical frequency at production $f_*$ can be estimated using the parameter $\beta$, a typical rate of the phase transition. Remember that the false vacuum decay rate was proportional to $ e^{-S_{E}/T}$. We can expand the exponentiated quantity around a particular time $t'$, 
\se{
\frac{S_{E}}{T} = \left. \frac{S_{E}}{T} \right|_{t=t'} + (t-t') \underbrace{\dd{}{t}  \left. \frac{S_{E}}{T} \right|_{t=t'}}_{-\beta} + \mathcal{O} (t-t')^{2}
}
defining the parameter $\beta = d/dt S_{E}/T$, such that
\se{
\Gamma = \Gamma (t') e^{\beta (t-t')}.
}
If the time $t'$ is chosen to be some characteristic time during the transition (for example the nucleation time), $\beta$ is often said to be the rate of the phase transition, and $\beta^{-1}$ its typical duration.

Associating $f_{*} $ with $\beta$, $\beta \sim f_*$, we can write
\se{f_0 &\sim 7.8 \times 10^{-16} \pL\frac{106}{g_* (T_*)} \pR^{1/3} \pL \frac{100 \, \rm GeV}{T_*}\pR H_*
\frac{\beta}{H_*}
\\ &\sim 10^{-5} \pL\frac{g_* (T_*)}{106} \pR^{1/6} \pL \frac{T_*}{100 \, \rm GeV}\pR 
\frac{\beta}{H_*} \, \rm Hz
} 
using that $\beta$ is usually normalised to the Hubble rate at the time of the transition. 
A more detailed calculation takes into account wall velocities away from $v_{w}=1$, as well as the fraction of space-time volume which has not decayed yet to estimate the bubble radius at collision to be $R = (8 \pi)^{1/3}v_{w}/\beta  $ and $f \sim R^{-1}$, see \cite{Enqvist:1991xw,Guth:1981uk}.

The ratio $\beta/H$ can be calculated for a given transition, and is usually $\mathcal{O}(10-10^{3})$. Note that if $\beta \lesssim H $, the phase transition is not expected to complete. Fixing $\beta/H = 100 $, and using $g_{*} = 106 $ as in the SM (noting that $f_{0}$ does not depend on the latter strongly), we can estimate the frequencies of phase transitions happening in the early Universe. For example, a phase transition happening at the weak scale, $ T_{*} = 100$ GeV, implies a gravitational wave spectrum today peaking at $f_{0} \sim 10^{-3}$ Hz, whereas a phase transition happening at $T_{*} = 10^{6} $ GeV would peak at $f_{0} \sim 10 $ Hz.
We can compare this with the typical frequency windows of experiments from which you can see that weak-scale phase transitions are in the range of space-based interferometers, explaining the interest into probing electroweak baryogenesis with LISA \cite{LISACosmologyWorkingGroup:2022jok}. 
To explain the very recently confirmed low-frequency SGWB in pulsar timing arrays \cite{NANOGrav:2023gor,Antoniadis:2023ott,Reardon:2023gzh,Xu:2023wog}, the temperature of the phase transition would have to be much lower.  

In these notes, we will not attempt to derive the amplitude of the SGWB from phase transitions, referring instead to results of simulations. Assuming the sound wave spectrum dominates these give \cite{Hindmarsh:2017gnf,Weir:2017wfa}, 
\begin{align}
\label{eq:Omega_sw}
h^2 \Omega_{\rm sw} (f) &=
    8.5 \times 10^{-6} \left( \frac{100}{g_\ast} \right)^{1/3}
    \kappa_f^2\,
    \frac{\alpha^2}{(1+\alpha)^2}
    \left( \frac{H_{*}}{\beta} \right) 
    S_{\rm sw}(f)
    \left(1-\frac{1}{\sqrt{1+2H_{*} t_{\rm sw}}} \right)
    \;, \\
\kappa_f &= \frac{\alpha}{\alpha+0.083 \sqrt{\alpha}+0.73}
    \;,
\end{align}
where $S_{\rm sw}(f)$ encodes the frequency dependence;
at the peak, $S_{\rm sw}(f)=1$. Here $t_{\rm sw} $ denotes the timescale on which acoustic waves are active, as eventually the plasma shells decay into turbulent motion. One would expect this to happen on timescales~\cite{Ellis:2019oqb,Guo:2020grp}%
\begin{equation}
H_{*} t_{\rm sw} = \frac{2(8 \pi )^{1/3}  \sqrt{1+\alpha }}{\sqrt{3\alpha \, \kappa_{f}}\, \beta/H_{*} }
\;,
\end{equation}
such that the faster the transition, the smaller the bubbles are at coalescence, and the faster the decay of the motion in the plasma into turbulence. 

Lastly, about the spectral shape, we can make the following comments, borrowing an argument from \cite{Caprini:2009fx}. 
We expect the energy density to go like $\rho_{\rm GW} \propto \vev{ \dot{h}_{ij}\dot{h}_{ij}} \propto f^{2} P_{s}(f) $, where we have assumed $\dot{h} \simeq f h $ and where $P_{s}(f)$ is some power spectrum. 
At frequencies smaller than the inverse of a characteristic length of the phase transition (for instance the inverse size of the bubbles at collision), $ f <  f_{\rm peak} $, we do not expect the power spectrum $P_{s}(f)$ to scale with $f$. Then, we find
\se{ \Omega_{\rm GW} (f) \propto \dd{\rho_{\rm GW}}{\log f} \propto f^{3} \quad \quad \quad f \ll f_{\rm peak}.
}  
At frequencies larger than the characteristic length, $P_{s}(f)$ has a negative power, to keep the total energy in gravitational waves finite. 
How quickly $\Omega_{\rm GW} (f)$ falls off depends on how coherent the source is, and typically differs for gravitational waves generated by plasma motion and scalar field shells.

\section{Conclusions and outlook}
The prospect of detecting gravitational wave signals from the pre-BBN Universe has reinvigorated research into cosmological first order phase transitions and related models of baryogenesis. These are mature research directions, with seminal works originating in the 1970s, 80s, and 90s. However, many significant technical challenges still lay ahead in characterising primordial gravitational wave spectra accurately, some of which we have identified in these notes. 

These notes started with an introductory overview of thermal field theory and a relatively in-depth discussion (in comparison to other topics in this work, certainly not in comparison with the literature) of false vacuum decay. 
I have found that studying these topics helped me develop a deeper understanding and greater appreciation of field theory. In my opinion, there are still great opportunities for technical developments in these areas, in particular in situations where the usual perturbative expansions are not reliable. This technical progress is essential to complement the many model-building efforts that are already flourishing, as it is required to make accurate phenomenological predictions. The same is true for calculations of the baryon asymmetry generated during first order phase transitions, though we spend less time on such in-depth calculations here. 

As gravitational waves were not introduced by any other lecturer at TASI 2022, I included a quick introduction to the topic. This was meant to develop some intuition for the types of events that could generate gravitational waves, before moving on to stochastic backgrounds such as generated by primordial sources.
Research on the gravitational wave phenomenology of first order phase transitions has to great extend been driven by enormous progress on hydrodynamical simulations of early-Universe bubble dynamics in the last decade. These notes have not focused on this aspect of the predictions -- primarily because this is not my area of expertise -- but there is a substantial amount of important ongoing research in this field, including on accurately capturing magnetohydrodynamic turbulence. 

In conclusion, the study of gravitational waves from phase transitions is a rapidly expanding field. The significant technical challenges are matched by the promise of this program, as it has the potential to provide profound insights into our understanding of the universe and its history. I hope that these notes are useful to anyone interested in contributing to this active area of research.

\acknowledgments
I am very thankful to the scientific organisers of the 2022 TASI summer school (JiJi Fan, Stefania Gori, Lian-Tao Wang) for inviting me to give these lectures and encouraging me to write up these lecture notes. I am also grateful to the local TASI organisers (Tom DeGrand, Oliver DeWolfe, and Ethan Neil) for ensuring the school ran smoothly. I thank the students attending TASI 2022 for their enthusiasm and many questions. 

I am indebted to my teachers and collaborators -- who I will not list individually, for fear of accidental omissions -- for teaching me many aspects relevant to these lectures over the years. Specifically, I am grateful to Oli Gould, Nell Hall, and Philipp Schicho, who provided insightful comments on these notes. Lastly, I am grateful to my partner for accompanying me to Boulder to give these lectures a couple of months before our son was born.

\bibliographystyle{JHEP}
\bibliography{refs}

\providecommand{\href}[2]{#2}\begingroup\raggedright\begin{thebibliography}{100}

\bibitem{Croon:2020cgk}
D.~Croon, O.~Gould, P.~Schicho, T.V.I.~Tenkanen and G.~White,
  \emph{{Theoretical uncertainties for cosmological first-order phase
  transitions}}, \href{https://doi.org/10.1007/JHEP04(2021)055}{\emph{JHEP}
  {\bfseries 04} (2021) 055}
  [\href{https://arxiv.org/abs/2009.10080}{{\ttfamily 2009.10080}}].

\bibitem{Ginsparg:1980ef}
P.H.~Ginsparg, \emph{{First Order and Second Order Phase Transitions in Gauge
  Theories at Finite Temperature}},
  \href{https://doi.org/10.1016/0550-3213(80)90418-6}{\emph{Nucl. Phys. B}
  {\bfseries 170} (1980) 388}.

\bibitem{Asadi:2022njl}
P.~Asadi et~al., \emph{{Early-Universe Model Building}},
  \href{https://arxiv.org/abs/2203.06680}{{\ttfamily 2203.06680}}.

\bibitem{Kapusta:2006pm}
J.I.~Kapusta and C.~Gale, \emph{{Finite-temperature field theory: Principles
  and applications}}, Cambridge Monographs on Mathematical Physics, Cambridge
  University Press (2011),
  \href{https://doi.org/10.1017/CBO9780511535130}{10.1017/CBO9780511535130}.

\bibitem{Quiros:1999jp}
M.~Quiros, \emph{{Finite temperature field theory and phase transitions}},  in
  \emph{{ICTP Summer School in High-Energy Physics and Cosmology}},
  pp.~187--259, 1, 1999 [\href{https://arxiv.org/abs/hep-ph/9901312}{{\ttfamily
  hep-ph/9901312}}].

\bibitem{Laine:2016hma}
M.~Laine and A.~Vuorinen, \emph{{Basics of Thermal Field Theory}}, vol.~925,
  Springer (2016),
  \href{https://doi.org/10.1007/978-3-319-31933-9}{10.1007/978-3-319-31933-9},
  [\href{https://arxiv.org/abs/1701.01554}{{\ttfamily 1701.01554}}].

\bibitem{das1997finite}
A.~Das, \emph{Finite temperature field theory}, World scientific (1997).

\bibitem{Bellac:2011kqa}
M.L.~Bellac, \emph{{Thermal Field Theory}}, Cambridge Monographs on
  Mathematical Physics, Cambridge University Press (3, 2011),
  \href{https://doi.org/10.1017/CBO9780511721700}{10.1017/CBO9780511721700}.

\bibitem{keldysh1965diagram}
L.V.~Keldysh et~al., \emph{Diagram technique for nonequilibrium processes},
  {\emph{Sov. Phys. JETP} {\bfseries 20} (1965) 1018}.

\bibitem{Schwinger:1960qe}
J.S.~Schwinger, \emph{{Brownian motion of a quantum oscillator}},
  \href{https://doi.org/10.1063/1.1703727}{\emph{J. Math. Phys.} {\bfseries 2}
  (1961) 407}.

\bibitem{Kubo:1957mj}
R.~Kubo, \emph{{Statistical mechanical theory of irreversible processes. 1.
  General theory and simple applications in magnetic and conduction problems}},
  \href{https://doi.org/10.1143/JPSJ.12.570}{\emph{J. Phys. Soc. Jap.}
  {\bfseries 12} (1957) 570}.

\bibitem{Martin:1959jp}
P.C.~Martin and J.S.~Schwinger, \emph{{Theory of many particle systems. 1.}},
  \href{https://doi.org/10.1103/PhysRev.115.1342}{\emph{Phys. Rev.} {\bfseries
  115} (1959) 1342}.

\bibitem{Matsubara:1955ws}
T.~Matsubara, \emph{{A New approach to quantum statistical mechanics}},
  \href{https://doi.org/10.1143/PTP.14.351}{\emph{Prog. Theor. Phys.}
  {\bfseries 14} (1955) 351}.

\bibitem{Dolan:1973qd}
L.~Dolan and R.~Jackiw, \emph{{Symmetry Behavior at Finite Temperature}},
  \href{https://doi.org/10.1103/PhysRevD.9.3320}{\emph{Phys. Rev. D} {\bfseries
  9} (1974) 3320}.

\bibitem{Jackiw:1974cv}
R.~Jackiw, \emph{{Functional evaluation of the effective potential}},
  \href{https://doi.org/10.1103/PhysRevD.9.1686}{\emph{Phys. Rev. D} {\bfseries
  9} (1974) 1686}.

\bibitem{Kajantie:1995dw}
K.~Kajantie, M.~Laine, K.~Rummukainen and M.E.~Shaposhnikov, \emph{{Generic
  rules for high temperature dimensional reduction and their application to the
  standard model}},
  \href{https://doi.org/10.1016/0550-3213(95)00549-8}{\emph{Nucl. Phys. B}
  {\bfseries 458} (1996) 90}
  [\href{https://arxiv.org/abs/hep-ph/9508379}{{\ttfamily hep-ph/9508379}}].

\bibitem{Linde:1980ts}
A.D.~Linde, \emph{{Infrared Problem in Thermodynamics of the Yang-Mills Gas}},
  \href{https://doi.org/10.1016/0370-2693(80)90769-8}{\emph{Phys. Lett. B}
  {\bfseries 96} (1980) 289}.

\bibitem{Kirzhnits:1976ts}
D.A.~Kirzhnits and A.D.~Linde, \emph{{Symmetry Behavior in Gauge Theories}},
  \href{https://doi.org/10.1016/0003-4916(76)90279-7}{\emph{Annals Phys.}
  {\bfseries 101} (1976) 195}.

\bibitem{Parwani:1991gq}
R.R.~Parwani, \emph{{Resummation in a hot scalar field theory}},
  \href{https://doi.org/10.1103/PhysRevD.45.4695}{\emph{Phys. Rev. D}
  {\bfseries 45} (1992) 4695}
  [\href{https://arxiv.org/abs/hep-ph/9204216}{{\ttfamily hep-ph/9204216}}].

\bibitem{Arnold:1992rz}
P.B.~Arnold and O.~Espinosa, \emph{{The Effective potential and first order
  phase transitions: Beyond leading-order}},
  \href{https://doi.org/10.1103/PhysRevD.47.3546}{\emph{Phys. Rev. D}
  {\bfseries 47} (1993) 3546}
  [\href{https://arxiv.org/abs/hep-ph/9212235}{{\ttfamily hep-ph/9212235}}].

\bibitem{Appelquist:1981vg}
T.~Appelquist and R.D.~Pisarski, \emph{{High-Temperature Yang-Mills Theories
  and Three-Dimensional Quantum Chromodynamics}},
  \href{https://doi.org/10.1103/PhysRevD.23.2305}{\emph{Phys. Rev. D}
  {\bfseries 23} (1981) 2305}.

\bibitem{Nadkarni:1982kb}
S.~Nadkarni, \emph{{Dimensional Reduction in Hot QCD}},
  \href{https://doi.org/10.1103/PhysRevD.27.917}{\emph{Phys. Rev. D} {\bfseries
  27} (1983) 917}.

\bibitem{Landsman:1989be}
N.P.~Landsman, \emph{{Limitations to Dimensional Reduction at High
  Temperature}},
  \href{https://doi.org/10.1016/0550-3213(89)90424-0}{\emph{Nucl. Phys. B}
  {\bfseries 322} (1989) 498}.

\bibitem{Farakos:1994kx}
K.~Farakos, K.~Kajantie, K.~Rummukainen and M.E.~Shaposhnikov, \emph{{3-D
  physics and the electroweak phase transition: Perturbation theory}},
  \href{https://doi.org/10.1016/0550-3213(94)90173-2}{\emph{Nucl. Phys.}
  {\bfseries B425} (1994) 67}
  [\href{https://arxiv.org/abs/hep-ph/9404201}{{\ttfamily hep-ph/9404201}}].

\bibitem{Braaten:1995cm}
E.~Braaten and A.~Nieto, \emph{{Effective field theory approach to high
  temperature thermodynamics}},
  \href{https://doi.org/10.1103/PhysRevD.51.6990}{\emph{Phys.\ Rev.} {\bfseries
  D51} (1995) 6990} [\href{https://arxiv.org/abs/hep-ph/9501375}{{\ttfamily
  hep-ph/9501375}}].

\bibitem{Braaten:1995jr}
E.~Braaten and A.~Nieto, \emph{{Free energy of QCD at high temperature}},
  \href{https://doi.org/10.1103/PhysRevD.53.3421}{\emph{Phys. Rev. D}
  {\bfseries 53} (1996) 3421}
  [\href{https://arxiv.org/abs/hep-ph/9510408}{{\ttfamily hep-ph/9510408}}].

\bibitem{Gould:2022ran}
O.~Gould, S.~G\"uyer and K.~Rummukainen, \emph{{First-order electroweak phase
  transitions: A nonperturbative update}},
  \href{https://doi.org/10.1103/PhysRevD.106.114507}{\emph{Phys. Rev. D}
  {\bfseries 106} (2022) 114507}
  [\href{https://arxiv.org/abs/2205.07238}{{\ttfamily 2205.07238}}].

\bibitem{Schicho:2021gca}
P.M.~Schicho, T.V.I.~Tenkanen and J.~\"Osterman, \emph{{Robust approach to
  thermal resummation: Standard Model meets a singlet}},
  \href{https://doi.org/10.1007/JHEP06(2021)130}{\emph{JHEP} {\bfseries 06}
  (2021) 130} [\href{https://arxiv.org/abs/2102.11145}{{\ttfamily
  2102.11145}}].

\bibitem{Farakos:1994xh}
K.~Farakos, K.~Kajantie, K.~Rummukainen and M.E.~Shaposhnikov, \emph{{3d
  physics and the electroweak phase transition: A Framework for lattice Monte
  Carlo analysis}},
  \href{https://doi.org/10.1016/0550-3213(95)80129-4}{\emph{Nucl. Phys.}
  {\bfseries B442} (1995) 317}
  [\href{https://arxiv.org/abs/hep-lat/9412091}{{\ttfamily hep-lat/9412091}}].

\bibitem{Kajantie:1995kf}
K.~Kajantie, M.~Laine, K.~Rummukainen and M.E.~Shaposhnikov, \emph{{The
  Electroweak phase transition: A Nonperturbative analysis}},
  \href{https://doi.org/10.1016/0550-3213(96)00052-1}{\emph{Nucl. Phys. B}
  {\bfseries 466} (1996) 189}
  [\href{https://arxiv.org/abs/hep-lat/9510020}{{\ttfamily hep-lat/9510020}}].

\bibitem{Kajantie:1996qd}
K.~Kajantie, M.~Laine, K.~Rummukainen and M.E.~Shaposhnikov, \emph{{A
  Nonperturbative analysis of the finite T phase transition in ${\rm
  SU(2)}\times{\rm U}(1)$ electroweak theory}},
  \href{https://doi.org/10.1016/S0550-3213(97)00164-8}{\emph{Nucl. Phys.}
  {\bfseries B493} (1997) 413}
  [\href{https://arxiv.org/abs/hep-lat/9612006}{{\ttfamily hep-lat/9612006}}].

\bibitem{Laine:2018lgj}
M.~Laine, P.~Schicho and Y.~Schr\"oder, \emph{{Soft thermal contributions to
  3-loop gauge coupling}},
  \href{https://doi.org/10.1007/JHEP05(2018)037}{\emph{JHEP} {\bfseries 05}
  (2018) 037} [\href{https://arxiv.org/abs/1803.08689}{{\ttfamily
  1803.08689}}].

\bibitem{Niemi:2021qvp}
L.~Niemi, P.~Schicho and T.V.I.~Tenkanen, \emph{{Singlet-assisted electroweak
  phase transition at two loops}},
  \href{https://doi.org/10.1103/PhysRevD.103.115035}{\emph{Phys. Rev. D}
  {\bfseries 103} (2021) 115035}
  [\href{https://arxiv.org/abs/2103.07467}{{\ttfamily 2103.07467}}].

\bibitem{Gould:2021oba}
O.~Gould and T.V.I.~Tenkanen, \emph{{On the perturbative expansion at high
  temperature and implications for cosmological phase transitions}},
  \href{https://doi.org/10.1007/JHEP06(2021)069}{\emph{JHEP} {\bfseries 06}
  (2021) 069} [\href{https://arxiv.org/abs/2104.04399}{{\ttfamily
  2104.04399}}].

\bibitem{Ekstedt:2022zro}
A.~Ekstedt, O.~Gould and J.~L\"ofgren, \emph{{Radiative first-order phase
  transitions to next-to-next-to-leading order}},
  \href{https://doi.org/10.1103/PhysRevD.106.036012}{\emph{Phys. Rev. D}
  {\bfseries 106} (2022) 036012}
  [\href{https://arxiv.org/abs/2205.07241}{{\ttfamily 2205.07241}}].

\bibitem{Ekstedt:2022bff}
A.~Ekstedt, P.~Schicho and T.V.I.~Tenkanen, \emph{{DRalgo: A package for
  effective field theory approach for thermal phase transitions}},
  \href{https://doi.org/10.1016/j.cpc.2023.108725}{\emph{Comput. Phys. Commun.}
  {\bfseries 288} (2023) 108725}
  [\href{https://arxiv.org/abs/2205.08815}{{\ttfamily 2205.08815}}].

\bibitem{Wetterich:1992yh}
C.~Wetterich, \emph{{Exact evolution equation for the effective potential}},
  \href{https://doi.org/10.1016/0370-2693(93)90726-X}{\emph{Phys. Lett. B}
  {\bfseries 301} (1993) 90}
  [\href{https://arxiv.org/abs/1710.05815}{{\ttfamily 1710.05815}}].

\bibitem{Dupuis:2020fhh}
N.~Dupuis, L.~Canet, A.~Eichhorn, W.~Metzner, J.M.~Pawlowski, M.~Tissier
  et~al., \emph{{The nonperturbative functional renormalization group and its
  applications}},  \href{https://arxiv.org/abs/2006.04853}{{\ttfamily
  2006.04853}}.

\bibitem{Croon:2021vtc}
D.~Croon, E.~Hall and H.~Murayama, \emph{{Non-perturbative methods for false
  vacuum decay}},  \href{https://arxiv.org/abs/2104.10687}{{\ttfamily
  2104.10687}}.

\bibitem{Coleman:1977py}
S.R.~Coleman, \emph{{The Fate of the False Vacuum. 1. Semiclassical Theory}},
  \href{https://doi.org/10.1103/PhysRevD.16.1248}{\emph{Phys. Rev. D}
  {\bfseries 15} (1977) 2929}.

\bibitem{Callan:1977pt}
C.G.~Callan, Jr. and S.R.~Coleman, \emph{{The Fate of the False Vacuum. 2.
  First Quantum Corrections}},
  \href{https://doi.org/10.1103/PhysRevD.16.1762}{\emph{Phys. Rev. D}
  {\bfseries 16} (1977) 1762}.

\bibitem{Andreassen:2016cff}
A.~Andreassen, D.~Farhi, W.~Frost and M.D.~Schwartz, \emph{{Direct Approach to
  Quantum Tunneling}},
  \href{https://doi.org/10.1103/PhysRevLett.117.231601}{\emph{Phys. Rev. Lett.}
  {\bfseries 117} (2016) 231601}
  [\href{https://arxiv.org/abs/1602.01102}{{\ttfamily 1602.01102}}].

\bibitem{Andreassen:2016cvx}
A.~Andreassen, D.~Farhi, W.~Frost and M.D.~Schwartz, \emph{{Precision decay
  rate calculations in quantum field theory}},
  \href{https://doi.org/10.1103/PhysRevD.95.085011}{\emph{Phys. Rev. D}
  {\bfseries 95} (2017) 085011}
  [\href{https://arxiv.org/abs/1604.06090}{{\ttfamily 1604.06090}}].

\bibitem{Athron:2023xlk}
P.~Athron, C.~Bal\'azs, A.~Fowlie, L.~Morris and L.~Wu, \emph{{Cosmological
  phase transitions: from perturbative particle physics to gravitational
  waves}},  \href{https://arxiv.org/abs/2305.02357}{{\ttfamily 2305.02357}}.

\bibitem{Ai:2019dqr}
W.-Y.~Ai, \emph{{Aspects of false vacuum decay}}, Ph.D. thesis, Munich, Tech.
  U., 2019.

\bibitem{Devoto:2022qen}
F.~Devoto, S.~Devoto, L.~Di~Luzio and G.~Ridolfi, \emph{{False vacuum decay: an
  introductory review}},
  \href{https://doi.org/10.1088/1361-6471/ac7f24}{\emph{J. Phys. G} {\bfseries
  49} (2022) 103001} [\href{https://arxiv.org/abs/2205.03140}{{\ttfamily
  2205.03140}}].

\bibitem{Ai:2019fri}
W.-Y.~Ai, B.~Garbrecht and C.~Tamarit, \emph{{Functional methods for false
  vacuum decay in real time}},
  \href{https://doi.org/10.1007/JHEP12(2019)095}{\emph{JHEP} {\bfseries 12}
  (2019) 095} [\href{https://arxiv.org/abs/1905.04236}{{\ttfamily
  1905.04236}}].

\bibitem{Coleman:1977th}
S.R.~Coleman, V.~Glaser and A.~Martin, \emph{{Action Minima Among Solutions to
  a Class of Euclidean Scalar Field Equations}},
  \href{https://doi.org/10.1007/BF01609421}{\emph{Commun. Math. Phys.}
  {\bfseries 58} (1978) 211}.

\bibitem{LANGER1967108}
J.~Langer, \emph{Theory of the condensation point},
  \href{https://doi.org/https://doi.org/10.1016/0003-4916(67)90200-X}{\emph{Annals
  of Physics} {\bfseries 41} (1967) 108}.

\bibitem{Affleck:1980ac}
I.~Affleck, \emph{{Quantum Statistical Metastability}},
  \href{https://doi.org/10.1103/PhysRevLett.46.388}{\emph{Phys. Rev. Lett.}
  {\bfseries 46} (1981) 388}.

\bibitem{Gould:2021ccf}
O.~Gould and J.~Hirvonen, \emph{{Effective field theory approach to thermal
  bubble nucleation}},
  \href{https://doi.org/10.1103/PhysRevD.104.096015}{\emph{Phys. Rev. D}
  {\bfseries 104} (2021) 096015}
  [\href{https://arxiv.org/abs/2108.04377}{{\ttfamily 2108.04377}}].

\bibitem{Ekstedt:2022tqk}
A.~Ekstedt, \emph{{Bubble Nucleation to All Orders}},
  \href{https://arxiv.org/abs/2201.07331}{{\ttfamily 2201.07331}}.

\bibitem{Kamionkowski:1993fg}
M.~Kamionkowski, A.~Kosowsky and M.S.~Turner, \emph{{Gravitational radiation
  from first order phase transitions}},
  \href{https://doi.org/10.1103/PhysRevD.49.2837}{\emph{Phys. Rev. D}
  {\bfseries 49} (1994) 2837}
  [\href{https://arxiv.org/abs/astro-ph/9310044}{{\ttfamily
  astro-ph/9310044}}].

\bibitem{Bodeker:2009qy}
D.~Bodeker and G.D.~Moore, \emph{{Can electroweak bubble walls run away?}},
  \href{https://doi.org/10.1088/1475-7516/2009/05/009}{\emph{JCAP} {\bfseries
  05} (2009) 009} [\href{https://arxiv.org/abs/0903.4099}{{\ttfamily
  0903.4099}}].

\bibitem{Bodeker:2017cim}
D.~Bodeker and G.D.~Moore, \emph{{Electroweak Bubble Wall Speed Limit}},
  \href{https://doi.org/10.1088/1475-7516/2017/05/025}{\emph{JCAP} {\bfseries
  05} (2017) 025} [\href{https://arxiv.org/abs/1703.08215}{{\ttfamily
  1703.08215}}].

\bibitem{Hoche:2020ysm}
S.~H\"oche, J.~Kozaczuk, A.J.~Long, J.~Turner and Y.~Wang, \emph{{Towards an
  all-orders calculation of the electroweak bubble wall velocity}},
  \href{https://doi.org/10.1088/1475-7516/2021/03/009}{\emph{JCAP} {\bfseries
  03} (2021) 009} [\href{https://arxiv.org/abs/2007.10343}{{\ttfamily
  2007.10343}}].

\bibitem{Azatov:2020ufh}
A.~Azatov and M.~Vanvlasselaer, \emph{{Bubble wall velocity: heavy physics
  effects}}, \href{https://doi.org/10.1088/1475-7516/2021/01/058}{\emph{JCAP}
  {\bfseries 01} (2021) 058}
  [\href{https://arxiv.org/abs/2010.02590}{{\ttfamily 2010.02590}}].

\bibitem{Gouttenoire:2021kjv}
Y.~Gouttenoire, R.~Jinno and F.~Sala, \emph{{Friction pressure on relativistic
  bubble walls}}, \href{https://doi.org/10.1007/JHEP05(2022)004}{\emph{JHEP}
  {\bfseries 05} (2022) 004}
  [\href{https://arxiv.org/abs/2112.07686}{{\ttfamily 2112.07686}}].

\bibitem{DeCurtis:2022hlx}
S.~De~Curtis, L.D.~Rose, A.~Guiggiani, A.G.~Muyor and G.~Panico, \emph{{Bubble
  wall dynamics at the electroweak phase transition}},
  \href{https://doi.org/10.1007/JHEP03(2022)163}{\emph{JHEP} {\bfseries 03}
  (2022) 163} [\href{https://arxiv.org/abs/2201.08220}{{\ttfamily
  2201.08220}}].

\bibitem{Lewicki:2022nba}
M.~Lewicki, V.~Vaskonen and H.~Veerm\"ae, \emph{{Bubble dynamics in fluids with
  N-body simulations}},  \href{https://arxiv.org/abs/2205.05667}{{\ttfamily
  2205.05667}}.

\bibitem{Laurent:2022jrs}
B.~Laurent and J.M.~Cline, \emph{{First principles determination of bubble wall
  velocity}}, \href{https://doi.org/10.1103/PhysRevD.106.023501}{\emph{Phys.
  Rev. D} {\bfseries 106} (2022) 023501}
  [\href{https://arxiv.org/abs/2204.13120}{{\ttfamily 2204.13120}}].

\bibitem{Kajantie:1996mn}
K.~Kajantie, M.~Laine, K.~Rummukainen and M.E.~Shaposhnikov, \emph{{Is there a~
  hot electroweak phase transition at $m_H \gtrsim m_W$?}},
  \href{https://doi.org/10.1103/PhysRevLett.77.2887}{\emph{Phys. Rev. Lett.}
  {\bfseries 77} (1996) 2887}
  [\href{https://arxiv.org/abs/hep-ph/9605288}{{\ttfamily hep-ph/9605288}}].

\bibitem{Laine:1998jb}
M.~Laine and K.~Rummukainen, \emph{{What's new with the electroweak phase
  transition?}},
  \href{https://doi.org/10.1016/S0920-5632(99)85017-8}{\emph{Nucl. Phys. B
  Proc. Suppl.} {\bfseries 73} (1999) 180}
  [\href{https://arxiv.org/abs/hep-lat/9809045}{{\ttfamily hep-lat/9809045}}].

\bibitem{Laine:2000xu}
M.~Laine, \emph{{Electroweak phase transition beyond the standard model}},  in
  \emph{{4th International Conference on Strong and Electroweak Matter}},
  pp.~58--69, 6, 2000, \href{https://doi.org/10.1142/9789812799913_0005}{DOI}
  [\href{https://arxiv.org/abs/hep-ph/0010275}{{\ttfamily hep-ph/0010275}}].

\bibitem{Buchmuller:1985jz}
W.~Buchm{\"u}ller and D.~Wyler, \emph{{Effective Lagrangian Analysis of New
  Interactions and Flavor Conservation}},
  \href{https://doi.org/10.1016/0550-3213(86)90262-2}{\emph{Nucl. Phys.}
  {\bfseries B268} (1986) 621}.

\bibitem{Grzadkowski:2010es}
B.~Grzadkowski, M.~Iskrzynski, M.~Misiak and J.~Rosiek, \emph{{Dimension-Six
  Terms in the Standard Model Lagrangian}},
  \href{https://doi.org/10.1007/JHEP10(2010)085}{\emph{JHEP} {\bfseries 10}
  (2010) 085} [\href{https://arxiv.org/abs/1008.4884}{{\ttfamily 1008.4884}}].

\bibitem{deBlas:2014mba}
J.~de~Blas, M.~Chala, M.~Perez-Victoria and J.~Santiago, \emph{{Observable
  Effects of General New Scalar Particles}},
  \href{https://doi.org/10.1007/JHEP04(2015)078}{\emph{JHEP} {\bfseries 04}
  (2015) 078} [\href{https://arxiv.org/abs/1412.8480}{{\ttfamily 1412.8480}}].

\bibitem{Marzocca:2020jze}
D.~Marzocca et~al., \emph{{BSM Benchmarks for Effective Field Theories in Higgs
  and Electroweak Physics}}, {\emph{{}} (2020) }
  [\href{https://arxiv.org/abs/2009.01249}{{\ttfamily 2009.01249}}].

\bibitem{Grojean:2004xa}
C.~Grojean, G.~Servant and J.D.~Wells, \emph{{First-order electroweak phase
  transition in the standard model with a low cutoff}},
  \href{https://doi.org/10.1103/PhysRevD.71.036001}{\emph{Phys.\ Rev. D}
  {\bfseries 71} (2005) 036001}
  [\href{https://arxiv.org/abs/hep-ph/0407019}{{\ttfamily hep-ph/0407019}}].

\bibitem{Bodeker:2004ws}
D.~B{\"o}deker, L.~Fromme, S.J.~Huber and M.~Seniuch, \emph{{The Baryon
  asymmetry in the standard model with a low cut-off}},
  \href{https://doi.org/10.1088/1126-6708/2005/02/026}{\emph{JHEP} {\bfseries
  02} (2005) 026} [\href{https://arxiv.org/abs/hep-ph/0412366}{{\ttfamily
  hep-ph/0412366}}].

\bibitem{Delaunay:2007wb}
C.~Delaunay, C.~Grojean and J.D.~Wells, \emph{{Dynamics of Non-renormalizable
  Electroweak Symmetry Breaking}},
  \href{https://doi.org/10.1088/1126-6708/2008/04/029}{\emph{JHEP} {\bfseries
  04} (2008) 029} [\href{https://arxiv.org/abs/0711.2511}{{\ttfamily
  0711.2511}}].

\bibitem{Cai:2017tmh}
R.-G.~Cai, M.~Sasaki and S.-J.~Wang, \emph{{The gravitational waves from the
  first-order phase transition with a dimension-six operator}},
  \href{https://doi.org/10.1088/1475-7516/2017/08/004}{\emph{JCAP} {\bfseries
  08} (2017) 004} [\href{https://arxiv.org/abs/1707.03001}{{\ttfamily
  1707.03001}}].

\bibitem{Chala:2018ari}
M.~Chala, C.~Krause and G.~Nardini, \emph{{Signals of the electroweak phase
  transition at colliders and gravitational wave observatories}},
  \href{https://doi.org/10.1007/JHEP07(2018)062}{\emph{JHEP} {\bfseries 07}
  (2018) 062} [\href{https://arxiv.org/abs/1802.02168}{{\ttfamily
  1802.02168}}].

\bibitem{Borsanyi:2020fev}
S.~Borsanyi, Z.~Fodor, J.N.~Guenther, R.~Kara, S.D.~Katz, P.~Parotto et~al.,
  \emph{{QCD Crossover at Finite Chemical Potential from Lattice Simulations}},
  \href{https://doi.org/10.1103/PhysRevLett.125.052001}{\emph{Phys. Rev. Lett.}
  {\bfseries 125} (2020) 052001}
  [\href{https://arxiv.org/abs/2002.02821}{{\ttfamily 2002.02821}}].

\bibitem{Pisarski:1983ms}
R.D.~Pisarski and F.~Wilczek, \emph{{Remarks on the Chiral Phase Transition in
  Chromodynamics}}, \href{https://doi.org/10.1103/PhysRevD.29.338}{\emph{Phys.
  Rev. D} {\bfseries 29} (1984) 338}.

\bibitem{Croon:2019iuh}
D.~Croon, R.~Houtz and V.~Sanz, \emph{{Dynamical Axions and Gravitational
  Waves}}, \href{https://doi.org/10.1007/JHEP07(2019)146}{\emph{JHEP}
  {\bfseries 07} (2019) 146}
  [\href{https://arxiv.org/abs/1904.10967}{{\ttfamily 1904.10967}}].

\bibitem{Bak:1976zza}
P.~Bak, S.~Krinsky and D.~Mukamel, \emph{{First-Order Transitions, Symmetry,
  and the epsilon Expansion}},
  \href{https://doi.org/10.1103/PhysRevLett.36.52}{\emph{Phys. Rev. Lett.}
  {\bfseries 36} (1976) 52}.

\bibitem{Iwasaki:1995ij}
Y.~Iwasaki, K.~Kanaya, S.~Sakai and T.~Yoshie, \emph{{Chiral phase transition
  in lattice QCD with Wilson quarks}},
  \href{https://doi.org/10.1007/s002880050179}{\emph{Z. Phys. C} {\bfseries 71}
  (1996) 337} [\href{https://arxiv.org/abs/hep-lat/9504019}{{\ttfamily
  hep-lat/9504019}}].

\bibitem{Cuteri:2021ikv}
F.~Cuteri, O.~Philipsen and A.~Sciarra, \emph{{On the order of the QCD chiral
  phase transition for different numbers of quark flavours}},
  \href{https://doi.org/10.1007/JHEP11(2021)141}{\emph{JHEP} {\bfseries 11}
  (2021) 141} [\href{https://arxiv.org/abs/2107.12739}{{\ttfamily
  2107.12739}}].

\bibitem{HotQCD:2018pds}
{\scshape HotQCD} collaboration, \emph{{Chiral crossover in QCD at zero and
  non-zero chemical potentials}},
  \href{https://doi.org/10.1016/j.physletb.2019.05.013}{\emph{Phys. Lett. B}
  {\bfseries 795} (2019) 15}
  [\href{https://arxiv.org/abs/1812.08235}{{\ttfamily 1812.08235}}].

\bibitem{Aoki:2009sc}
Y.~Aoki, S.~Borsanyi, S.~Durr, Z.~Fodor, S.D.~Katz, S.~Krieg et~al., \emph{{The
  QCD transition temperature: results with physical masses in the continuum
  limit II.}}, \href{https://doi.org/10.1088/1126-6708/2009/06/088}{\emph{JHEP}
  {\bfseries 06} (2009) 088} [\href{https://arxiv.org/abs/0903.4155}{{\ttfamily
  0903.4155}}].

\bibitem{Aoki:2006we}
Y.~Aoki, G.~Endrodi, Z.~Fodor, S.D.~Katz and K.K.~Szabo, \emph{{The Order of
  the quantum chromodynamics transition predicted by the standard model of
  particle physics}}, \href{https://doi.org/10.1038/nature05120}{\emph{Nature}
  {\bfseries 443} (2006) 675}
  [\href{https://arxiv.org/abs/hep-lat/0611014}{{\ttfamily hep-lat/0611014}}].

\bibitem{Ipek:2018lhm}
S.~Ipek and T.M.P.~Tait, \emph{{Early Cosmological Period of QCD Confinement}},
  \href{https://doi.org/10.1103/PhysRevLett.122.112001}{\emph{Phys. Rev. Lett.}
  {\bfseries 122} (2019) 112001}
  [\href{https://arxiv.org/abs/1811.00559}{{\ttfamily 1811.00559}}].

\bibitem{Croon:2019ugf}
D.~Croon, J.N.~Howard, S.~Ipek and T.M.P.~Tait, \emph{{QCD baryogenesis}},
  \href{https://doi.org/10.1103/PhysRevD.101.055042}{\emph{Phys. Rev. D}
  {\bfseries 101} (2020) 055042}
  [\href{https://arxiv.org/abs/1911.01432}{{\ttfamily 1911.01432}}].

\bibitem{Berger:2020maa}
D.~Berger, S.~Ipek, T.M.P.~Tait and M.~Waterbury, \emph{{Dark Matter Freeze Out
  during an Early Cosmological Period of QCD Confinement}},
  \href{https://doi.org/10.1007/JHEP07(2020)192}{\emph{JHEP} {\bfseries 07}
  (2020) 192} [\href{https://arxiv.org/abs/2004.06727}{{\ttfamily
  2004.06727}}].

\bibitem{Svetitsky:1982gs}
B.~Svetitsky and L.G.~Yaffe, \emph{{Critical Behavior at Finite Temperature
  Confinement Transitions}},
  \href{https://doi.org/10.1016/0550-3213(82)90172-9}{\emph{Nucl. Phys. B}
  {\bfseries 210} (1982) 423}.

\bibitem{Kaczmarek:1999mm}
O.~Kaczmarek, F.~Karsch, E.~Laermann and M.~Lutgemeier, \emph{{Heavy quark
  potentials in quenched QCD at high temperature}},
  \href{https://doi.org/10.1103/PhysRevD.62.034021}{\emph{Phys. Rev. D}
  {\bfseries 62} (2000) 034021}
  [\href{https://arxiv.org/abs/hep-lat/9908010}{{\ttfamily hep-lat/9908010}}].

\bibitem{Alexandrou:1998wv}
C.~Alexandrou, A.~Borici, A.~Feo, P.~de~Forcrand, A.~Galli, F.~Jegerlehner
  et~al., \emph{{The Deconfinement phase transition in one flavor QCD}},
  \href{https://doi.org/10.1103/PhysRevD.60.034504}{\emph{Phys. Rev. D}
  {\bfseries 60} (1999) 034504}
  [\href{https://arxiv.org/abs/hep-lat/9811028}{{\ttfamily hep-lat/9811028}}].

\bibitem{Saito:2011fs}
{\scshape WHOT-QCD} collaboration, \emph{{Phase structure of finite temperature
  QCD in the heavy quark region}},
  \href{https://doi.org/10.1103/PhysRevD.85.079902}{\emph{Phys. Rev. D}
  {\bfseries 84} (2011) 054502}
  [\href{https://arxiv.org/abs/1106.0974}{{\ttfamily 1106.0974}}].

\bibitem{Croon:2018erz}
D.~Croon, V.~Sanz and G.~White, \emph{{Model Discrimination in Gravitational
  Wave spectra from Dark Phase Transitions}},
  \href{https://doi.org/10.1007/JHEP08(2018)203}{\emph{JHEP} {\bfseries 08}
  (2018) 203} [\href{https://arxiv.org/abs/1806.02332}{{\ttfamily
  1806.02332}}].

\bibitem{Fields:2019pfx}
B.D.~Fields, K.A.~Olive, T.-H.~Yeh and C.~Young, \emph{{Big-Bang
  Nucleosynthesis after Planck}},
  \href{https://doi.org/10.1088/1475-7516/2020/03/010}{\emph{JCAP} {\bfseries
  03} (2020) 010} [\href{https://arxiv.org/abs/1912.01132}{{\ttfamily
  1912.01132}}].

\bibitem{Planck:2018vyg}
{\scshape Planck} collaboration, \emph{{Planck 2018 results. VI. Cosmological
  parameters}},
  \href{https://doi.org/10.1051/0004-6361/201833910}{\emph{Astron. Astrophys.}
  {\bfseries 641} (2020) A6}
  [\href{https://arxiv.org/abs/1807.06209}{{\ttfamily 1807.06209}}].

\bibitem{Riotto:1999yt}
A.~Riotto and M.~Trodden, \emph{{Recent progress in baryogenesis}},
  \href{https://doi.org/10.1146/annurev.nucl.49.1.35}{\emph{Ann. Rev. Nucl.
  Part. Sci.} {\bfseries 49} (1999) 35}
  [\href{https://arxiv.org/abs/hep-ph/9901362}{{\ttfamily hep-ph/9901362}}].

\bibitem{Dine:2003ax}
M.~Dine and A.~Kusenko, \emph{{The Origin of the matter - antimatter
  asymmetry}}, \href{https://doi.org/10.1103/RevModPhys.76.1}{\emph{Rev. Mod.
  Phys.} {\bfseries 76} (2003) 1}
  [\href{https://arxiv.org/abs/hep-ph/0303065}{{\ttfamily hep-ph/0303065}}].

\bibitem{Cline:2018fuq}
J.M.~Cline, \emph{{TASI Lectures on Early Universe Cosmology: Inflation,
  Baryogenesis and Dark Matter}}, {\emph{PoS} {\bfseries TASI2018} (2019) 001}
  [\href{https://arxiv.org/abs/1807.08749}{{\ttfamily 1807.08749}}].

\bibitem{Jarlskog:1985ht}
C.~Jarlskog, \emph{{Commutator of the Quark Mass Matrices in the Standard
  Electroweak Model and a Measure of Maximal $CP$~Nonconservation}},
  \href{https://doi.org/10.1103/PhysRevLett.55.1039}{\emph{Phys. Rev. Lett.}
  {\bfseries 55} (1985) 1039}.

\bibitem{Morrissey:2012db}
D.E.~Morrissey and M.J.~Ramsey-Musolf, \emph{{Electroweak baryogenesis}},
  \href{https://doi.org/10.1088/1367-2630/14/12/125003}{\emph{New J. Phys.}
  {\bfseries 14} (2012) 125003}
  [\href{https://arxiv.org/abs/1206.2942}{{\ttfamily 1206.2942}}].

\bibitem{Kainulainen:2001cn}
K.~Kainulainen, T.~Prokopec, M.G.~Schmidt and S.~Weinstock, \emph{{First
  principle derivation of semiclassical force for electroweak baryogenesis}},
  \href{https://doi.org/10.1088/1126-6708/2001/06/031}{\emph{JHEP} {\bfseries
  06} (2001) 031} [\href{https://arxiv.org/abs/hep-ph/0105295}{{\ttfamily
  hep-ph/0105295}}].

\bibitem{Prokopec:2003pj}
T.~Prokopec, M.G.~Schmidt and S.~Weinstock, \emph{{Transport equations for
  chiral fermions to order h bar and electroweak baryogenesis. Part 1}},
  \href{https://doi.org/10.1016/j.aop.2004.06.002}{\emph{Annals Phys.}
  {\bfseries 314} (2004) 208}
  [\href{https://arxiv.org/abs/hep-ph/0312110}{{\ttfamily hep-ph/0312110}}].

\bibitem{Prokopec:2004ic}
T.~Prokopec, M.G.~Schmidt and S.~Weinstock, \emph{{Transport equations for
  chiral fermions to order h-bar and electroweak baryogenesis. Part II}},
  \href{https://doi.org/10.1016/j.aop.2004.06.001}{\emph{Annals Phys.}
  {\bfseries 314} (2004) 267}
  [\href{https://arxiv.org/abs/hep-ph/0406140}{{\ttfamily hep-ph/0406140}}].

\bibitem{Huet:1994jb}
P.~Huet and E.~Sather, \emph{{Electroweak baryogenesis and standard model CP
  violation}}, \href{https://doi.org/10.1103/PhysRevD.51.379}{\emph{Phys. Rev.
  D} {\bfseries 51} (1995) 379}
  [\href{https://arxiv.org/abs/hep-ph/9404302}{{\ttfamily hep-ph/9404302}}].

\bibitem{Huet:1995sh}
P.~Huet and A.E.~Nelson, \emph{{Electroweak baryogenesis in supersymmetric
  models}}, \href{https://doi.org/10.1103/PhysRevD.53.4578}{\emph{Phys. Rev. D}
  {\bfseries 53} (1996) 4578}
  [\href{https://arxiv.org/abs/hep-ph/9506477}{{\ttfamily hep-ph/9506477}}].

\bibitem{Riotto:1995hh}
A.~Riotto, \emph{{Towards a nonequilibrium quantum field theory approach to
  electroweak baryogenesis}},
  \href{https://doi.org/10.1103/PhysRevD.53.5834}{\emph{Phys. Rev. D}
  {\bfseries 53} (1996) 5834}
  [\href{https://arxiv.org/abs/hep-ph/9510271}{{\ttfamily hep-ph/9510271}}].

\bibitem{Lee:2004we}
C.~Lee, V.~Cirigliano and M.J.~Ramsey-Musolf, \emph{{Resonant relaxation in
  electroweak baryogenesis}},
  \href{https://doi.org/10.1103/PhysRevD.71.075010}{\emph{Phys. Rev. D}
  {\bfseries 71} (2005) 075010}
  [\href{https://arxiv.org/abs/hep-ph/0412354}{{\ttfamily hep-ph/0412354}}].

\bibitem{Postma:2019scv}
M.~Postma and J.~van~de Vis, \emph{{Source terms for electroweak baryogenesis
  in the vev-insertion approximation beyond leading order}},
  \href{https://doi.org/10.1007/JHEP02(2020)090}{\emph{JHEP} {\bfseries 02}
  (2020) 090} [\href{https://arxiv.org/abs/1910.11794}{{\ttfamily
  1910.11794}}].

\bibitem{Cline:2020jre}
J.M.~Cline and K.~Kainulainen, \emph{{Electroweak baryogenesis at high bubble
  wall velocities}},
  \href{https://doi.org/10.1103/PhysRevD.101.063525}{\emph{Phys. Rev. D}
  {\bfseries 101} (2020) 063525}
  [\href{https://arxiv.org/abs/2001.00568}{{\ttfamily 2001.00568}}].

\bibitem{Postma:2021zux}
M.~Postma, \emph{{A different perspective on the vev insertion approximation
  for electroweak baryogenesis}},
  \href{https://doi.org/10.1007/JHEP09(2021)055}{\emph{JHEP} {\bfseries 09}
  (2021) 055} [\href{https://arxiv.org/abs/2107.05971}{{\ttfamily
  2107.05971}}].

\bibitem{Cline:2021dkf}
J.M.~Cline and B.~Laurent, \emph{{Electroweak baryogenesis from light fermion
  sources: A critical study}},
  \href{https://doi.org/10.1103/PhysRevD.104.083507}{\emph{Phys. Rev. D}
  {\bfseries 104} (2021) 083507}
  [\href{https://arxiv.org/abs/2108.04249}{{\ttfamily 2108.04249}}].

\bibitem{Kainulainen:2021oqs}
K.~Kainulainen, \emph{{CP-violating transport theory for electroweak
  baryogenesis with thermal corrections}},
  \href{https://doi.org/10.1088/1475-7516/2021/11/042}{\emph{JCAP} {\bfseries
  11} (2021) 042} [\href{https://arxiv.org/abs/2108.08336}{{\ttfamily
  2108.08336}}].

\bibitem{Postma:2022dbr}
M.~Postma, J.~van~de Vis and G.~White, \emph{{Resummation and cancellation of
  the VIA source in electroweak baryogenesis}},
  \href{https://doi.org/10.1007/JHEP12(2022)121}{\emph{JHEP} {\bfseries 12}
  (2022) 121} [\href{https://arxiv.org/abs/2206.01120}{{\ttfamily
  2206.01120}}].

\bibitem{Cohen:1987vi}
A.G.~Cohen and D.B.~Kaplan, \emph{{Thermodynamic Generation of the Baryon
  Asymmetry}}, \href{https://doi.org/10.1016/0370-2693(87)91369-4}{\emph{Phys.
  Lett. B} {\bfseries 199} (1987) 251}.

\bibitem{Sagunski:2023ynd}
L.~Sagunski, P.~Schicho and D.~Schmitt, \emph{{Supercool exit: Gravitational
  waves from QCD-triggered conformal symmetry breaking}},
  \href{https://doi.org/10.1103/PhysRevD.107.123512}{\emph{Phys. Rev. D}
  {\bfseries 107} (2023) 123512}
  [\href{https://arxiv.org/abs/2303.02450}{{\ttfamily 2303.02450}}].

\bibitem{affleck1985new}
I.~Affleck and M.~Dine, \emph{A new mechanism for baryogenesis}, {\emph{Nuclear
  Physics B} {\bfseries 249} (1985) 361}.

\bibitem{Harvey:1990qw}
J.A.~Harvey and M.S.~Turner, \emph{{Cosmological baryon and lepton number in
  the presence of electroweak fermion number violation}},
  \href{https://doi.org/10.1103/PhysRevD.42.3344}{\emph{Phys. Rev. D}
  {\bfseries 42} (1990) 3344}.

\bibitem{Fukugita:1986hr}
M.~Fukugita and T.~Yanagida, \emph{{Baryogenesis Without Grand Unification}},
  \href{https://doi.org/10.1016/0370-2693(86)91126-3}{\emph{Phys. Lett. B}
  {\bfseries 174} (1986) 45}.

\bibitem{Luty:1992un}
M.A.~Luty, \emph{{Baryogenesis via leptogenesis}},
  \href{https://doi.org/10.1103/PhysRevD.45.455}{\emph{Phys. Rev. D} {\bfseries
  45} (1992) 455}.

\bibitem{Buchmuller:2004nz}
W.~Buchmuller, P.~Di~Bari and M.~Plumacher, \emph{{Leptogenesis for
  pedestrians}}, \href{https://doi.org/10.1016/j.aop.2004.02.003}{\emph{Annals
  Phys.} {\bfseries 315} (2005) 305}
  [\href{https://arxiv.org/abs/hep-ph/0401240}{{\ttfamily hep-ph/0401240}}].

\bibitem{Davidson:2008bu}
S.~Davidson, E.~Nardi and Y.~Nir, \emph{{Leptogenesis}},
  \href{https://doi.org/10.1016/j.physrep.2008.06.002}{\emph{Phys. Rept.}
  {\bfseries 466} (2008) 105}
  [\href{https://arxiv.org/abs/0802.2962}{{\ttfamily 0802.2962}}].

\bibitem{Flanagan:2005yc}
E.E.~Flanagan and S.A.~Hughes, \emph{{The Basics of gravitational wave
  theory}}, \href{https://doi.org/10.1088/1367-2630/7/1/204}{\emph{New J.
  Phys.} {\bfseries 7} (2005) 204}
  [\href{https://arxiv.org/abs/gr-qc/0501041}{{\ttfamily gr-qc/0501041}}].

\bibitem{Maggiore:2007ulw}
M.~Maggiore, \emph{{Gravitational Waves. Vol. 1: Theory and Experiments}},
  Oxford Master Series in Physics, Oxford University Press (2007).

\bibitem{Witten:1984rs}
E.~Witten, \emph{{Cosmic Separation of Phases}},
  \href{https://doi.org/10.1103/PhysRevD.30.272}{\emph{Phys. Rev. D} {\bfseries
  30} (1984) 272}.

\bibitem{Hogan:1986qda}
C.J.~Hogan, \emph{{Gravitational radiation from cosmological phase
  transitions}}, {\emph{Mon. Not. Roy. Astron. Soc.} {\bfseries 218} (1986)
  629}.

\bibitem{Jinno:2016vai}
R.~Jinno and M.~Takimoto, \emph{{Gravitational waves from bubble collisions: An
  analytic derivation}},
  \href{https://doi.org/10.1103/PhysRevD.95.024009}{\emph{Phys. Rev. D}
  {\bfseries 95} (2017) 024009}
  [\href{https://arxiv.org/abs/1605.01403}{{\ttfamily 1605.01403}}].

\bibitem{Hindmarsh:2016lnk}
M.~Hindmarsh, \emph{{Sound shell model for acoustic gravitational wave
  production at a first-order phase transition in the early Universe}},
  \href{https://doi.org/10.1103/PhysRevLett.120.071301}{\emph{Phys. Rev. Lett.}
  {\bfseries 120} (2018) 071301}
  [\href{https://arxiv.org/abs/1608.04735}{{\ttfamily 1608.04735}}].

\bibitem{Hindmarsh:2019phv}
M.~Hindmarsh and M.~Hijazi, \emph{{Gravitational waves from first order
  cosmological phase transitions in the Sound Shell Model}},
  \href{https://doi.org/10.1088/1475-7516/2019/12/062}{\emph{JCAP} {\bfseries
  12} (2019) 062} [\href{https://arxiv.org/abs/1909.10040}{{\ttfamily
  1909.10040}}].

\bibitem{Cai:2023guc}
R.-G.~Cai, S.-J.~Wang and Z.-Y.~Yuwen, \emph{{Hydrodynamic sound shell model}},
  \href{https://doi.org/10.1103/PhysRevD.108.L021502}{\emph{Phys. Rev. D}
  {\bfseries 108} (2023) L021502}
  [\href{https://arxiv.org/abs/2305.00074}{{\ttfamily 2305.00074}}].

\bibitem{Hindmarsh:2013xza}
M.~Hindmarsh, S.J.~Huber, K.~Rummukainen and D.J.~Weir, \emph{{Gravitational
  waves from the sound of a first order phase transition}},
  \href{https://doi.org/10.1103/PhysRevLett.112.041301}{\emph{Phys. Rev. Lett.}
  {\bfseries 112} (2014) 041301}
  [\href{https://arxiv.org/abs/1304.2433}{{\ttfamily 1304.2433}}].

\bibitem{Hindmarsh:2015qta}
M.~Hindmarsh, S.J.~Huber, K.~Rummukainen and D.J.~Weir, \emph{{Numerical
  simulations of acoustically generated gravitational waves at a first order
  phase transition}},
  \href{https://doi.org/10.1103/PhysRevD.92.123009}{\emph{Phys. Rev. D}
  {\bfseries 92} (2015) 123009}
  [\href{https://arxiv.org/abs/1504.03291}{{\ttfamily 1504.03291}}].

\bibitem{Hindmarsh:2017gnf}
M.~Hindmarsh, S.J.~Huber, K.~Rummukainen and D.J.~Weir, \emph{{Shape of the
  acoustic gravitational wave power spectrum from a first order phase
  transition}}, \href{https://doi.org/10.1103/PhysRevD.96.103520}{\emph{Phys.\
  Rev.} {\bfseries D96} (2017) 103520}
  [\href{https://arxiv.org/abs/1704.05871}{{\ttfamily 1704.05871}}].

\bibitem{Enqvist:1991xw}
K.~Enqvist, J.~Ignatius, K.~Kajantie and K.~Rummukainen, \emph{{Nucleation and
  bubble growth in a first order cosmological electroweak phase transition}},
  \href{https://doi.org/10.1103/PhysRevD.45.3415}{\emph{Phys. Rev. D}
  {\bfseries 45} (1992) 3415}.

\bibitem{Guth:1981uk}
A.H.~Guth and E.J.~Weinberg, \emph{{Cosmological Consequences of a First Order
  Phase Transition in the SU(5) Grand Unified Model}},
  \href{https://doi.org/10.1103/PhysRevD.23.876}{\emph{Phys. Rev. D} {\bfseries
  23} (1981) 876}.

\bibitem{LISACosmologyWorkingGroup:2022jok}
{\scshape LISA Cosmology Working Group} collaboration, \emph{{Cosmology with
  the Laser Interferometer Space Antenna}},
  \href{https://arxiv.org/abs/2204.05434}{{\ttfamily 2204.05434}}.

\bibitem{NANOGrav:2023gor}
{\scshape NANOGrav} collaboration, \emph{{The NANOGrav 15-year Data Set:
  Evidence for a Gravitational-Wave Background}},
  \href{https://arxiv.org/abs/2306.16213}{{\ttfamily 2306.16213}}.

\bibitem{Antoniadis:2023ott}
J.~Antoniadis et~al., \emph{{The second data release from the European Pulsar
  Timing Array III. Search for gravitational wave signals}},
  \href{https://arxiv.org/abs/2306.16214}{{\ttfamily 2306.16214}}.

\bibitem{Reardon:2023gzh}
D.J.~Reardon et~al., \emph{{Search for an isotropic gravitational-wave
  background with the Parkes Pulsar Timing Array}},
  \href{https://arxiv.org/abs/2306.16215}{{\ttfamily 2306.16215}}.

\bibitem{Xu:2023wog}
H.~Xu et~al., \emph{{Searching for the nano-Hertz stochastic gravitational wave
  background with the Chinese Pulsar Timing Array Data Release I}},
  \href{https://arxiv.org/abs/2306.16216}{{\ttfamily 2306.16216}}.

\bibitem{Weir:2017wfa}
D.J.~Weir, \emph{{Gravitational waves from a first order electroweak phase
  transition: a brief review}},
  \href{https://doi.org/10.1098/rsta.2017.0126}{\emph{Phil. Trans. Roy. Soc.
  Lond. A} {\bfseries 376} (2018) 20170126}
  [\href{https://arxiv.org/abs/1705.01783}{{\ttfamily 1705.01783}}].

\bibitem{Ellis:2019oqb}
J.~Ellis, M.~Lewicki, J.M.~No and V.~Vaskonen, \emph{{Gravitational wave energy
  budget in strongly supercooled phase transitions}},
  \href{https://doi.org/10.1088/1475-7516/2019/06/024}{\emph{JCAP} {\bfseries
  06} (2019) 024} [\href{https://arxiv.org/abs/1903.09642}{{\ttfamily
  1903.09642}}].

\bibitem{Guo:2020grp}
H.-K.~Guo, K.~Sinha, D.~Vagie and G.~White, \emph{{Phase Transitions in an
  Expanding Universe: Stochastic Gravitational Waves in Standard and
  Non-Standard Histories}},
  \href{https://doi.org/10.1088/1475-7516/2021/01/001}{\emph{JCAP} {\bfseries
  01} (2021) 001} [\href{https://arxiv.org/abs/2007.08537}{{\ttfamily
  2007.08537}}].

\bibitem{Caprini:2009fx}
C.~Caprini, R.~Durrer, T.~Konstandin and G.~Servant, \emph{{General Properties
  of the Gravitational Wave Spectrum from Phase Transitions}},
  \href{https://doi.org/10.1103/PhysRevD.79.083519}{\emph{Phys. Rev. D}
  {\bfseries 79} (2009) 083519}
  [\href{https://arxiv.org/abs/0901.1661}{{\ttfamily 0901.1661}}].

\end{thebibliography}\endgroup

\end{document}